\documentclass[12pt]{article}

\usepackage{subcaption}
\usepackage[dvipsnames]{xcolor}
\usepackage[totalheight = 21cm, totalwidth = 15cm]{geometry}
\usepackage{amssymb,amsmath,amsfonts,amsbsy,graphicx,ctable,colortbl,float}
\usepackage{bm,color}
\usepackage{cite}
\usepackage[colorlinks=true,urlcolor=MidnightBlue,linkcolor=MidnightBlue,
citecolor=OliveGreen,pdfpagelabels=true,hypertexnames=true,plainpages=false,
naturalnames=false]{hyperref}

\usepackage{epsfig}

\newcommand{\nn}{{\nonumber}}
\newcommand{\cellb}{{\bar{C}_\ell}}
\newcommand{\covl}{{\left(\text{Cov}_\ell\right)}}

\newcommand{\ltsima}{$\; \buildrel < \over \sim \;$}
\newcommand{\lsim}{\lower.5ex\hbox{\ltsima}}
\newcommand{\gsim}{\lower.5ex\hbox{\gtsima}}

\newcommand{\dd}{\mathrm{d}}

\newcommand{\be}{\begin{equation}}
\newcommand{\ee}{\end{equation}}
\newcommand{\bea}{\begin{eqnarray}}
\newcommand{\eea}{\end{eqnarray}}

\begin{document}

\begin{titlepage}

\begin{center}

{\Large  \bf Constraints on inf{l}ation with LSS surveys: features in the primordial power spectrum}

\vskip 1.0cm

{\large
Gonzalo A. Palma, Domenico Sapone and Spyros Sypsas
}

\vskip 0.5cm

{\it
Departamento de F\'isica, FCFM, Universidad de Chile, \\ Blanco Encalada 2008, Santiago, Chile
}

\vskip 1.2cm

\end{center}

\begin{abstract}

We analyse the efficiency of future large scale structure surveys to unveil the presence of scale dependent features in the primordial spectrum --resulting from cosmic inflation-- imprinted in the distribution of galaxies. Features may appear as a consequence of non-trivial dynamics during cosmic inflation, in which one or more background quantities experienced small but rapid deviations from their characteristic slow-roll evolution. We consider two families of features: localised features and oscillatory extended features. To characterise them we employ various possible templates parametrising their scale dependence and provide forecasts on the constraints on these parametrisations for LSST like surveys. We perform a Fisher matrix analysis for three observables: cosmic microwave background (CMB), galaxy clustering and weak lensing. We find that the combined data set of these observables will be able to limit the presence of features down to levels that are more restrictive than current constraints coming from CMB observations only. In particular, we address the possibility of gaining information on currently known deviations from scale invariance inferred from CMB data, such as the feature appearing at the $\ell \sim 20$ multipole (which is the main contribution to the low-$\ell$ deficit) and another one around $\ell \sim 800$.

\end{abstract}

\end{titlepage}

\hypersetup{linktocpage}
%\hrulefill
\noindent\makebox[\linewidth]{\rule{\textwidth}{1.3pt}}
%\vspace{-20pt}
\tableofcontents
\noindent\makebox[\linewidth]{\rule{\textwidth}{1.3pt}}
%\hrulefill

%\newpage

\renewcommand{\theequation}{\arabic{section}.\arabic{equation}}
\setcounter{page}{1}

\section{Introduction} \label{s1:Intro}

The next generation of large scale structure surveys (LSS) will give us the chance to test our ideas about the primordial universe in a way complementary to cosmic microwave background (CMB) observations. At the moment, cosmic inflation~\cite{Guth:1980zm,Starobinsky:1980te} constitutes our best theoretical framework to understand the origin of the primordial curvature fluctuations that gave rise to both the observed CMB temperature anisotropies and the LSS of our universe. The simplest version of cosmic inflation, known as single-field slow-roll inflation~\cite{Linde:1981mu,Albrecht:1982wi}, predicts primordial curvature perturbations following an almost Gaussian distribution parametrised by a nearly scale-invariant power spectrum. So far, these characteristics have been found to be fully consistent with both CMB and LSS observations~\cite{planck,Ade:2015ava}.

A prevailing hope within the study of inflation is that future observations involving primordial perturbations will reveal definite information about the ultra-violet (UV) complete theory that hosts the inflationary phase as a consistent mechanism. For instance, a detection of B-modes in the CMB polarisation in the near future would reveal a new energy scale not too far from the grand unification scale, inviting us to speculate about the structure of a more fundamental theory accounting for both inflation and the standard model of particle physics. Other observations ---such as departures from scale invariance and/or Gaussianity--- would force us to leave behind the simplest models of inflation, and would reveal nontrivial information about the laws of physics that took place during the early universe by introducing further new energy scales. 

Since the first versions of single-field slow-roll inflation were introduced, there has been some progress in understanding the variety of realisations of inflation within fundamental theories ---such as string theory and supergravity--- that bring together general relativity and quantum field theory. All of these theoretical realisations have in common the existence of a large number of scalar degrees of freedom that could have been dynamically relevant during inflation. It is now well understood that the interaction between the inflaton field and other degrees of freedom may lead to particular events during the inflationary history resulting in the appearance of features, i.e. departures from scale invariance, in the spectra of primordial fluctuations. Roughly speaking, these features could be categorised in two families: (a) localised features, namely, scale invariance departures that privilege a particular scale, and (b) extended oscillatory features, which are departures that spread along the entire observable spectra.

Examples of models leaving features in the spectra include: (1) models where the slow-roll parameters experience brief deviations from slow-roll, coming from features in the potential (e.g. steps and bumps in the potential)~\cite{Starobinsky:1992ts,Adams:2001vc,Gong:2005jr,Chen:2006xjb, Hazra:2014goa, Hazra:2014jka, Hazra:2016fkm ,Cadavid:2015iya, GallegoCadavid:2017bzb, GallegoCadavid:2017pol, GallegoCadavid:2016wcz}; (2) multi-field models of inflation with a meandering inflationary trajectory~\cite{Achucarro:2010jv, Achucarro:2010da,Chen:2012ge,Pi:2012gf,Noumi:2013cfa,Burgess:2012dz,Battefeld:2013xka,Konieczka:2014zja,Mizuno:2014jja,Saito:2012pd,Gao:2013ota,Cespedes:2012hu,Shiu:2011qw,Gao:2012uq}; (3) particle production during inflation for a brief period of time~\cite{Chung:1999ve, Elgaroy:2003hp, Mathews:2004vu, Romano:2008rr, Barnaby:2009mc, Barnaby:2009dd, Barnaby:2010ke}; (4) $P(X)$-models with features in the kinetic terms of the inflaton field~\cite{Mooij:2015cxa}; (5) string models of inflation~\cite{Kobayashi:2012kc,Ashoorioon:2008qr,Cai:2015xla,Bean:2008na}; (6) string and gauge theory models involving axions~\cite{Wang:2002hf,Flauger:2009ab,Gariazzo:2017akm}. In addition, efforts have been made to study the appearance of features in the spectra in a model independent manner~\cite{Achucarro:2013cva, Achucarro:2014msa, Mooij:2016dsi, Palma:2016wqu, Gong:2014spa, Gong:2017vve, Gao:2015aba, Gong:2017yih, Gariazzo:2016blm, Gariazzo:2015qea, Appleby:2015bpw}, see Ref.~\cite{Chluba:2015bqa} for a review.

The latest Planck results marginally support the existence of deviations from scale-invariance localised at different multipoles in the angular power spectrum~\cite{Hazra:2013nca,Hazra:2014jwa,Hunt:2015iua}, that could be due to features in the primordial power spectrum. Features in the primordial spectra are excesses over the $\Lambda$CDM line, very much like the bumps over the standard model background seen in the LHC data. They are present in both Planck and WMAP, and can be thought of as indicators of new degrees of freedom at the inflationary energy scale~\cite{Arkani-Hamed:2015bza}. Therefore, in order to increase the statistical significance of these bumps we have to look at different channels, just like in particle physics, where one looks at different decay amplitudes. In cosmology, these correspond to power spectra and bispectra of primordial (CMB) and late-time fields (LSS).

However, the sensitivity of CMB surveys is limited by several factors. Apart from cosmic variance, the CMB contains information emitted from the two dimensional surface of last scattering, limiting the number of modes that we have access to, hence limiting the size of the sample with which we are able to perform statistical analysis. Nonetheless, forthcoming LSS surveys such as Euclid~\cite{Euclid-web} and LSST~\cite{Lsst-web}, will give us access to a three dimensional volume of modes which is expected to improve upon the current CMB results.
 
The purpose of this article is to anticipate how LSS surveys, specifically LSST-like ones, will help us constrain, or even unveil, the existence of features in the primordial spectrum. Similarly to the CMB angular power spectrum, the matter power spectrum that parametrises the distribution of galaxies in our universe results from the initial conditions of the Hot Big-Bang universe delivered by inflation. Thus, if features were present in the primordial power spectrum, traces of them should be present in LSS observables. In view of the coming surveys, Refs.~\cite{Huang:2012mr,Chen:2016vvw,Ballardini:2016hpi,Fard:2017oex} have recently initiated the study of features in the LSS signal of Euclid- and LSST-like surveys\footnote{See also \cite{Xu:2016kwz,Chen:2016zuu,Pourtsidou:2016ctq} for proposals of constraining features using the 21cm signal.}. In this work we extend this line of research by forecasting constraints on inflationary features obtained by LSST-like surveys. More specifically, we use feature templates of the primordial power spectrum available in the literature and constrain them using three observables, namely galaxy clustering (GC), weak lensing (WL) and CMB power spectra. We construct Fisher matrices for each one and forecast errors for the amplitude, position, width, frequency and phase of primordial inflationary features, while marginalising over the $\Lambda$CDM parameters.  We find that the addition of WL to the CMB+GC forecasts of Ref.~\cite{Chen:2016vvw} considerably improves constraints on the various parameters involved in the feature templates.

The present paper is organised as follows: In Section~\ref{s2:Templates} we present the templates used to model features in the primordial power spectrum. In Secction~\ref{s3:probes} we discuss the CMB and LSS probes we use to forecast errors in Section~\ref{s4:forecasts}. Finally, we offer some concluding remarks in Section~\ref{s5:conclusions}.

\setcounter{equation}{0} 
\section{Features in the primordial spectra} \label{s2:Templates}

The possible signatures of new physics in the primordial spectra may be roughly divided into two classes:  features localised around certain multipoles, and oscillatory features that extend over a range of wavenumbers. In this section we present the five templates that will be used for the forecasts.

In many relevant contexts, features in the primordial spectra are the result of rapid variations in the evolution of background quantities during inflation, that are communicated to cosmological perturbations around horizon crossing. A fairly powerful framework to study the appearance of this type of features, in a model independent way, consists of the effective field theory of inflation~\cite{Cheung:2007st}. In this scheme, the evolution of perturbations is sensitive to two important background quantities that effectively capture the background dynamics of a large variety of UV complete models:\footnote{Here, by UV complete models we mean models that have a UV cutoff scale much larger than the Hubble expansion rate, and not necessarily complete all the way up to the more fundamental Planck scale.} The first parameter is the Hubble expansion rate $H$, and slow-roll parameters derived from it, such as $\epsilon \equiv -\dot H / H^2$ and $\eta \equiv \dot \epsilon / (H \epsilon)$. The second parameter is the sound speed $c_s$, and quantities derived from it such as its running $s \equiv \dot c_s / (H c_s)$ (the value $c_s=1$ corresponds to single-field canonical inflation). In the EFT of inflation, features in the primordial power spectrum may be traced back to the sudden time variation of $H$ and $c_s$. In order to study the emergence of features for a given model, the time dependence of these parameters must be obtained by solving the background equations of motions. However, for the sake of simplicity, one may follow the simpler (but more limited approach) of assuming a specific time dependence. 

If $H$ and $c_s$ experience rapid sudden variations, but in such a way that the amplitudes of these variations remain small, it is possible to derive a simple relation linking background quantities and the features in the power spectrum $\Delta \mathcal P$, given by~\cite{Palma:2014hra}:
\be
\frac{\Delta \mathcal P}{\mathcal P_0 } (k) = \frac{1}{8 k} \int_{-\infty}^{+\infty} \!\!\!\!\! d \tau \, \left[ (1 - c_s^2)''  - \frac{2}{ \tau}  \eta '    \right] \, \sin (2 k \tau) \, , \label{features-power-time}
\ee
where primes $(')$ represent derivatives with respect to conformal time $\tau$. In the previous expression, ${\mathcal P_0 }$ denotes the featureless power spectrum obtained in standard single-field inflation:
\be 
{\mathcal P_0 } = A_s \left( \frac{k}{k_0}\right)^{n_s-1}.
\ee
The formula \eqref{features-power-time} offers a simple way to connect features in the power spectrum and the behaviour of the background felt by the perturbations during inflation. Analytic expressions alternative to (\ref{features-power-time}), leading to essentially the same link between features and the variation of background quantities can be found in the literature, with slight differences coming from the formalism employed to derive them. For instance, the method to derive (\ref{features-power-time}) has been dubbed the slow-roll Fourier transform (SRFT) approximation, and was first used in Ref.~\cite{Achucarro:2012fd} to derive an expression for $\Delta \mathcal P / \mathcal P_0$ due to variations of $c_s$ only. Alternatively, other works have analysed the generation of features in the power spectrum using the so called generalised slow-roll (GSR) method \cite{Choe:2004zg}. For example, in~\cite{Dvorkin:2009ne,Appleby:2015bpw} the GSR method was used to derive an analytical expression for $\Delta \mathcal P / \mathcal P_0$ due to sudden variations of $H$ that would allow to study the features generated by a step in the potential~\cite{Adshead:2011jq}. On the other hand, in Ref.~\cite{Achucarro:2014msa} it was used to derive an expression capturing the effects of sudden variations of $c_s$. A thorough discussion comparing the two methods (SRFT and GRS) can be found in Ref.~\cite{Achucarro:2014msa} for the particular case where features are produced by changes in the sound speed $c_s$. In that work, the authors conclude that both methods agree to a high degree of accuracy. 

Note however, that the features studied with these methods cannot be too sharp. In such a case, the EFT starts being invalid since the slow-roll parameters grow in amplitude and the decoupling limit cannot be implemented anymore. GSR holds slightly beyond this regime but for an extremely sharp feature perturbation theory in general breaks down and both approaches fail to capture the dynamics of perturbations on such a background.

Another context in which features in the power spectrum may appear, and that cannot be treated in terms of a single-field EFT description, is related to the sudden production of particles during inflation able to backreact on the evolution of the inflaton perturbations~\cite{Chung:1999ve}. For example, if the inflaton is coupled to a second field that suddenly becomes massless during inflation, the latter can generate bursts of quanta that may backreact on the dynamics of inflaton perturbations~\cite{Barnaby:2009mc}. In this case, obtaining an analytic expression for the power spectrum such as~(\ref{features-power-time}) is less trivial. However, generically the features resulting from particle production correspond to bumps, which are easy to fit with simple functions of the scale.

In what follows we proceed to present five templates for $\Delta \mathcal P / \mathcal P_0$ deduced by different authors, studying different types of phenomena giving rise to features. Most of them come from an effective description in which perturbations are affected by the sudden evolution of background quantities, as expressed in Eq.~(\ref{features-power-time}). The third template, appears within the context of particle production during inflation. These templates will be used in the following sections as reference models to analyse the search of features in an LSST-like survey.

Notice that three out of these five templates (III-V) have already been considered\footnote{Template II was also analysed in \cite{Chen:2016vvw}. We have kept the same fiducial values, consistent with the Planck constraints. In the present work though, we have added a correction to the template (See Eq.~\eqref{step} and the discussion around it) and we have also considered another set of fiducial values that corresponds to a sharper feature. This template has also appeared in the analysis \cite{Ballardini:2017qwq}, and their results agree with our CMB+GC constraints.} in Ref.~\cite{Chen:2016vvw} for the particular case of CMB+GC. Before discussing them in detail, let us clarify that the fiducial amplitudes we have used are the Planck best fits~\cite{Ade:2015lrj}. %In some cases though, following Ref.~\cite{Chen:2016vvw}, we have used values that have been excluded by Planck at high confidence level\footnote{For example, the $1\%$ relative amplitude we have used for the $k_d=0.1 \;\text{Mpc}^{-1}$ case of Template III (c.f. Eq.~\eqref{bump-fid}), translates to a $0.5\%$ amplitude around $\ell \sim 1000$ in multipole space. This is $5\sigma$ away from the Planck corresponding constraint.}.
In order to discuss the improvement of forecasts upon adding WL data to CMB+GC, we have mostly kept the same fiducial values for the parameters of the corresponding templates as in Ref.~\cite{Chen:2016vvw}. As we shall see, our CMB+GC results correctly reproduce those cases, while CMB+GC+WL considerably reduces the forecasted errorbars for some features.

%Finally, before passing to the discussion of specific templates let us clarify that the fiducial amplitudes we have used are comparable to the Planck best fits~\cite{Ade:2015lrj}. In some cases though we use values that have been excluded by Planck at high confidence level\footnote{For example, the $1\%$ relative amplitude we have used for the $k\sim0.1 \;\text{Mpc}^{-1}$ case of Template III, translates to a $0.5\%$ amplitude around $\ell \sim 1000$ in multipole space. This is five times the Planck $0.1\%$ precision in this area.}. Had we chosen values strictly within the Planck errorbars, the forecasted errors would have been slightly worse, without, however, invalidating our argument, since this would affect GC and WL in the same way.

\subsection{Template I}

\begin{figure}[h]
\centering
\epsfig{figure=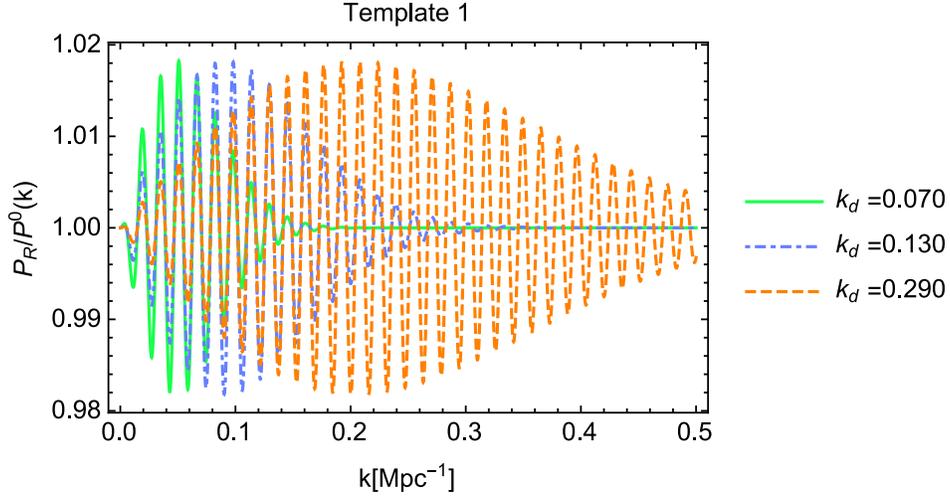,width=5in}
\caption{Plot of template 1 of Eq.~\eqref{temp1} for the values quoted in Eq.~\eqref{temp1-fid}.}
\label{fig:temp1}
\end{figure}

The authors of Ref.~\cite{Achucarro:2014msa} considered various classes of time variations of the sound speed $c_s$ to deduce useful templates that would allow the study of features in the CMB~\cite{Torrado:2016sls}. One particular case of interest corresponds to the situation where the sound speed varies with a Gaussian profile of the form:
\be
1 - c_s^2  = \Delta c \times \exp (- \beta^2 \left[\ln (\tau / \tau_f) \right]^2 ) .
\ee
Here the time dependent behavior for $c_s$ is dictated by three parameters: $\Delta c$ tells us how large is the departure of $c_s$ from unity, $\tau_f$ corresponds to the conformal time at which the departure becomes maximal, and $\beta^{-1}$ gives the duration of the departure in units of $H^{-1}$. Using this behavior, the authors of Ref.~\cite{Achucarro:2014msa} were able to deduce a power spectrum of the form
\be \label{temp1}
\frac{\Delta \mathcal P}{\mathcal P_0 } (k) =  \left[ \sin(2 k / k_f) + \frac{k_f}{k} \cos(2 k / k_f) \right] D (k) - \frac{1}{2}  \frac{k_f^2}{k} \sin(2 k / k_f)  \frac{d}{d  k} D (k) ,
\ee
where $D (k)$ is an envelope function (centered at $k = 0$) given by
\be
D (k) = \frac{4\sqrt{\pi}C}{36} \frac{ k}{k_d} \exp \left\{ -  \frac{ k^2}{k_d^2} \right\} ,
\ee
with $k_d = |\beta / \tau_f|$. In the previous expressions, the parameter $C$ determines the amplitude of the feature. On the other hand, $k_f = 1 / |\tau_f|$ parametrises the lengthscale of the oscillatory behavior and determines the scale at which the feature starts to appear in the spectrum. Finally, the parameter $k_d$ parametrises the length of the tail of the feature. When $k_f \ll k_d$, then $k_d$ roughly determines the location at which the feature peaks. We will use the following fiducial values~\cite{Achucarro:2014msa} for the parameters:
\be \label{temp1-fid}
C= 0.215, \quad k_d \;= (0.07\,;\,0.13\,;\, 0.29) \; {\rm Mpc}^{-1}, \quad k_f = 1/200 \; {\rm Mpc}^{-1} \; ,
\ee
for which we plot the template in Fig.~\ref{fig:temp1}. The $k_d$'s are chosen so that the template peaks around the positions of the expected features at $k=(0.05\,;\,0.1\,;\, 0.2) \; {\rm Mpc}^{-1}$, as can be seen in Fig.~\ref{fig:temp1}.

The authors of Ref.~\cite{Achucarro:2014msa} also considered other possible behaviors for $c_s$, leading to other classes of templates, e.g. localised oscillatory ones. Such a signal also belongs to the class of standard clock templates~\cite{Chen:2014joa,Chen:2014cwa}, constrained in~\cite{Chen:2016vvw}, which is however much broader, since it interpolates between sharp and resonant clock features. In order to not overload our results we choose to study the one written in Eq.~\eqref{temp1}.

\subsection{Template II}

The second template to consider models the effects of a step in the inflationary potential on the power spectrum. In Ref.~\cite{Adshead:2011jq} the authors studied the effect of a step in the potential of the form
\be \label{Vstep}
V(\phi) = V_0(\phi) \left[ 1 - C_0 \tanh \left(\frac{\phi_f - \phi}{d} \right) \right] ,
\ee
where $C$ and $d$ denote the hight and length of the step, respectively. On the other hand, $\phi_f$ denotes the location of the step in field space. The authors of Ref.~\cite{Adshead:2011jq} were able to integrate the background equations of motion to show that with such a step-feature, the slow-roll parameter $\epsilon = - \dot H / H^2$ is given by
\be
\epsilon (\tau) = \epsilon_0 (\tau)  +  \frac{ 6 C_0}{2}  \left( \frac{\tau}{\tau_f} \right)^3  \left[ 1 - \tanh \left( \frac{\sqrt{2\epsilon_0}M_{\rm Pl}}{d} \ln \frac{\tau}{\tau_f}\right) \right]\,, \label{epsilon-tau-1}
\ee
where $\tau_f$ is the conformal time at which the inflaton encounters the steps (determined by the value of $\phi_f$). Then, using the generalised slow-roll formalism it is possible to find a power spectrum with features of the form \cite{Miranda:2013wxa}
\be
\frac{\Delta \mathcal P}{\mathcal P_0 } (k) = \exp \left\{ f_0(k) \right\}(1+f_1(k)^2) -1, \label{step}
\ee
where $f_0(k)$ and $f_1(k)$ are given by
\be
f_0(k)  = \left[ \left( -3+\frac{9 k_f^2}{k^2 } \right)\cos(2 k /k_f) + \left(15 - \frac{9 k_f^2}{k^2} \right)\frac{\sin(2k / k_f)}{2k /k_f}  \right]  D(k) ,
\ee
\bea
f_1(k)  &= \dfrac{1}{\sqrt{2}} \Bigg[ \dfrac{\pi}{2}(1-n_s) -3\dfrac{k_f^3}{k^3}\left( \dfrac{ k}{k_f } \cos( k /k_f) - \sin( k /k_f)  \right)  \nn\\& \left(3 \dfrac{ k}{k_f } \cos( k /k_f) +\left( 2\dfrac{ k^2}{k_f^2 }-3 \right) \sin( k /k_f) \right)  \Bigg]  D(k) ,
\eea
while $D(k)$ is an envelope function:
\be
D(k) = C \frac{k /k_d}{\sinh(k /k_d)} .
\ee 
$f_1$ is a second order correction in the GSR expansion, which is subleading with respect to $f_0$. However, as shown in \cite{Miranda:2013wxa}, its inclusion is necessary given Planck's precision.

\begin{figure}[h]
\centering
\epsfig{figure=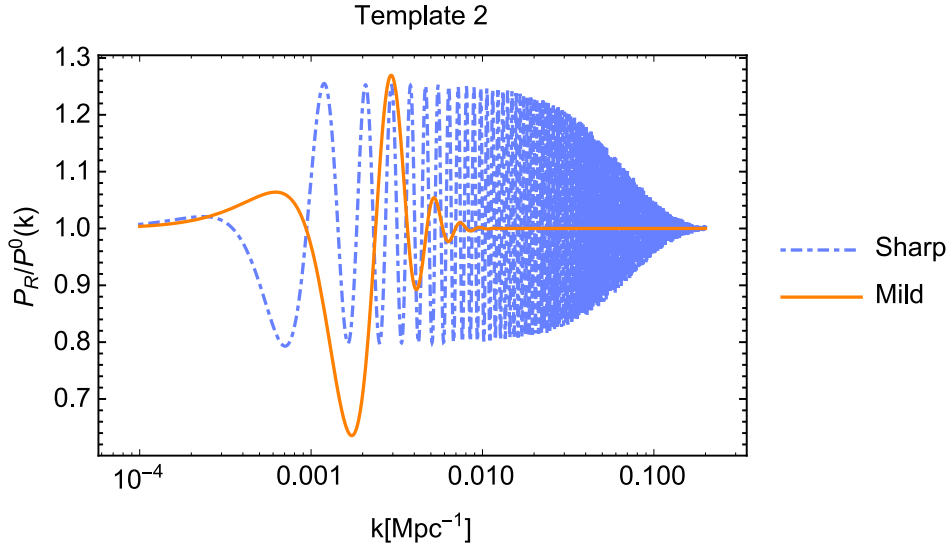,width=5in}
\caption{Plot of template 2 of Eq.~\eqref{step} for the two sets of fiducial values quoted in Eqs.~\eqref{step-fid-m} (mild) and \eqref{step-fid-s} (sharp).}
\label{fig:temp2}
\end{figure}

The advantage of (\ref{step}) is that it allows one to study features with fairly large amplitudes (up to $C \simeq 2/3$). In the case where\footnote{See Ref.~\cite{Miranda:2015cea} for templates valid for larger amplitudes.} $C \ll 2/3$ one recovers $\frac{\Delta \mathcal P}{\mathcal P_0 } (k) = f(k)$, which agrees with the analytical expression~(\ref{features-power-time}) after $\eta(\tau)$ derived from (\ref{epsilon-tau-1}) is inserted. The roles of $C$, $k_f$ and $k_d$ are essentially the same as in our first template: $C = 6 C_0 / \epsilon_0$ is the amplitude of the feature, $k_f = 1 / |\tau_f|$ parametrises the lengthscale of the oscillatory pattern, and $k_d =  \sqrt{2 \epsilon_0} M_{\rm Pl} / (\pi d |\tau_f|)$ determines the length of the tail. In our analysis we will consider two sets of fiducial values for these parameters: one is the best fit of Planck~\cite{Ade:2015lrj} and matches the one considered in Ref.~\cite{Chen:2016vvw}:
\be \label{step-fid-m}
C= 0.218, \quad  k_f= 0.00069\;\text{Mpc}^{-1}, \quad k_d= 0.0011\;\text{Mpc}^{-1} ,
\ee
chosen to reproduce the low-$\ell$ deficit found in the CMB angular power spectrum\footnote{The feature around $\ell=20$ has been also studied in the context of bouncing cosmology in~\cite{Cai:2017pga}.}. The other set is taken from~\cite{Miranda:2013wxa} and reads
\be \label{step-fid-s}
C= 0.075, \quad  k_f= 2.7\times10^{-4}\;\text{Mpc}^{-1}, \quad k_d= 2.8\times10^{-2}\;\text{Mpc}^{-1}.
\ee
The reason for the two sets is the following: the sharpness of the feature is controled by the ratio $k_d/k_f$; the higher the value, the more drastic is the change of the background parameters within an e-fold. The fiducial values from the Planck collaboration correspond to a feature which is not sharp enough, in the sense that the Hubble parameters change considerably at a scale of an e-fold. In this case of a mild feature, the analytical template and the exact numerical solution of the Mukhanov-Sasaki equation deviate considerably from each other~\cite{Miranda:2013wxa,Achucarro:2014msa}. This means that in this regime, the template~\eqref{step} has little to do with the theoretical model (Eqs.~\eqref{Vstep} and \eqref{epsilon-tau-1}) that is supposed to lie behind the appearance of such a feature. However, whatever the theory might be Planck uses this template and constrains it. It is a challenge for theorists to find a mechanism that produces a mild feature, which could be reliably modelled by such a template. The other set of fiducial values corresponds to a truly sharp feature, which can be safely attributed to a step in the inflaton potential.
%\textbf{Planck values from 1502.02114:}
%%
%\be \label{step-fid-P}
%C= 0.374, \quad  k_f= 0.00079\;\text{Mpc}^{-1}, \quad k_d= 0.0011\;\text{Mpc}^{-1} .
%\ee
%%
We plot the template obtained with these values in Fig.~\ref{fig:temp2}. 

\subsection{Template III}

The third template that we consider was derived in Ref.~\cite{Barnaby:2009mc} in order to model features resulting from particle production during inflation. As already explained, in this case features are not directly generated by variations of the background, as with templates I and II. Instead, they are sourced by the sudden production of quanta of a field coupled to the inflaton. In general, the features generated by particle production are bumpy, and are well fit by the following scale dependent function:
\be \label{bump}
\frac{\Delta \mathcal P}{\mathcal P_0 } (k) = C\left(\frac{\pi e}{3}\right)^{\frac{3}{2}}\left(\frac{k}{k_d}\right)^3 e^{-\frac{\pi}{2}\frac{k^2}{k_d^2}}.
\ee
This template depends only on two parameters: $C$ denotes the amplitude of the feature and $k_d$ its position in $k$-space. Notice that these two parameters accomplish the same role as those introduced in the previous two templates, once we exclude $k_f$. Thus, the main difference between this template and the previous two, is the absence of oscillatory patterns, parametrised by $k_f$.  We choose fiducial values for the two parameters as:
\be \label{bump-fid}
C= 0.002, \; k_d = (0.05;0.1;0.2) \; \text{Mpc}^{-1}.
\ee
Note that this type of bumpy feature has a lower amplitude than the other cases. The reason is that around $k=0.1$ Mpc$^{-1}$, which corresponds to $\ell\sim1000$, Planck has an accuracy\footnote{We thank Vinicius Miranda and Eiichiro Komatsu for pointing this out.} of $0.1\%$. We thus need to make sure that our choise respects this accuracy in multipole space. We have checked that the chosen fiducial amplitude fulfills this requirement, while also staying inside the 1$\sigma$ errorbar of the polarisation spectra.

\begin{figure}[h]
\centering
\epsfig{figure=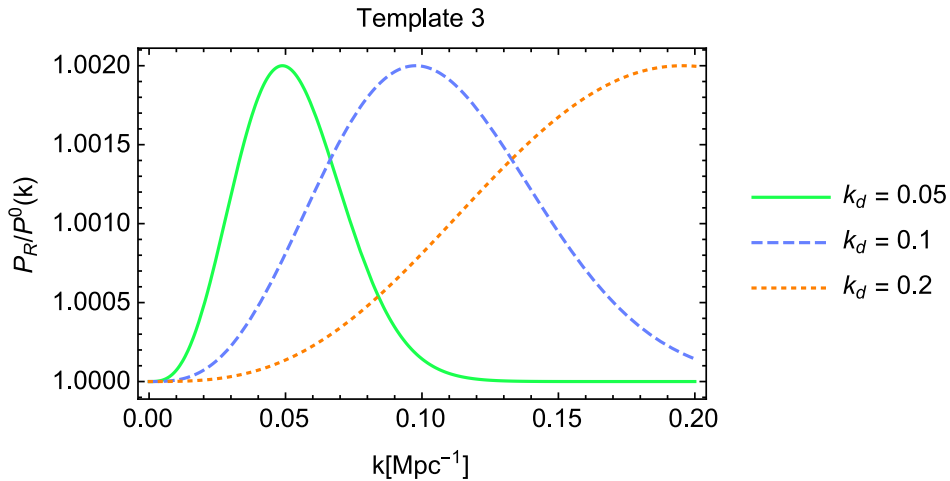,width=5in}
\caption{Plot of template 3 of Eq.~\eqref{bump} for the values quoted in Eq.~\eqref{bump-fid}.}
\label{fig:temp3}
\end{figure}

We plot the template resulting from these values in Fig.~\ref{fig:temp3}.

\subsection{Template IV}

\begin{figure}[h]
\centering
\epsfig{figure=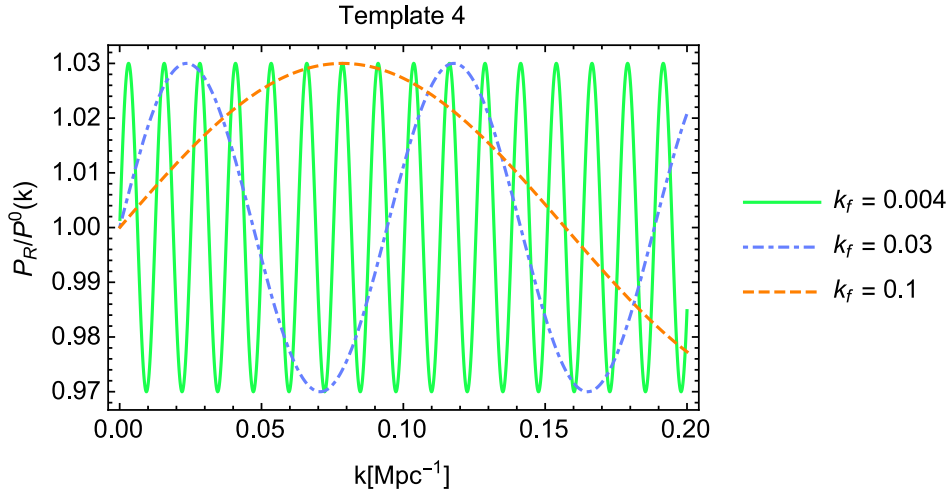,width=5in}
\caption{The template describing the oscillatory feature of Eq.~\eqref{sharp}.}
\label{fig:template-IV}
\end{figure}

The three previous templates have the common characteristic of being localised in $k-$space. However, the search of extended oscillatory features is also well justified. For instance, if the background quantity appearing in the square bracket of Eq.~\eqref{features-power-time} has a very sharp sudden variation that could be approximated by a Dirac delta-function $\propto \delta(\tau - \tau_0)$, then one would obtain a feature of the form
\be
\frac{\Delta \mathcal P}{\mathcal P_0 } (k)  \propto  \sin (2 k \tau_0 ).
\ee
This corresponds to a scale dependent feature spreading along the entire spectrum with a characteristic frequency determined by $\tau_0$. Such a situation could be achieved either by sharp variation of the sound speed $c_s$ (i.e. induced by a sudden turn in the multi-field inflationary trajectory) or by having a sharp variation of the inflationary background (i.e. induced by very sharp step in the inflationary potential). To model these type of oscillatory feature we will consider the following parametrisation:
\be
\frac{\Delta \mathcal P}{\mathcal P_0 }(k)  = C\sin\left[\frac{2k}{k_f}+\phi\right]. \label{sharp} 
\ee
This template was also considered in Refs.~\cite{Chen:2011zf,Fergusson:2014hya,Fergusson:2014tza,Meerburg:2015owa}. We will consider the following fiducial values for the parameters
\be
C= 0.03, \; k_f = (0.004;0.03;0.1)\;\text{Mpc}^{-1},\; \phi = 0, \label{sin-fid}
\ee
which are based on Refs.~\cite{Chen:2016vvw,Ade:2015lrj}. We plot the template of Eq.~\eqref{sharp} with these values in Fig.~\ref{fig:template-IV}.

\subsection{Template V}

\begin{figure}[h]
\centering
\epsfig{figure=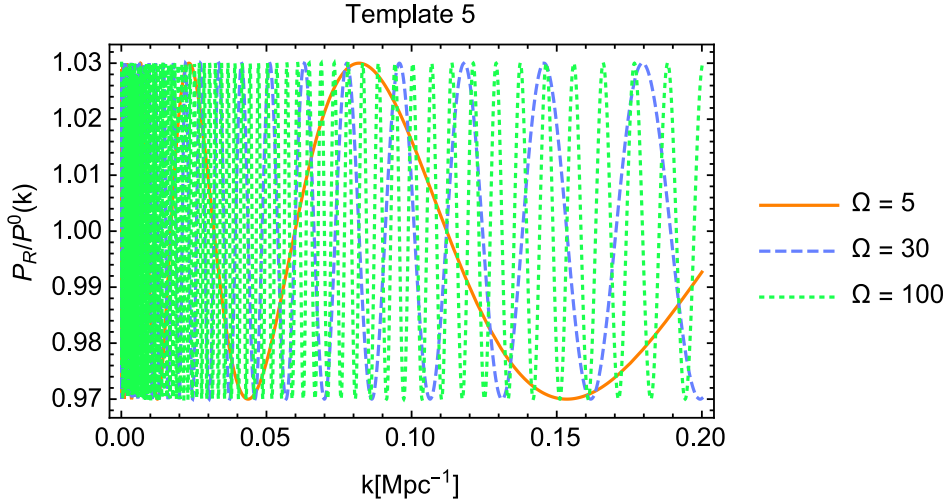,width=5in}
\caption{The oscillatory template of Eq.~\eqref{res}.}
\label{fig:template-V}
\end{figure}

Another class of oscillatory features appears when the background quantities appearing in Eq.~\eqref{features-power-time} have an oscillatory behavior themselves. More to the point, if the $c_s$ and/or $\epsilon$ have a sinusoidal behavior of the form $\propto \sin (\omega \ln \tau)$ then the primordial power spectrum is found to have the following resonant profile
\be
\frac{\Delta \mathcal P}{\mathcal P_0 } (k)  \propto  \sin ( \omega_0 \ln k + \varphi ) ,
\ee
where $\varphi$ is a given phase. A background behaving like this could appear in inflationary models with periodic effective potentials like those appearing in monodromy inflation~\cite{McAllister:2008hb} and natural inflation~\cite{Freese:1990rb}. We follow the parametrisation of this type of features used in Ref.~\cite{Chen:2008wn}, and write\footnote{This type of feature has been also studied in \cite{Miranda:2015cea}, where it was shown that features due to axionic dynamics can be accurately described by the template~\eqref{res} beyond the linear order.}:
\be
\frac{\Delta \mathcal P}{\mathcal P_0 } (k) = C\sin\left[\Omega\log(2k)+\phi\right]. \label{res}
\ee
In addition, we take the following fiducial values according to Refs.~\cite{Chen:2016vvw,Ade:2015lrj}:
\be
C= 0.03, \; \Omega = (5,30,100), \; \phi= 0 \label{sinlog-fid}.
\ee
We plot the template with these values in Fig.~\ref{fig:template-V}.

\setcounter{equation}{0} 
\section{Cosmological Probes} \label{s3:probes}

In this section we discuss the three observables that will be used to forecast constraints on the cosmological parameters, having in mind the LSST setup. These are: galaxy power spectrum (GC) and weak lensing (WL) to which we add the CMB prior from Planck.

\subsection{Spectroscopic galaxy power spectrum}

Here we only show the expression of the observed power spectrum, which will be used  to compute our forecasts, following Refs.~\cite{SeoPU} and \cite{SaponeJN}, we write the observed galaxy power spectrum as
\be
P_{\gamma\gamma}^\mathrm{spec}(z,k_r,\mu_r) =
\frac{D_{A,\,r}^{2}(z)H(z)}{D_{A}^{2}(z)H_{r}(z)}b(z)^{2}\left(b\sigma_8(z)+f\sigma_8(z)\mu^{2}\right)^{2}\frac{\bar{P}_{\rm lin}(z;~k)}{\sigma^2_8(z)} +  P_\mathrm{shot} \,, \label{eq:pk}
\ee
where the subscript $r$ refers to the reference (or fiducial) cosmological model. 

Here $P_\mathrm{shot}$ is a scale-independent offset due to imperfect removal of shot-noise, $\mu = \vec{k}\ \cdot\hat{r}/k$ is the cosine of the angle of the wave mode with respect to the line of sight pointing into the direction $\hat{r}$, $\bar{P}_{\rm lin}$ is the fiducial matter power spectrum evaluated at different redshifts in which the effects of the features are included:
\be
\bar{P}_{\rm lin}(z;~k) = \Delta_{\rm f} \times P_{\rm m}(z;~k),
\ee
where we have used the shorthand notation $\Delta_{\rm f}\equiv\frac{\Delta \mathcal P}{\mathcal P_0 }$, to denote the feature part of the power spectrum, while $P_{\rm m}(z;~k)$ is the linear featureless matter power spectrum. We have evaluated $\bar{P}_{\rm lin}(z;~k)$ with CAMB\footnote{\url{http://camb.info/}}, see \cite{LewisBS}, including the feature modulation $\Delta_{\rm f}$.  The wavenumber $k$ and $\mu$ also have to be written in terms of the fiducial cosmology, see \cite{SeoPU, SaponeJN,AmendolaBE} for more details. 

The redshift space distortion term in Eq.~\eqref{eq:pk} has been paramterised by $b\sigma_8(z)$ and $f\sigma_8(z)$, i.e. the bias factor and the growth rate multiplied by $\sigma_8(z)$, respectively. 
The Alcock-Paczynski effect takes into account the change in volume by different cosmology and it is given by the Hubble parameter $H(z)$ and the angular diameter distance $D_{A}(z)$ . 

In this paper we follow the specifications reported in Table 2 of \cite{Chen:2016vvw}: the LSST experiment will have an area of about  $23000$ square degrees, to which corresponds  a fraction of the sky $f_{\rm sky} = 0.58$.  
The redshift range goes from $z = 0.2$ up to $z=3$ divided into $7$ redshift bins, each bin will have its corresponding galaxy density and $k_{\rm min}-k_{\rm max}$ range. The photometric redshift error will be $\sigma_z = 0.04(1 + z)$  
and the bias parameter is assumed to be redshift dependent paramaterised as $b = 1 + 0.84z$. The matter power spectrum, the r.m.s. amplitude of density fluctuations and the growth rate have been computed with the CAMB code.

In order to match the results with the WL signal we consider the direct derivatives of the galaxy power spectrum with respect to the cosmological parameter of interest. 

\subsection{Weak Lensing} \label{subsec:wlps}
To the galaxy clustering signal we add the weak lensing tomographic signal \cite{wl-tomo, wl-tomo-2, wl-tomo-3, wl-tomo-4}. Here, we only give the main equation expressing the weak lensing power spectrum, which is used for our forecasts, and we refer to the literature for further details.

The weak lensing convergence power spectrum (which in the linear regime is equal
to the ellipticity power spectrum) is a linear function of the matter power spectrum convoluted with the
lensing properties of space. For a $\Lambda$CDM cosmology it can be written as 
\begin{equation}
C_{\kappa,ij}(\ell) = 
\int_0^{\chi_H}\frac{\dd\chi}{\chi^2}\:W_{\kappa,i}(\chi)W_{\kappa,j}(\chi)\: P_\mathrm{\rm nl}(\chi;~k=\ell/\chi),
\label{eq:convergence-wl}
\end{equation}
where $\ell$ is the multipole number, $W_i$'s are the window functions and the function $\chi$ is the comoving distance between the objects and the lens. $P_{\rm nl}\left[\bar{P}_{\rm lin}(z;~k)\right]$ is the non-linear power spectrum at redshift $z$ obtained by correcting the linear featurefull matter power spectrum $\bar{P}_{\rm lin} (z;~k)$. However, the former needs to be modified. There is no easy way to modify the convergence power spectrum 
when we consider the non-linear scales. For WL signal we need to evaluate the correlation of the WL potential: $\langle \Phi^2\rangle$; the latter is equal to $\Delta_{\rm f}\langle P_{\rm m}\rangle$ only at linear scales. If we consider non-linear scales, the above expression, in principle, is no longer valid because the feature part is a modification to the linear matter power spectrum. 

On the other hand, we cannot strictly make use of the analytic expression of $P_{\rm nl}$, see e.g. \cite{halo-fit, halo-fit2}, because it is designed for featureless  models close to $\Lambda$CDM. Hence, if we want to take the features into account we need to use CAMB, which does not, however, compute $P_{\rm nl}(\Delta_{\rm f}P_{\rm m})$ but only $P_{\rm nl}(P_{\rm m})$. Therefore, in this paper, we make the approximation of $P_{\rm nl}(\Delta_{\rm f}P_{\rm m})\sim \Delta_{\rm f}P_{\rm nl}(P_{\rm m})$.  
We do not really know the error we commit when making this approximation, however we have tested it by evaluating the convergence power spectra for WL using the analytic expression for the nonlinear power spectrum~\cite{halo-fit} and we found a difference of less than $0.5\%$. 
As a consequence, we expect the errors to change by the same percentage.

The subscript ${ij}$ in the window functions refers to the redshift bins around $z_i$ and $z_j$, with
\bea
W_{\kappa,i}(\chi) & = & \frac{3\Omega_m}{2\chi_H^2}\frac{F_i(\chi)}{a}\chi, \\
F_i(\chi) & = & \int_\chi^{\chi_H}\dd\chi^\prime n(\chi^\prime)\:D_i(\chi^\prime)\:\frac{\chi^\prime-\chi}{\chi^\prime},
\eea
	where $D_i(z)$ is the radial distribution function of galaxies in the $i$-th $z$-bin, obtained as follows: Since in this paper we choose WL tomography, we divide the whole galaxy sample into 5 bins with equal number of galaxies. We then simply extract the binned galaxy distribution function $D_i(z)$ by binning the overall distribution $D(z)$ and convolving it with the photometric redshift distribution function \cite{AmendolaRR}. The overall radial distribution is chosen to be (see \cite{ZhanUWU} and references therein)
\be \label{eq:gal-distr}
D(z) = z^2\exp\left(-3.2 z\right)\,.
\ee
 %

% ---  --- %

% ---  --- %

% --- section: forecasts ---%

% --- section: forecasts ---%

\setcounter{equation}{0} 
\section{Statistical errors forecasts} \label{s4:forecasts}

In this section we estimate marginalised statistical errors on the parameters entering the featured power spectrum modelled by the five templates of Sec.~\ref{s2:Templates}. To perform our forecasts we use the Fisher information matrix \cite{TegmarkBZ} which, assuming a Gaussian likelihood and unbiased measurements, yields the best errors on the parameters given a specification.

\subsection{Spectroscopic galaxy distribution Fisher matrix}
The galaxy  power spectrum Fisher matrix is given by \cite{SeoPU}
\begin{equation}
F_{\alpha\beta}^\mathrm{GC} = 
\int_{k_{\rm min}}^{k_{\rm max}}\frac{k^{2}{\rm d}k}{4\pi^{2}}
\frac{\partial\ln P_{\gamma\gamma}^\mathrm{spec}\left(z;k,\mu\right)}{\partial\theta_\alpha}
\frac{\partial\ln P_{\gamma\gamma}^\mathrm{spec}\left(z;k,\mu\right)}{\partial\theta_\beta}
\times V_{\rm eff},
\label{eq:FisherMatrix}
\end{equation}
where GC stands for galaxy clustering, while the observed galaxy power spectrum $P_{\gamma\gamma}^\mathrm{spec}$ is given by Eq.~(\ref{eq:pk}) corrected with the featured power spectrum. The derivatives are evaluated at the parameter values of the fiducial model, $k_{\rm min}$ and $k_{\rm max}$ depend on the redshift bin. Finally, $V_{\rm eff}$ is the effective volume of the survey, given by
\begin{equation}
V_{\rm eff} \simeq   
\left(\frac{\bar n\,P_{\gamma\gamma}^\mathrm{spec}\left(k,\mu\right)}{\bar n\, P_{\gamma\gamma}^\mathrm{spec} \left(k,\mu\right)+1}\right)^{2}V_{\rm survey},
\label{eq:Volume}
\end{equation}
the latter equation holding for an average comoving number density $\bar n$. The number densities and further specifications can be found in Ref.~\cite{Chen:2016vvw}.

\subsection{Weak Lensing Fisher matrix}
The Fisher matrix for weak lensing is given by \cite{AmendolaRR}:
\be
F_{\alpha\beta}^{\rm WL} = 
f_{\rm sky}\sum_{\ell} \frac{\left(2\ell+1\right)}{2} 
\frac{\partial C_{\kappa\kappa,ij}(\ell) }{\partial \theta_\alpha}\bar{C}_{jk}^{-1}(\ell)
\frac{\partial C_{\kappa\kappa,km}(\ell) }{\partial \theta_\beta}\bar{C}_{mi}^{-1}(\ell),
\ee
where the sum runs from $\ell = 5$ to $\ell = 5000$, and summation over repeated indices is implied. To the weak lensing spectra we added a Poissonian shape noise term:
\be
\bar{C}_{\kappa\kappa,ij}(\ell) = 
C_{\kappa\kappa,ij}(\ell) + \delta_{ij}\frac{\langle \gamma_{\rm int}^{1/2}\rangle}{n_i},
\ee
where $\gamma_{\rm int}$ is the r.m.s. intrinsic ellipticity of galaxy images. In this work we use $\langle\gamma_{\rm int}^{1/2}\rangle$=0.2, see \cite{ZhanUWU}. The number of galaxies per steradians belonging to the $i$-th bin is $n_{i}=40$ per $\text{arcmin}^2$. Finally, we take a Gaussian shape of the covariance, however we report that  non-Gaussian contributions might have strong influence on the forecasts, see for instance \cite{TakadaFN}.

\subsection{CMB Fisher matrix} 

Finally, we impose a CMB prior to the LSS observables by adding to the GC and WL signals the Fisher matrix for the Planck mission. Following Refs.~\cite{knox, jungman, seljak,zaldarriagaXE,kamionkowskiKS}, we define the CMB Fisher matrix as:
\be
F_{\alpha\beta}^{\rm CMB} = f_\text{sky} \sum_\ell\frac{2\ell+1}{2}\sum_{X,\,Y} \frac{\partial C_\ell^{X}}{\partial \theta_\alpha}\left(\text{Cov}_\ell^{-1}\right)_{X\,Y}\frac{\partial C_\ell^{Y}}{\partial \theta_\beta} \,,
\ee
where $C_\ell^{X}$ is the CMB angular power spectrum at the $\ell^\text{th}$ multipole for the signals $X,\,Y = \left\{ TT,\,EE,\,BB\right\}$. For the CMB temperature and polarisation we add the instrumental noise and the beam as \cite{knox} 
\be
N_\ell^X = \left(\frac{\Delta_X}{T_\text{CMB}}\right)\exp\left[\ell(\ell+1)\frac{\theta_\text{FWHM}^2}{8\ln 2}\right],
\ee
where $\Delta_X$ is the detector noise for the different signals and $\theta_\text{FWHM}$ is the full width half maximum of the beam evaluated in radians. If we consider multiple frequency channels, the total noise is given by 
\be
N_\ell^X  = \left[\sum_\text{channels}\frac{1}{N_{\ell\,,i}^{X}}\right]^{-1},
\ee
with $i$ running over all channels. 

The non vanishing elements of the covariance matrix $\text{Cov}_\ell$ are:
\bea
\covl_{TTTT} &=& \left(\cellb^{TT}\right)^2-2\frac{\left(\cellb^{TE}\,\cellb^{TB}\right)^2}{\cellb^{EE}\,\cellb^{BB}}\,,\nn\\
\covl_{EEEE} &=& \left(\cellb^{EE}\right)^2\,, \nn\\
\covl_{TETE} &=& \frac{\left(\cellb^{TE}\right)^2+\cellb^{TT}\,\cellb^{EE}}{2}-\frac{\cellb^{EE}\left(\cellb^{TB}\right)^2}{2\cellb^{BB}}\,,\nn\\
\covl_{BBBB} &=& \left(\cellb^{BB}\right)^2\,, \nn \\
\covl_{TBTB} &=& \frac{\left(\cellb^{TB}\right)^2+\cellb^{TT}\,\cellb^{BB}}{2}-\frac{\cellb^{BB}\left(\cellb^{TE}\right)^2}{2\cellb^{EE}}\,,\nn\\
\covl_{TTEE} &=& \left(\cellb^{TE}\right)^2\,, \\
\covl_{TTTE} &=& \cellb^{TT}\,\cellb^{TE}-\frac{\cellb^{TE}\left(\cellb^{TB}\right)^2}{\cellb^{BB}}\,,\nn\\
\covl_{TTBB} &=& \left(\cellb^{TB}\right)^2\,, \nn\\
\covl_{TTTB} &=& \cellb^{TT}\,\cellb^{TB}-\frac{\cellb^{TB}\left(\cellb^{TE}\right)^2}{\cellb^{EE}}\,,\nn\\
\covl_{EETE} &=& \cellb^{EE}\cellb^{TE}\,, \nn\\
\covl_{BBTB} &=& \cellb^{BB}\cellb^{TB}\,\nn, 
\eea
where $\cellb^X=C_\ell^X+N_\ell^X$ and the noise for cross terms vanishes: $N_\ell^{TE}=N_\ell^{TB}=0$ .

We consider a Planck-like survey for which the fraction of sky covered is $f_\text{sky} = 0.75$ and multipoles range from $\ell = 2$ up to $\ell_\text{max} = 2500$. We use three frequencies for the signal $100$GHz, $143$GHz and $217$GHz to which correspond a sensitivity of $25\mu$K-arcmin in temperature and $50\mu$K-arcmin in polarisation with a Gaussian beam width of about $5$ arcmin, see \cite{AdamWUA}.
The theoretical angular power spectra of the CMB have been computed with the CAMB code.

% ---  --- %

% ---  --- %

% ---  --- %

\subsection{Forecasts}  

The parameters $\theta_\alpha$ for which we evaluate the Fisher matrices are the six cosmological parameters:
\be
\omega_b = \Omega_{b,0}h^2, \; h, \; \omega_m = \Omega_{m,0}h^2, \; n_s, \; \tau, \; A_s.
\ee
These correspond to the baryon density, reduced Hubble parameter, matter density, spectral index, reionisation and amplitude of scalar perturbations, respectively. To this set we add the subset of parameters  corresponding to each template modeling features. The fiducial values of the cosmological parameters are the best-fit coming from the Planck mission~\cite{planck}, while for the feature parameters we use the values quoted in Eqs.~\eqref{temp1-fid}, \eqref{step-fid-m}, \eqref{step-fid-s}, \eqref{bump-fid}, \eqref{sin-fid} and \eqref{sinlog-fid}, which we rewrite here for convenience:
\be
\{C\,, k_d[\text{Mpc}^{-1}]\,,k_f[\text{Mpc}^{-1}]  \} = \{0.215,\,(0.07\,;\,0.13\,;\, 0.29),\, 0.005 \},\nn
\ee 
for the localised oscillation template of Eq.~\eqref{temp1},
\be
\{C\,, k_d[\text{Mpc}^{-1}]\,,k_f[\text{Mpc}^{-1}]  \} = \{(0.218;0.075),\, (1.1\ \cdot10^{-3};2.8\cdot 10^{-2}),\, (6.9\ \cdot10^{-4};2.7\cdot 10^{-4})\; \},\nn
\ee 
for the step-template of Eq.~\eqref{step}, 
\be
\{C\,, k_d[\text{Mpc}^{-1}] \} = \{0.002,\, (0.05\,;\, 0.1\,;\, 0.2) \}, \nn
\ee
for the template of Eq.~\eqref{bump} modeling particle production,
\be
\{C\,,k_f[\text{Mpc}^{-1}]\,, \phi \} = \{ 0.03\,,(0.004;0.03;0.1)\,, 0\}, \nn
\ee
for the template of Eq.~\eqref{sharp}, modeling a sharp feature, and
\be
\{C\,,\Omega\,, \phi \} = \{ 0.03\,,(5;30;100)\,, 0\}, \nn
\ee
for the resonant feature template of Eq.~\eqref{res}.

We are interested in computing forecasts on the measurements of parameters involved in the feature part of the power spectra combining different 
observables. The probes are assumed to be uncorrelated, hence their Fisher matrices add:
\begin{equation}
F_{\alpha\beta} = F_{\alpha\beta}^\mathrm{GC} + F_{\alpha\beta}^{\rm WL} + F^{\rm CMB}_{\alpha\beta},
\end{equation}
and we derive confidence contours on the feature parameters and individual errors from this combined Fisher matrix, marginalising over all the other cosmological parameters considered in this analysis. Moreover, $F^\mathrm{GC}_{\alpha\beta}$ has been further marginalised over $P_\mathrm{shot}$.

We report the results for the errors upon the feature parameters using three data sets, i.e. CMB+GC, CMB+WL and CMB+GC+WL, in Tab.~\ref{tab:model5-forecasts}, Tab.~\ref{tab:model2-forecasts}, Tab.~\ref{tab:model4-forecasts}, Tab.~\ref{tab:model1-forecasts} and Tab.~\ref{tab:model3-forecasts} for the templates of Eqs.~\eqref{temp1}, \eqref{step}, \eqref{bump}, \eqref{sharp} and \eqref{res}, respectively. The lines corresponding to CMB+GC have already been forecasted in Ref.~\cite{Chen:2016vvw} for the templates III-V. We find that our results are in one order of magnitude agreement due to the different CMB Fisher matrix that we use. We have a more ideal approach giving us the best errors possible. However, the relative correction induced by the addition of weak lensing data is the same in both cases.
%
%TAB Mod5 temp1
\begin{table}[H]
\begin{centering}\begin{tabular}{cccccc}
\toprule
& & & \multicolumn{3}{c}{\textbf{ Template 1} }  \tabularnewline
 \toprule
%\rowcolor{gray} \multicolumn{5}{c}{}  \tabularnewline
%\hline 
\multicolumn{2}{c}{}& $k_d^{\rm fid}$ & $\delta C$ & $\delta  k_{d}$ & $\delta k_f$   \tabularnewline
% \hline
% \multicolumn{5}{|c|}{ }  \tabularnewline
%\hline
 \rowcolor{gray}\multicolumn{1}{l}{\bf{CMB} + {\bf GC} }& &\multicolumn{4}{c}{ }  \tabularnewline
% \hline
  & & 0.07 & $2.4891 \cdot 10^{-2}$ & $4.4269 \cdot 10^{-3}$ & $1.1253 \cdot 10^{-3}$ \\ \hline
    & & 0.13 & $1.0135 \cdot 10^{-2}$ & $4.7787 \cdot 10^{-2}$ & $3.5708 \cdot 10^{-4}$ \\ \hline
      & & 0.29 & $6.1381 \cdot 10^{-3}$ & $1.8326 \cdot 10^{-1}$ & $1.8232 \cdot 10^{-4}$
    \tabularnewline
%\hline 
% \multicolumn{5}{c}{ }  \tabularnewline
%\hline
 \rowcolor{gray} \multicolumn{1}{l}{\bf{CMB} + {\bf WL} }& &\multicolumn{4}{c}{ }  \tabularnewline
%\hline 
& &0.07 & $2.6417 \cdot 10^{-2}$ & $5.7033 \cdot 10^{-3}$ & $1.8863 \cdot 10^{-3}$ \\\hline
& &0.13 & $1.9040 \cdot 10^{-2}$ & $1.7779 \cdot 10^{-2}$ & $1.2746 \cdot 10^{-3}$ \\\hline
& &0.29 & $6.0371 \cdot 10^{-2}$ & $1.1458 \cdot 10^{-1}$ & $1.4117 \cdot 10^{-3}$ \tabularnewline
%\hline 
%\rowcolor{gray} \multicolumn{5}{c}{ }  \tabularnewline
%\hline
 \rowcolor{gray} \multicolumn{1}{l}{ \bf{CMB} + \bf{GC}+\bf{WL} }& &\multicolumn{4}{c}{ }  \tabularnewline
 %\hline
  & &0.07 & $1.6706 \cdot 10^{-2}$ & $2.7914 \cdot 10^{-3}$ & $8.8401 \cdot 10^{-4}$ \\\hline
    & &0.13 & $8.5889 \cdot 10^{-3}$ & $4.3085 \cdot 10^{-3}$ & $3.3254 \cdot 10^{-4}$ \\\hline
      & &0.29 & $6.0606 \cdot 10^{-3}$ & $1.5243 \cdot 10^{-2}$ & $1.7575 \cdot 10^{-4}$ 
  \tabularnewline
\bottomrule	
\end{tabular}\par\end{centering}
\caption{Constraints on the parameters of the template of Eq.~\eqref{temp1} for all the observables.
\label{tab:model5-forecasts}}
\end{table}
\begin{figure}[H]
\centering
\epsfig{figure=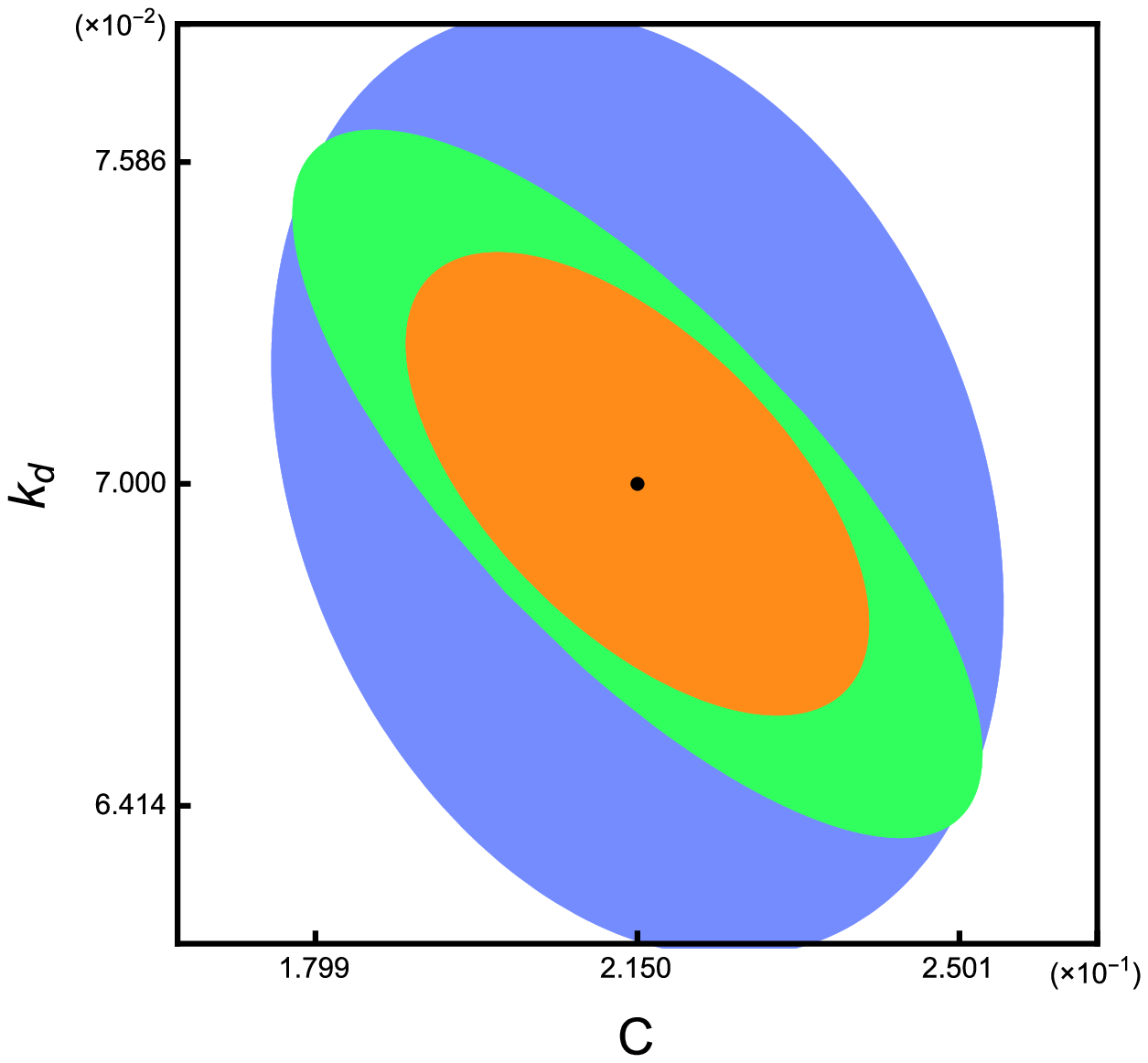,width=1.8in}
\epsfig{figure=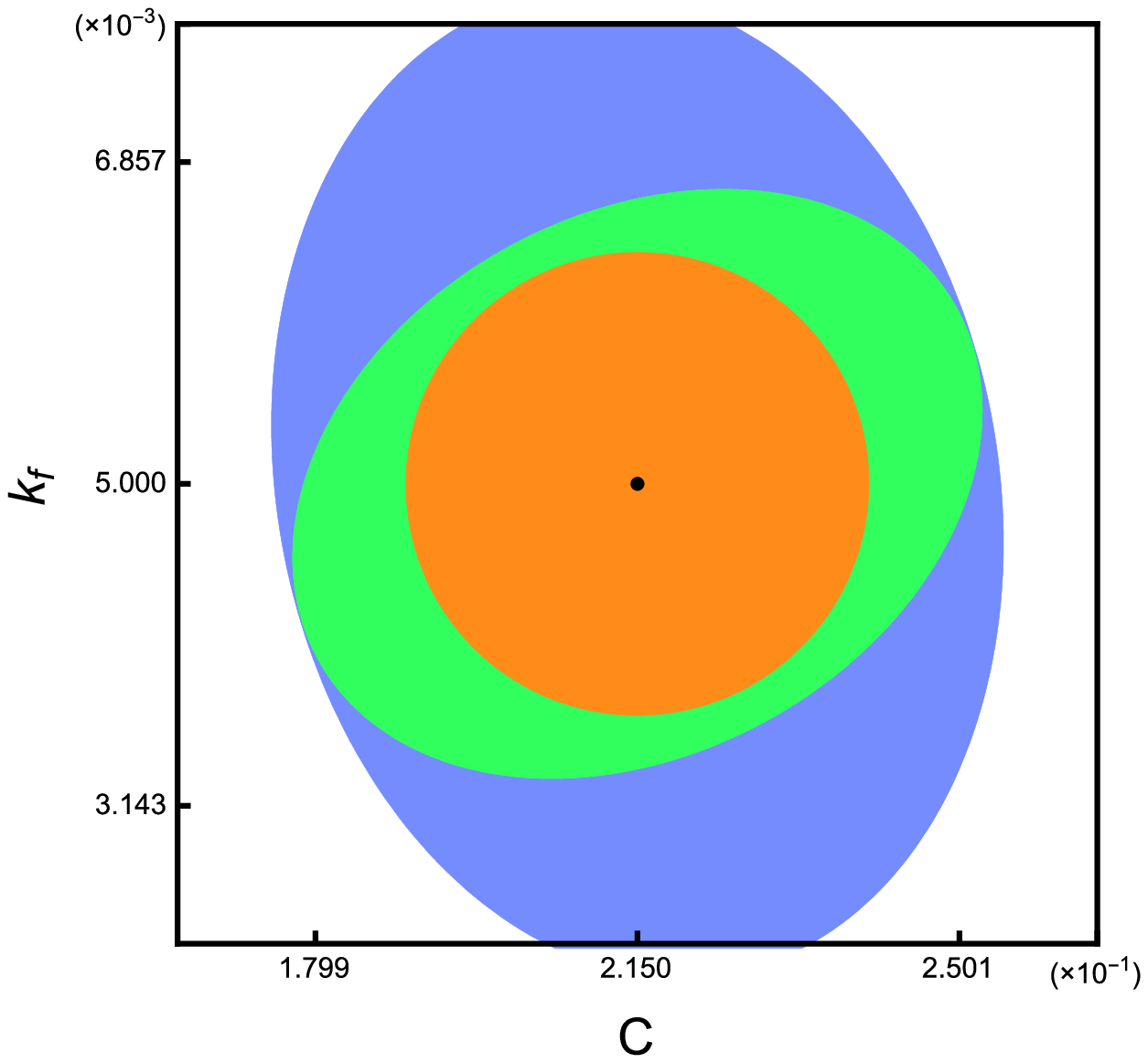,width=1.8in}
\epsfig{figure=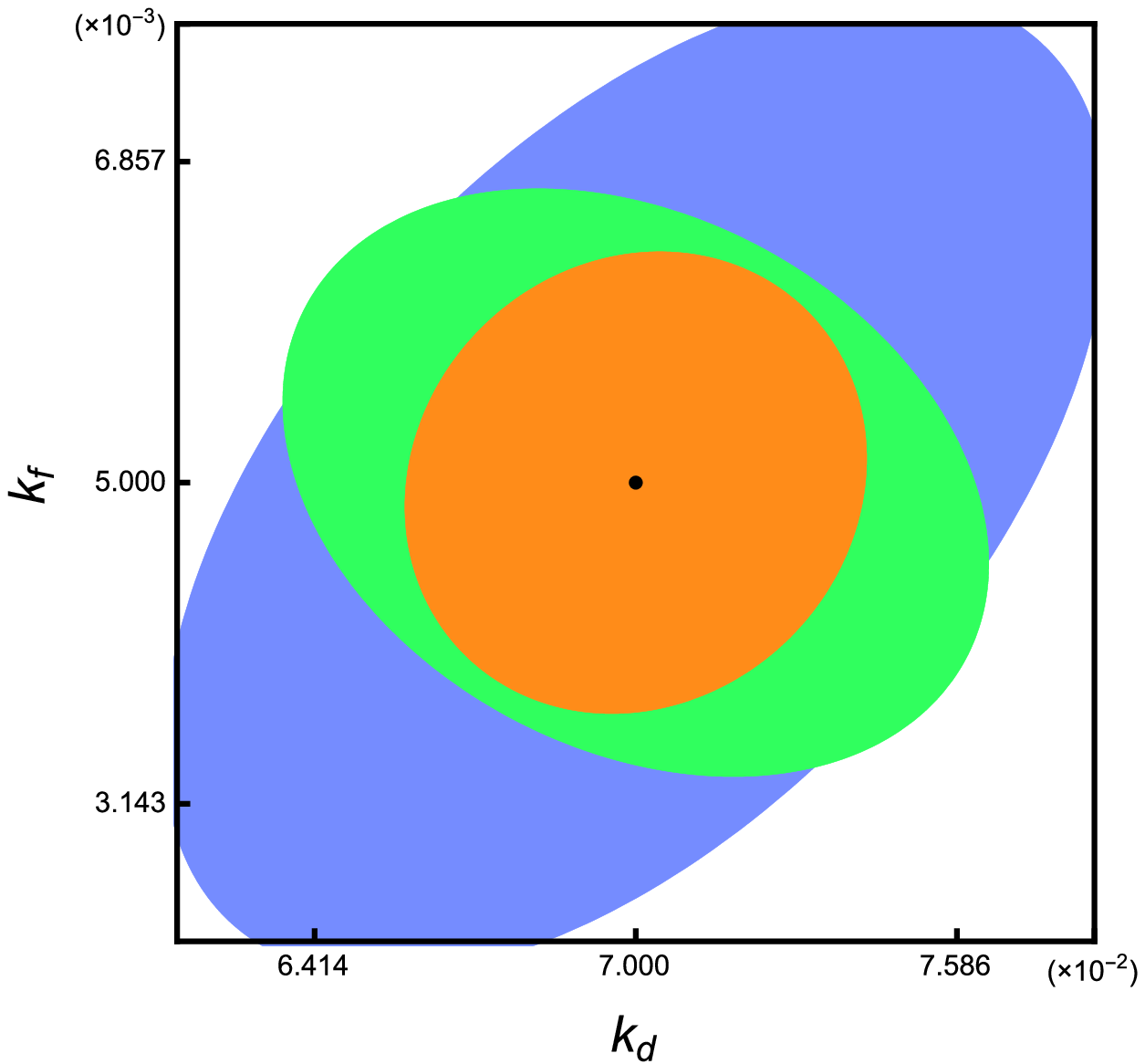,width=1.8in} \\
\epsfig{figure=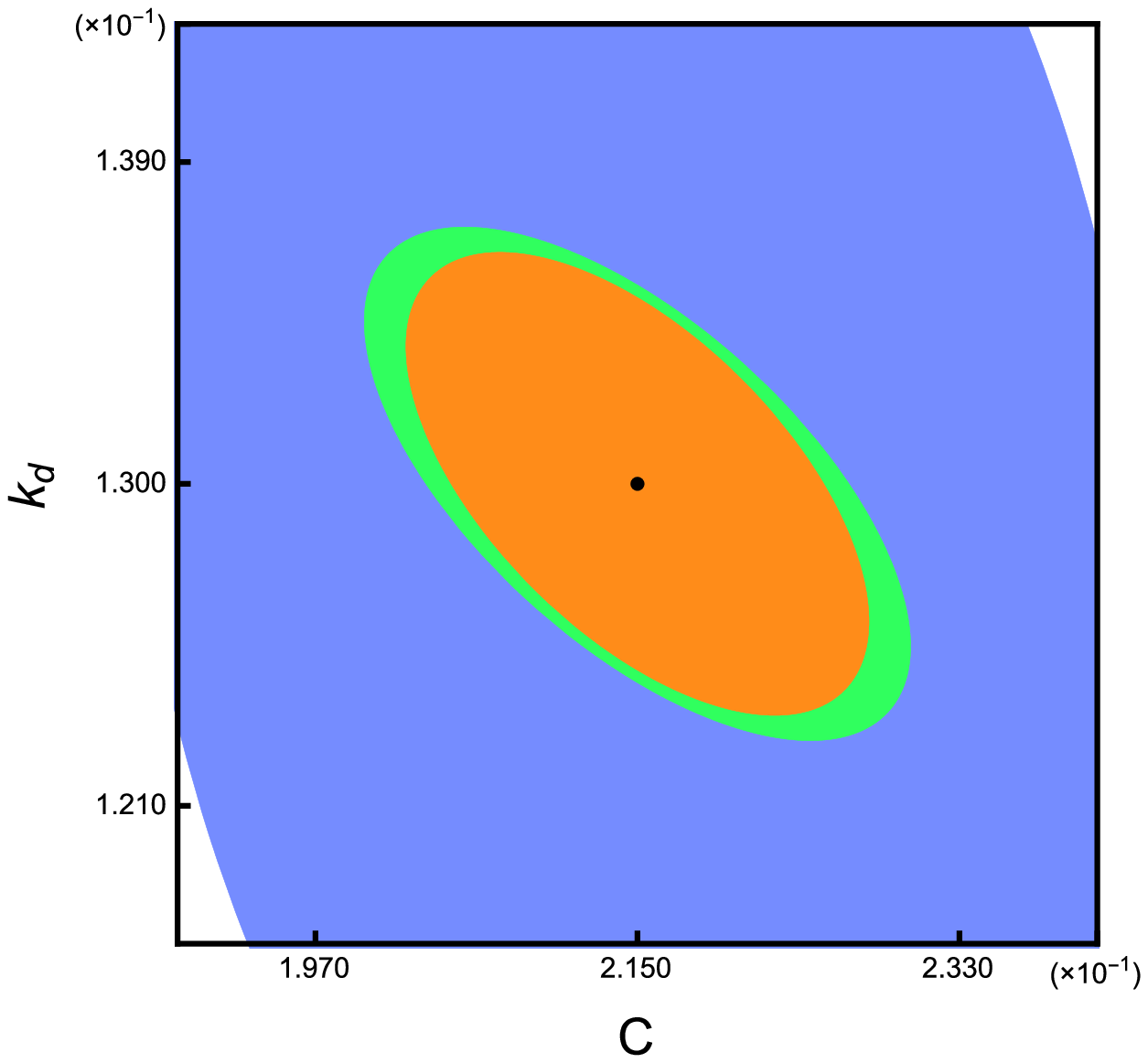,width=1.8in}
\epsfig{figure=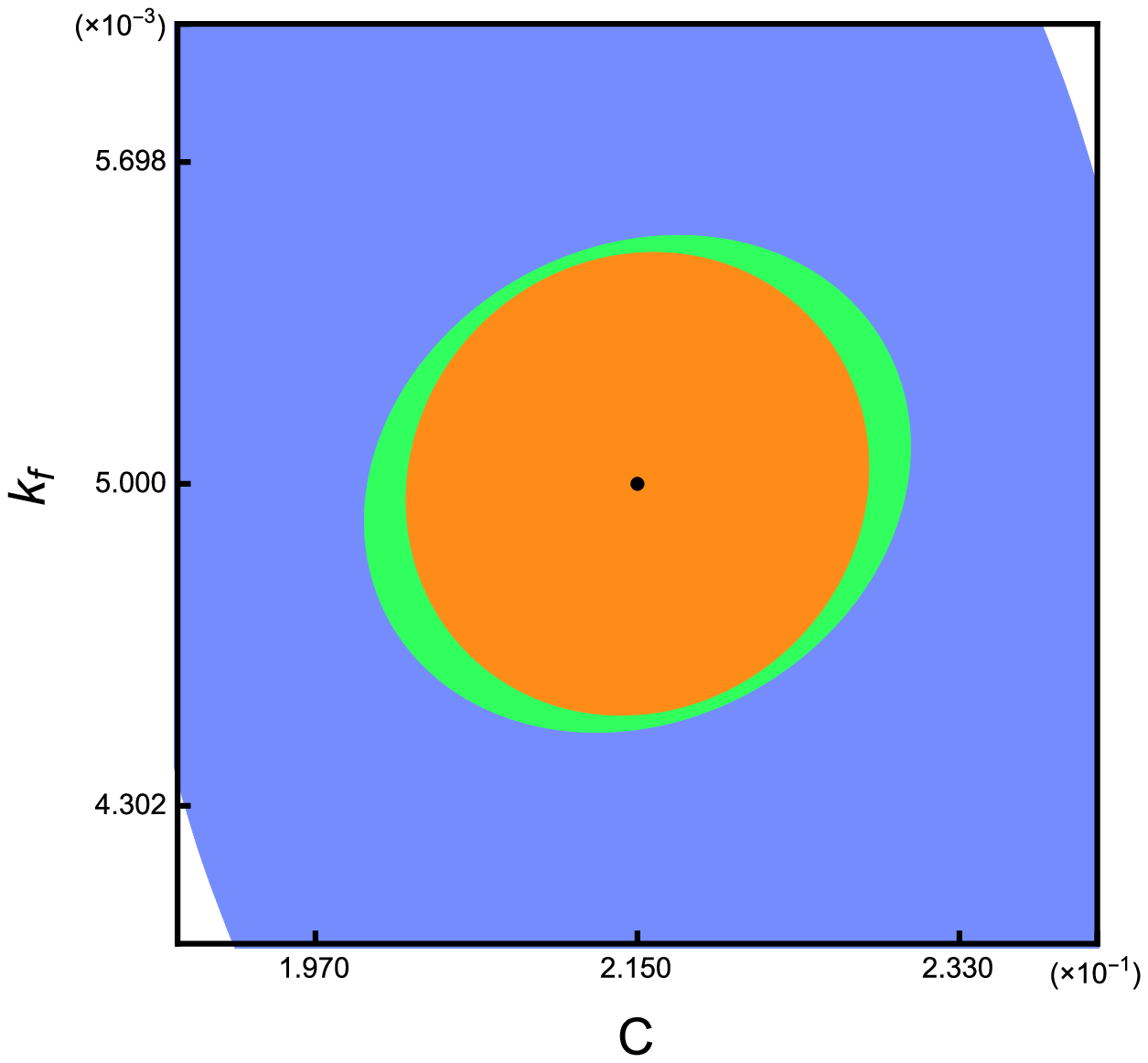,width=1.8in}
\epsfig{figure=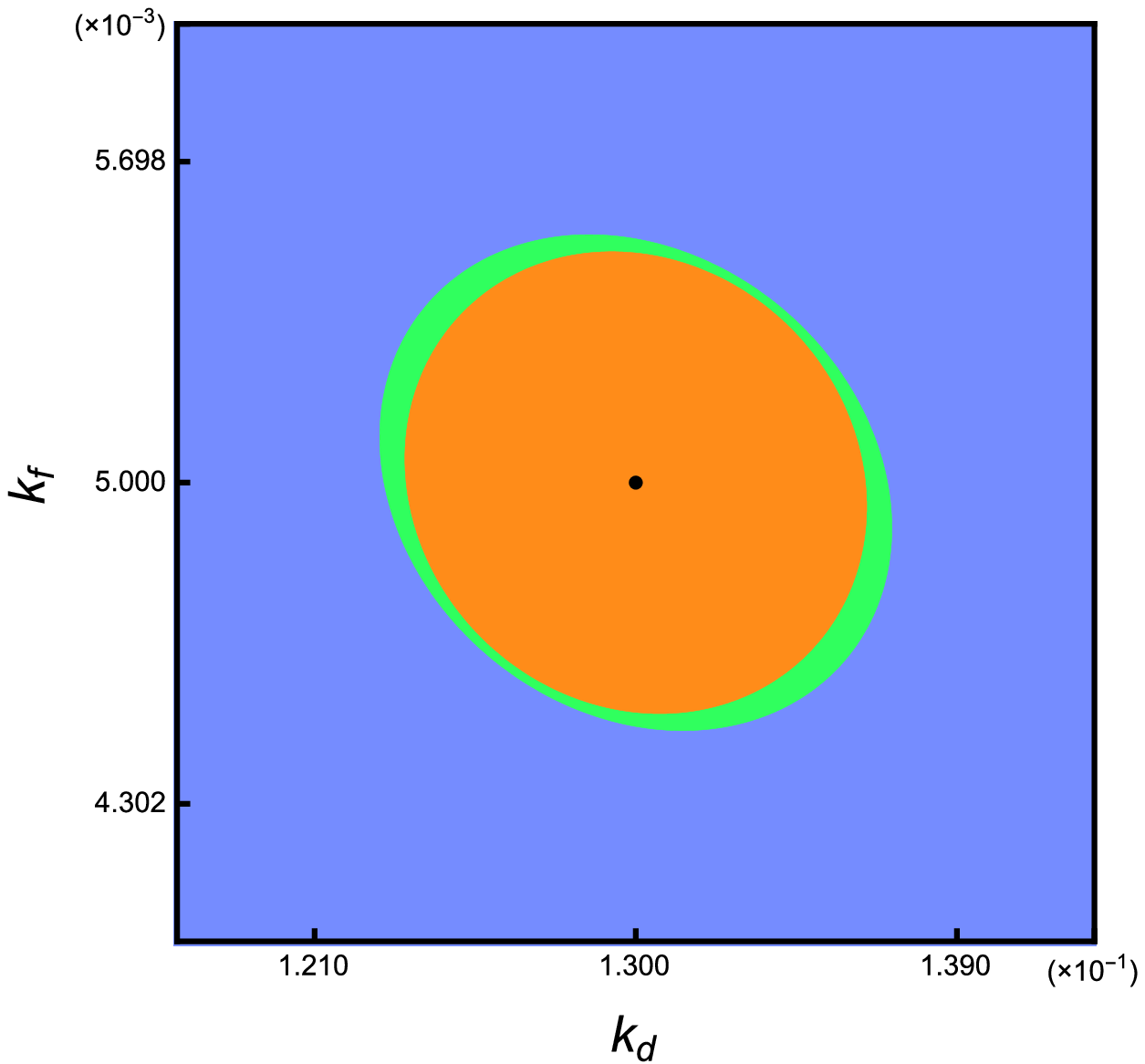,width=1.8in} \\
\epsfig{figure=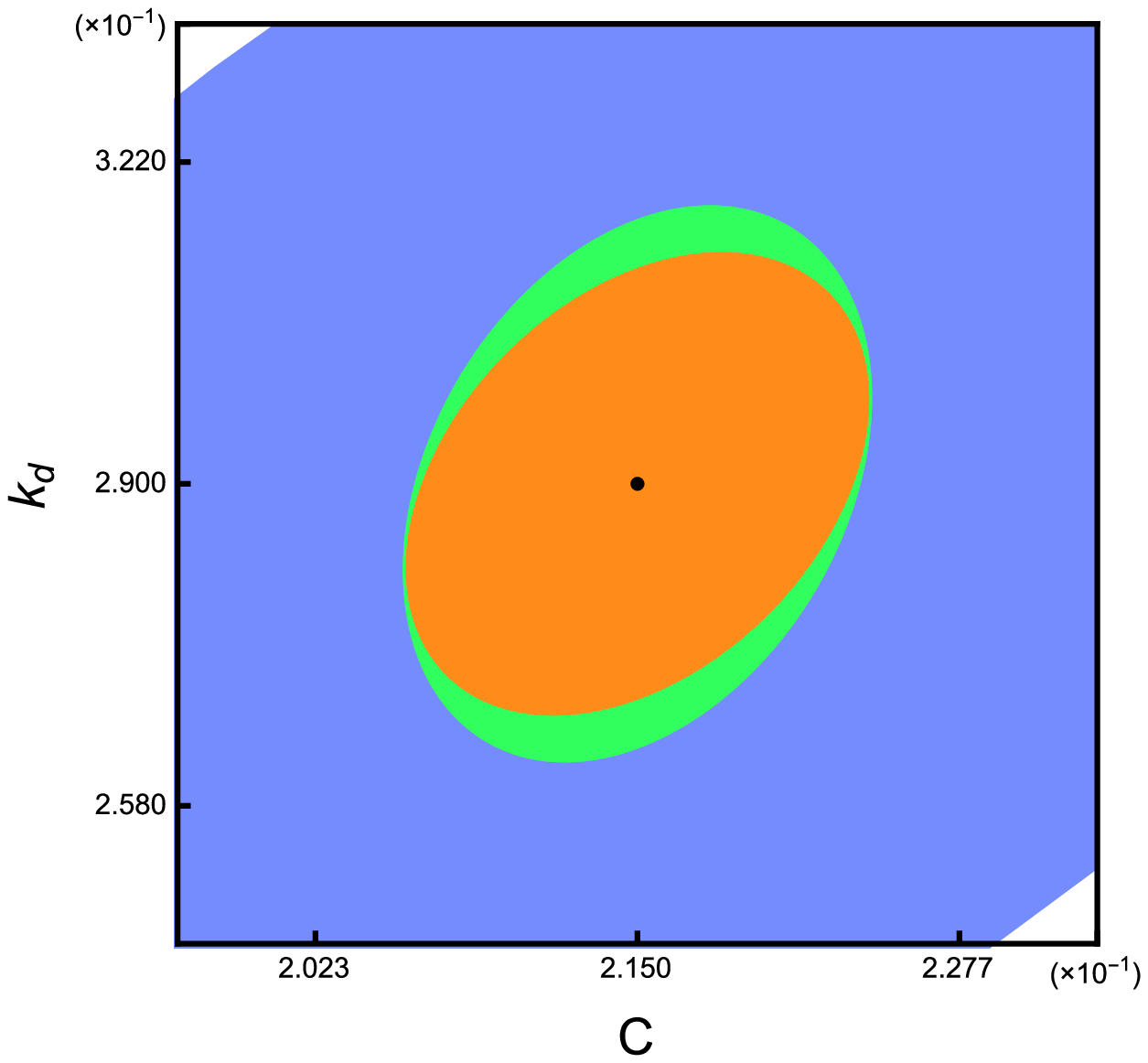,width=1.8in}
\epsfig{figure=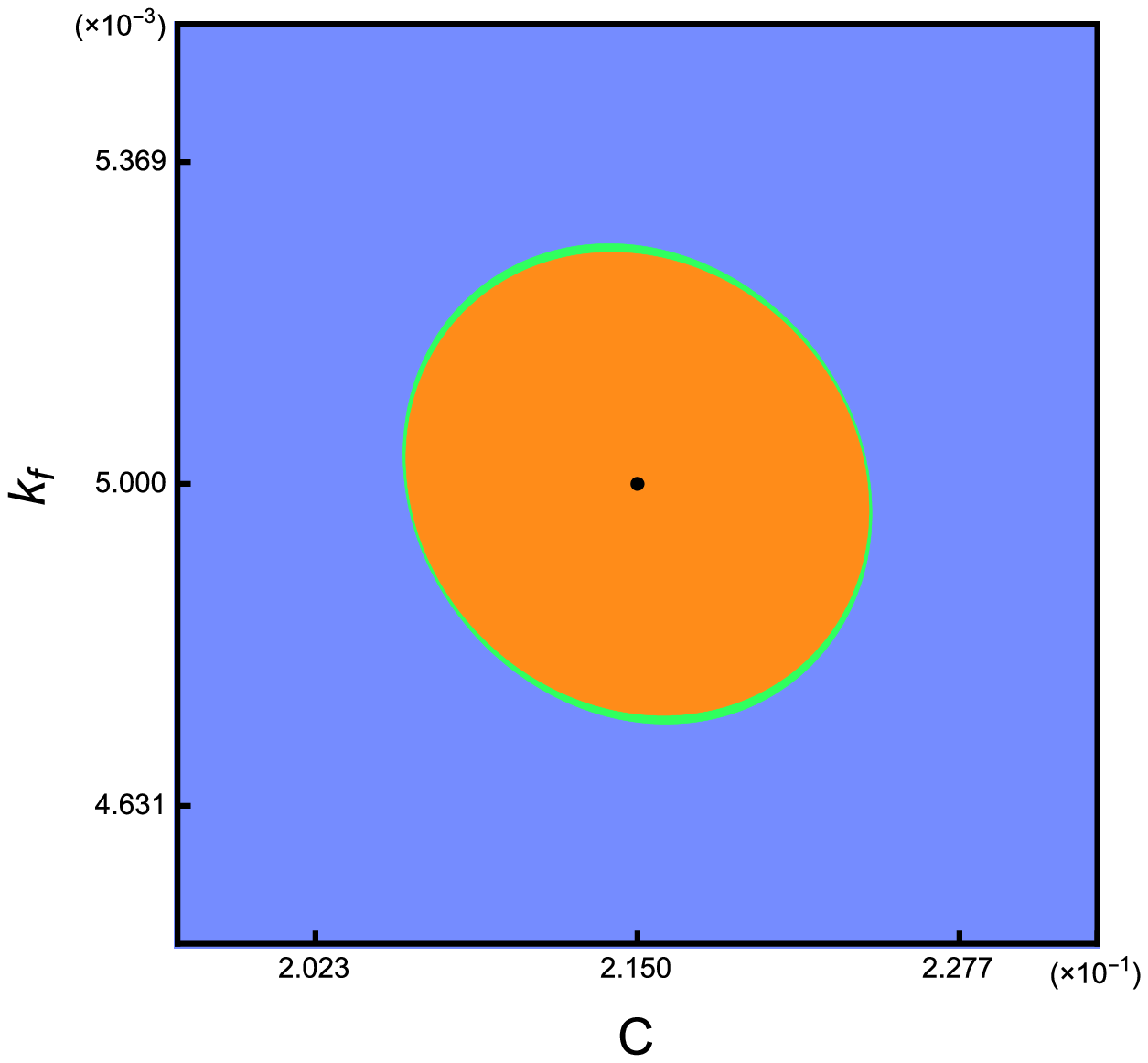,width=1.8in}
\epsfig{figure=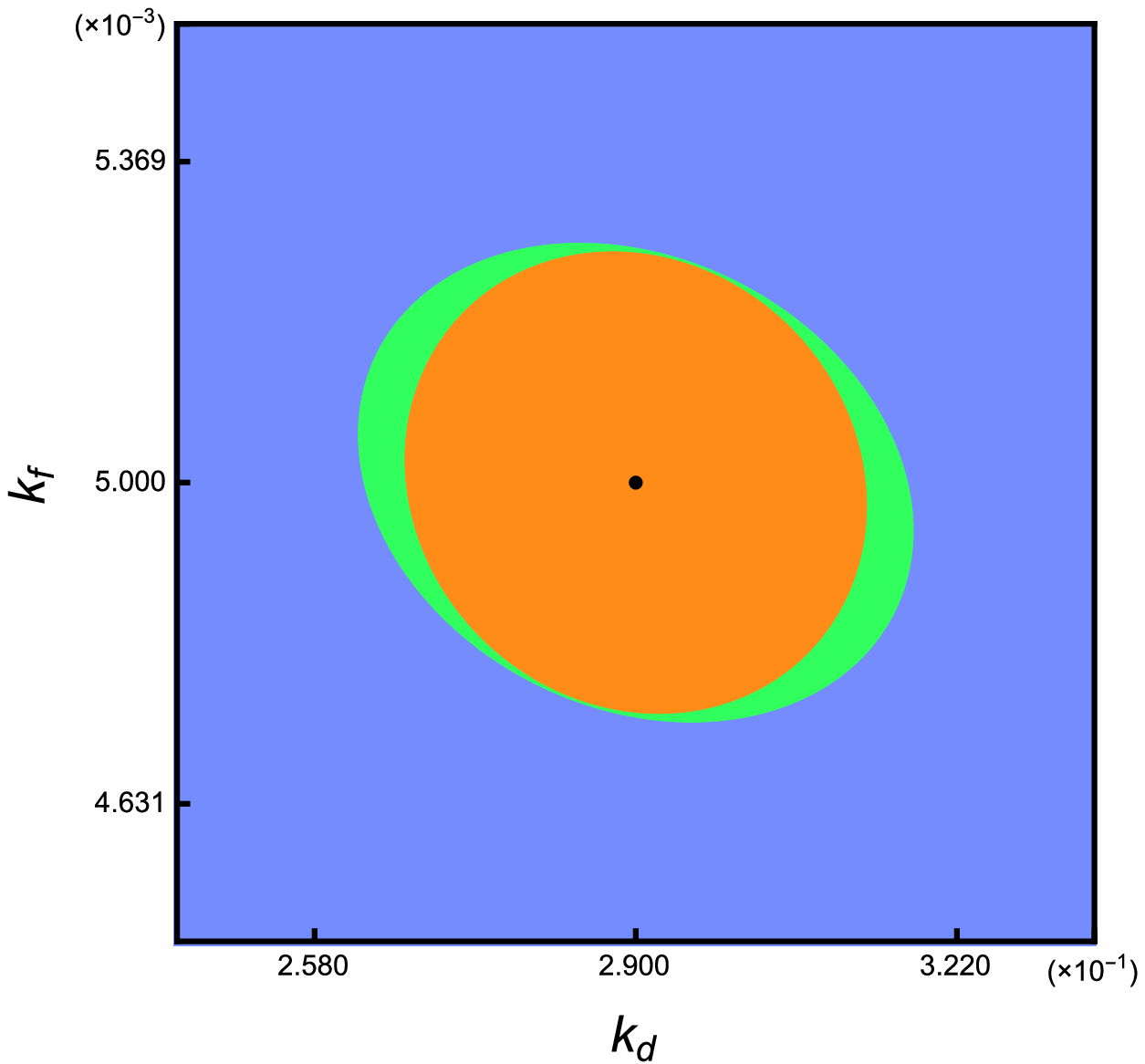,width=1.8in}
\caption{Here we show the sensitivity of the parameters entering in template I --Eq.~\eqref{temp1}-- for the three observables. In green we plot the constraints from CMB+GC, in light blue from CMB+WL and orange is the sum of the three observables.}
\label{fig:model5}
\end{figure}
%
%TAB Mod2 temp 2
\begin{table}[H]
\begin{centering}\begin{tabular}{cccccc}
\toprule
& & &\multicolumn{3}{c}{\textbf{ Template 2} }  \tabularnewline
 \toprule
%\rowcolor{gray} \multicolumn{5}{c}{}  \tabularnewline
%\hline 
\multicolumn{2}{c}{}&fiducial set & $\delta C$ & $\delta  k_{d}$ & $\delta k_f$   \tabularnewline
% \hline
% \multicolumn{5}{|c|}{ }  \tabularnewline
%\hline
 \rowcolor{gray}\multicolumn{1}{l}{\bf{CMB} + {\bf GC} }& &\multicolumn{4}{c}{ }  \tabularnewline
% \hline
  & & \small{m} & $4.0689 \cdot 10^{-2}$ & $ 2.1790 \cdot 10^{-4}$ & $1.5487 \cdot 10^{-6}$ \\
  \hline 
  &&\small{s}&$2.1302 \cdot 10^{-3}$&$6.6109 \cdot 10^{-4}$&$6.8453 \cdot 10^{-8}$\tabularnewline
%\hline 
% \multicolumn{5}{c}{ }  \tabularnewline
%\hline
 \rowcolor{gray} \multicolumn{1}{l}{\bf{CMB} + {\bf WL} }& &\multicolumn{4}{c}{ }  \tabularnewline
%\hline 
&  & \small{m} & $3.3584 \cdot 10^{-2}$ & $2.1149 \cdot 10^{-4}$ & $1.5044 \cdot 10^{-6}$ \\
\hline 
&&\small{s}&$5.1695 \cdot 10^{-2}$& $3.1393 \cdot 10^{-2}$ & $5.8353 \cdot 10^{-7}$\tabularnewline
%\hline 
%\rowcolor{gray} \multicolumn{5}{c}{ }  \tabularnewline
%\hline
 \rowcolor{gray} \multicolumn{1}{l}{ \bf{CMB} + \bf{GC}+\bf{WL} }& &\multicolumn{4}{c}{ }  \tabularnewline
 %\hline
 &  & \small{m} & $3.1808\cdot 10^{-2}$ & $1.8639 \cdot 10^{-4}$ & $1.4975 \cdot 10^{-6}$ \\
 \hline 
 &&\small{s}& $1.9700 \cdot 10^{-3}$ & $6.3405\cdot 10^{-4}$ & $6.8415\cdot 10^{-8}$
   \tabularnewline
\bottomrule	
\end{tabular}\par\end{centering}
\caption{Constraints on the parameters of the template of Eq.~\eqref{step} for all the observables. The fiducial parameter sets m and s refer to the values written in Eqs.~\eqref{step-fid-m} and \eqref{step-fid-s}, respectively.
\label{tab:model2-forecasts}}
\end{table}
\begin{figure}[H]
\centering
\epsfig{figure=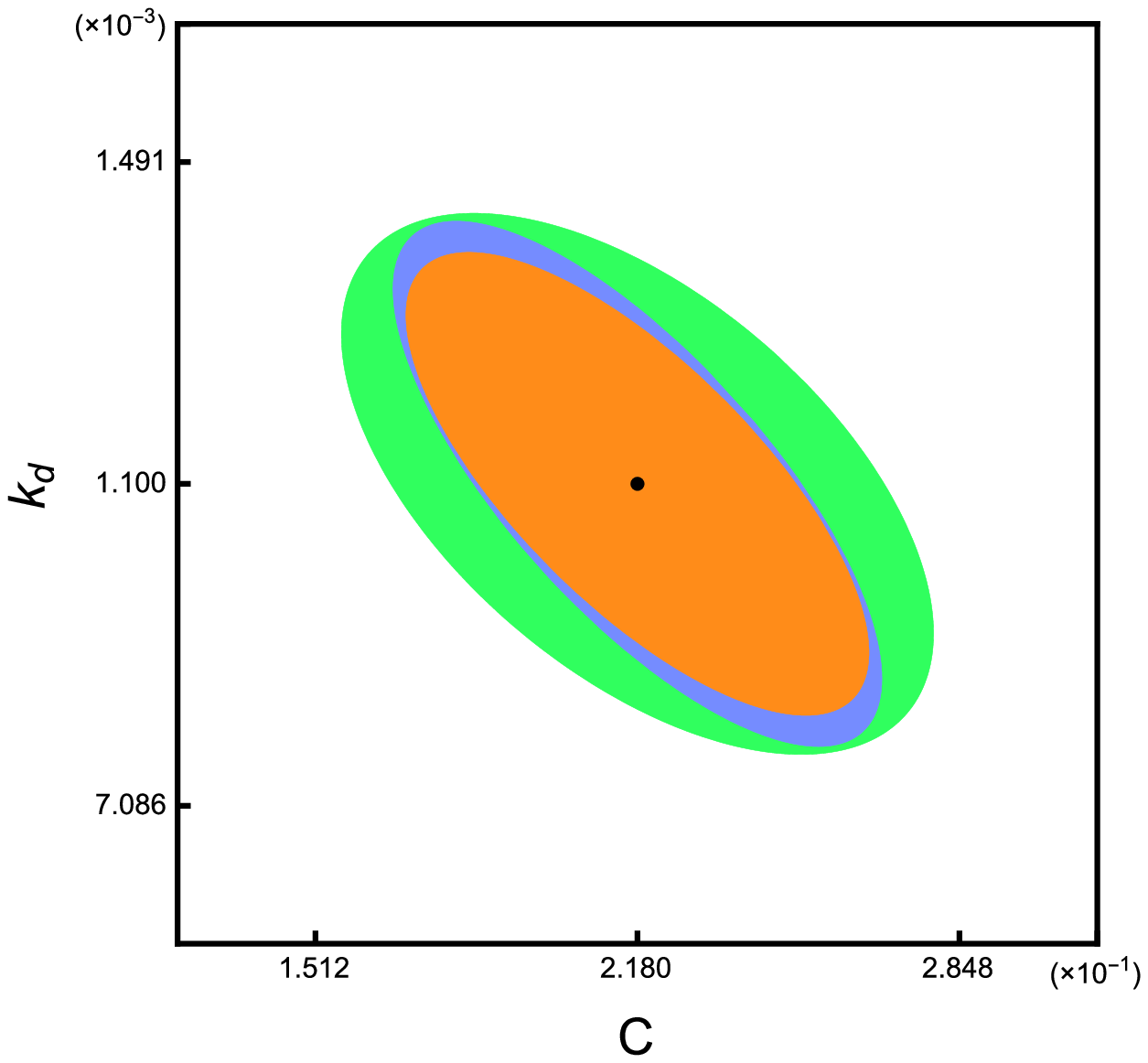,width=1.8in}
\epsfig{figure=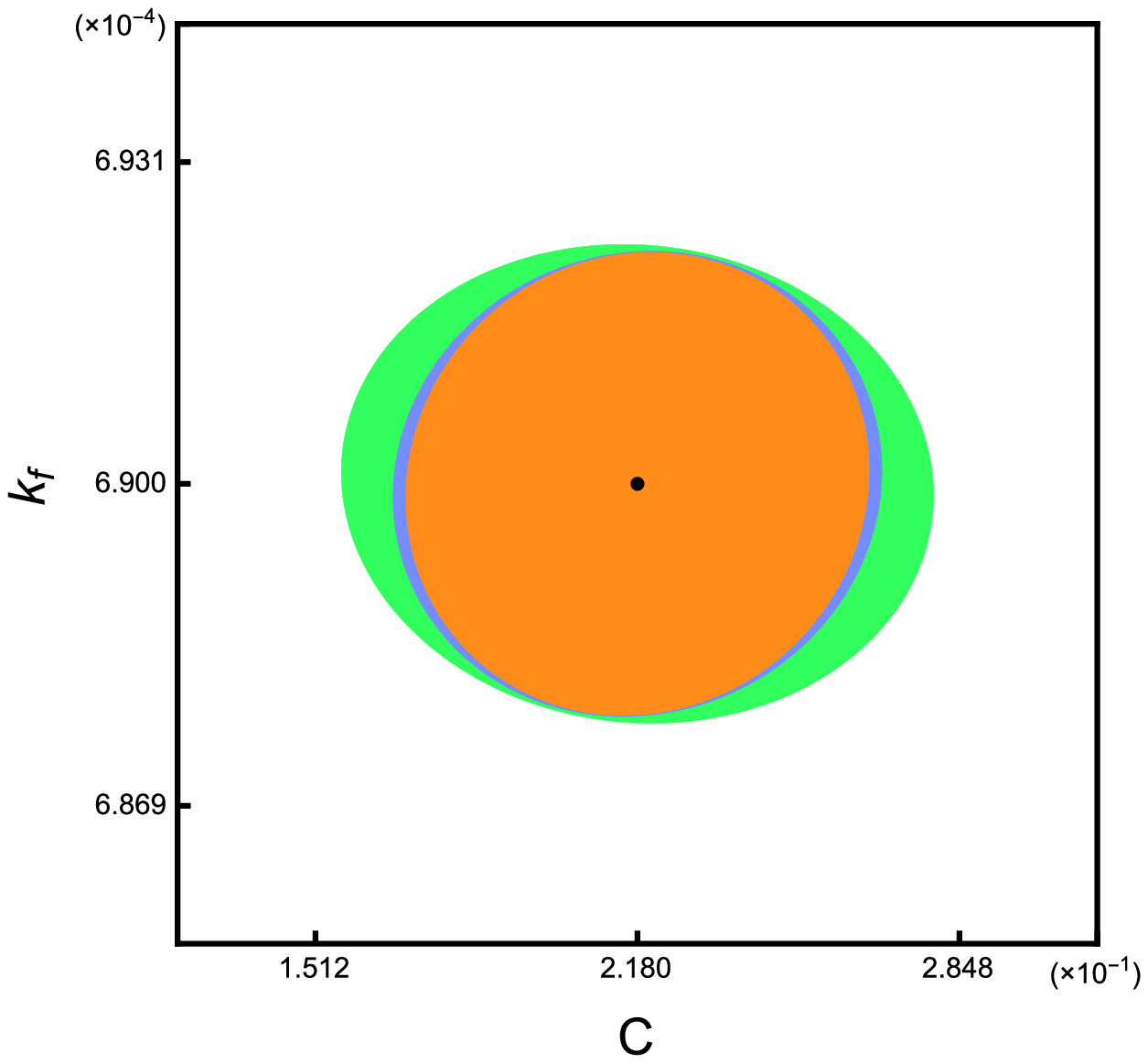,width=1.8in}
\epsfig{figure=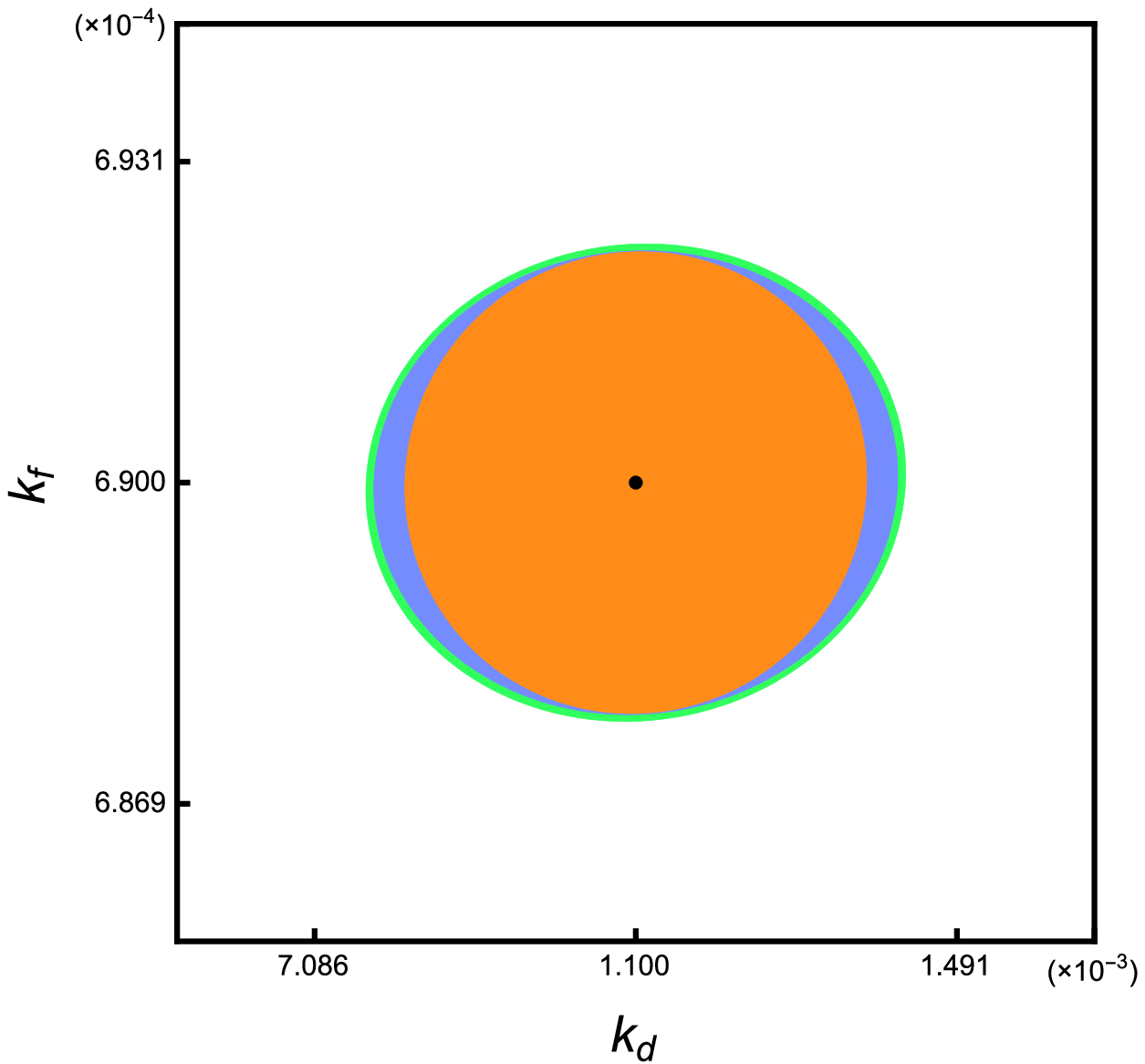,width=1.8in}
\epsfig{figure=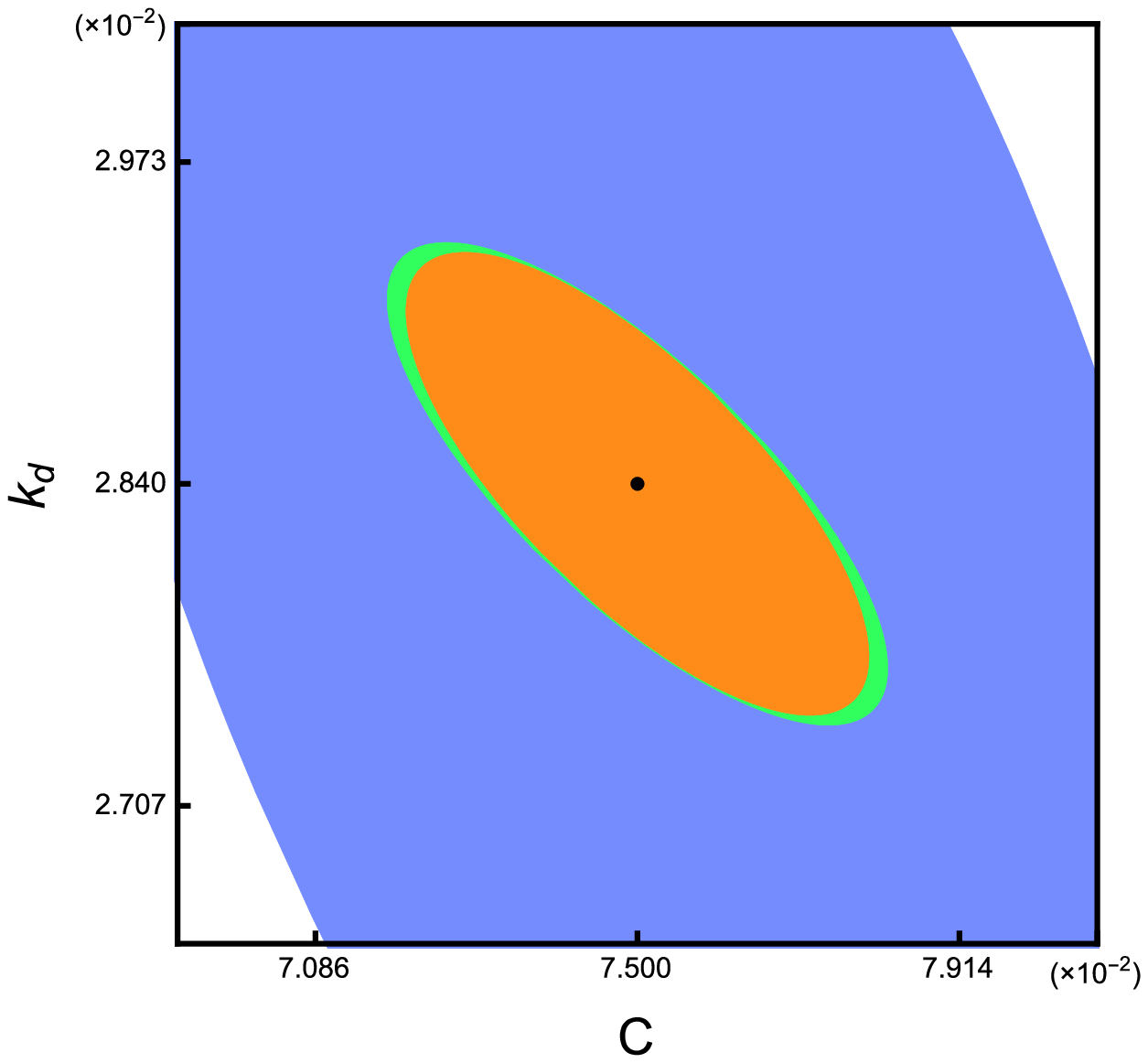,width=1.8in}
\epsfig{figure=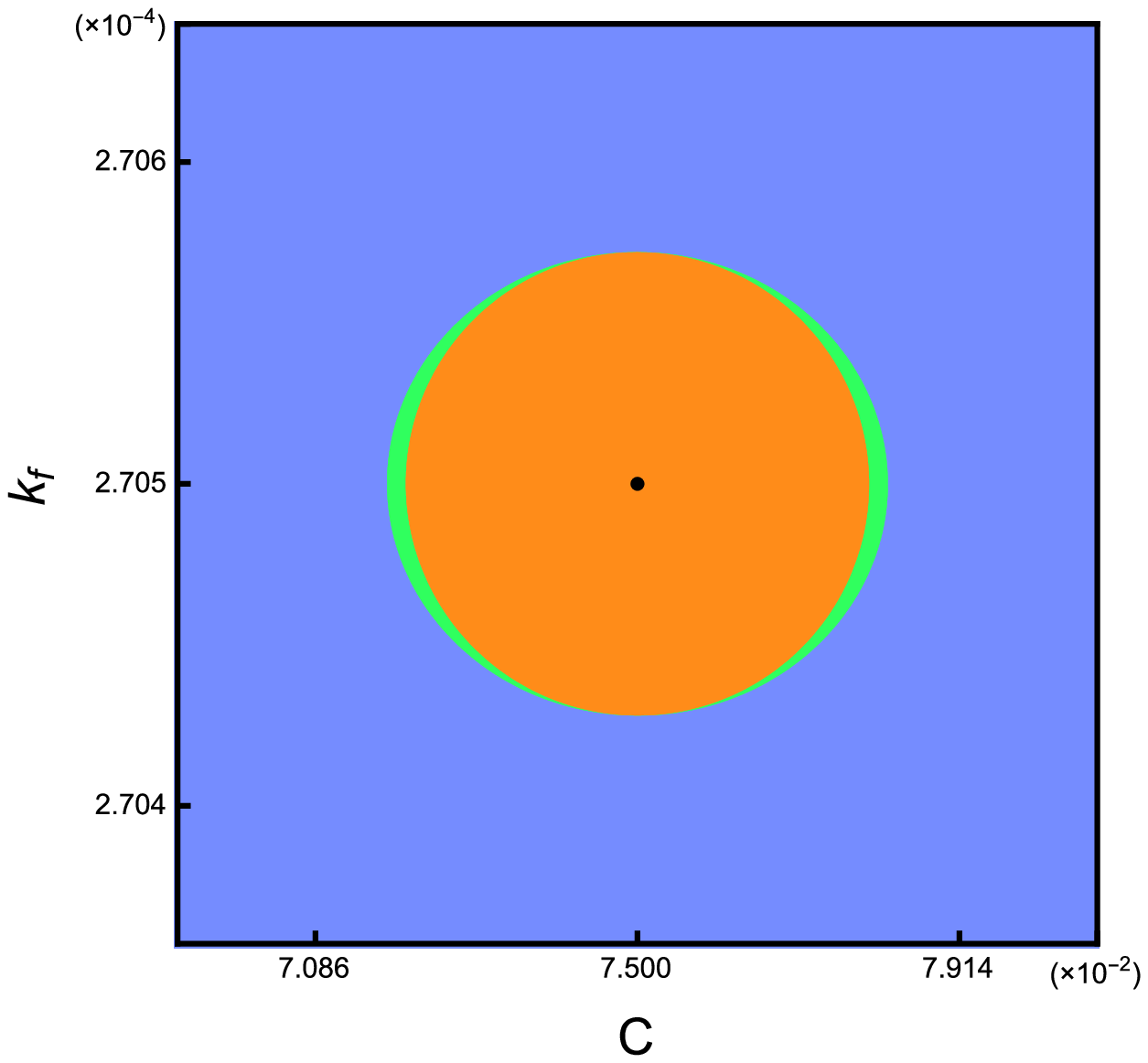,width=1.8in}
\epsfig{figure=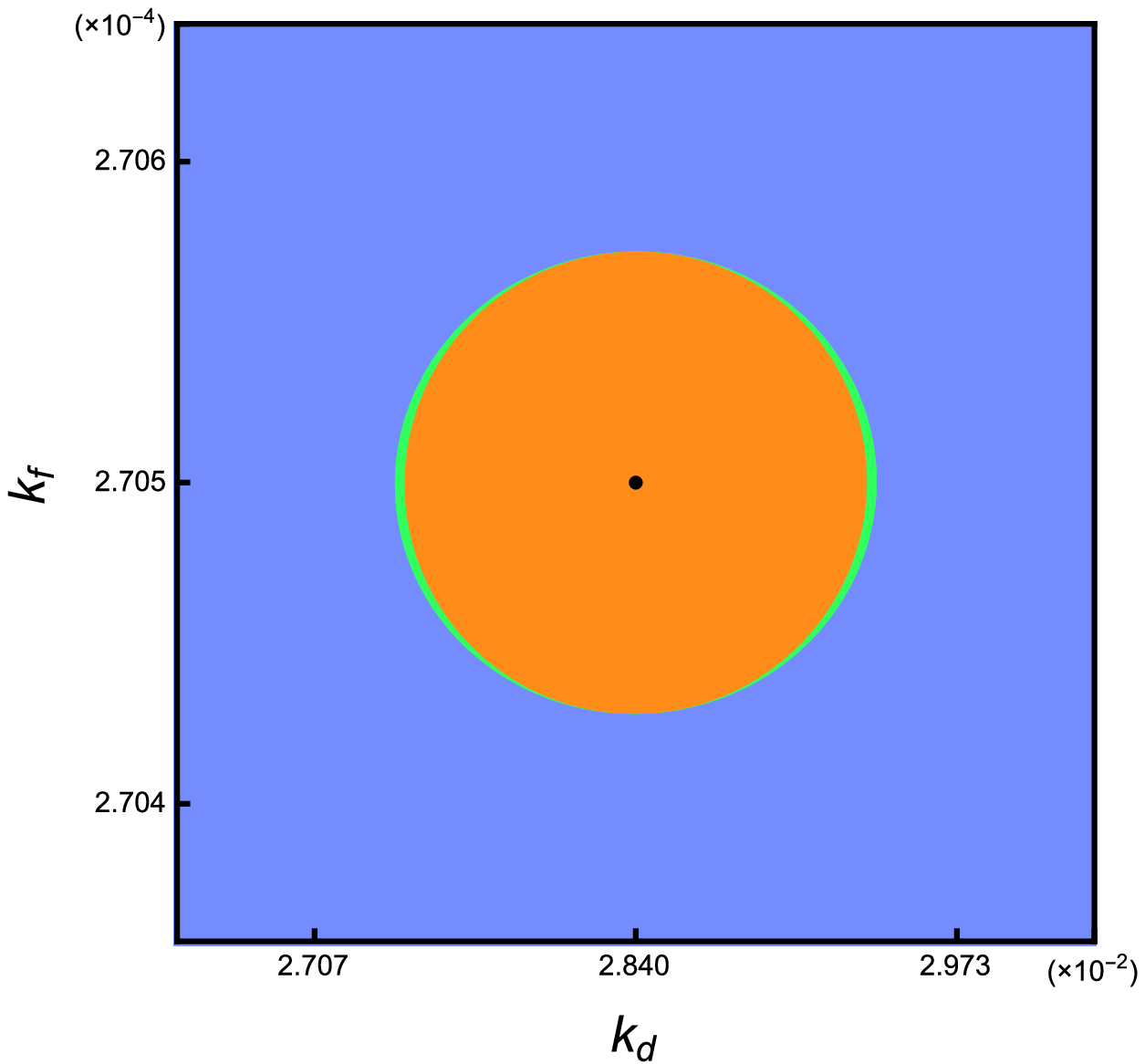,width=1.8in}
\caption{Here we show the sensitivity of the parameters entering in template II --Eq.~\eqref{step}-- for the three observables. In green we plot the constraints from CMB+GC, in light blue from CMB+WL and orange is the sum of the three observables.}
\label{fig:model2}
\end{figure}
%
%\definecolor{lgray}{gray}{0.95}
%\definecolor{dgray}{gray}{0.45}
%\definecolor{gray}{gray}{0.9}
%%%%%%%%%%%%%TAB Mod4 temp 3
\begin{table}[H]
\begin{centering}\begin{tabular}{lccc}
\toprule
%\hline 
% & & \multicolumn{2}{c|}{\textbf{Gaussian Bump} }  \tabularnewline
 & \multicolumn{3}{c}{\textbf{Template 3} }  \tabularnewline
\toprule 
 %\hline
%\rowcolor{gray} \multicolumn{4}{c}{ }  \tabularnewline
%\midrule 
\multicolumn{1}{l}{}&$k^{\rm fid}_{d}$ & $\delta C$  & $\delta k_{d}$   \tabularnewline
% \hline
 %\multicolumn{4}{|c|}{ }  \tabularnewline

 \rowcolor{gray}\multicolumn{1}{l}{\bf{CMB + GC} }& &\multicolumn{2}{c}{ }  \tabularnewline
 %\hline
  & $0.05$ &  $7.2641 \cdot 10^{-4}$ & $1.8118 \cdot 10^{-2}$ \\
  \hline
  & $0.1$  & $6.3169 \cdot 10^{-4}$ & $3.1640 \cdot 10^{-2}$\\
  \hline
  & $0.2$ & $7.4073 \cdot 10^{-4}$ & $ 7.4152 \cdot 10^{-2}$
  
  \tabularnewline
%\hline 
% \multicolumn{4}{c}{ }  \tabularnewline
%\hline
\rowcolor{gray} \multicolumn{1}{l}{\bf{CMB + WL} }& &\multicolumn{2}{c}{ }  \tabularnewline
%\hline
  & $0.05$ &  $6.2446 \cdot 10^{-4}$ & $1.5586 \cdot 10^{-2}$ \\
  \hline
  & $0.1$  & $5.2487 \cdot 10^{-4}$ & $2.6385 \cdot 10^{-2}$\\
  \hline
  & $0.2$ & $5.1719 \cdot 10^{-4}$ & $5.1932 \cdot 10^{-2}$
\tabularnewline
\rowcolor{gray} \multicolumn{1}{l}{\bf{CMB} + \bf{GC} + \bf{WL} }& &\multicolumn{2}{c}{ }  \tabularnewline
%\hline
  & $0.05$ &  $5.6087 \cdot 10^{-4}$ & $ 1.4004 \cdot 10^{-2}$ \\
  \hline
  & $0.1$  & $4.4374 \cdot 10^{-4}$ & $2.2304 \cdot 10^{-2}$\\
  \hline
  & $0.2$ & $4.3496 \cdot 10^{-4}$ & $4.3656 \cdot 10^{-2}$

  \tabularnewline
\bottomrule	
\end{tabular}\par\end{centering}
\caption{Constraints on the parameters of template of Eq.~\eqref{bump} for all the observables.
\label{tab:model4-forecasts}}
\end{table}
\begin{figure}[H]
\centering
\epsfig{figure=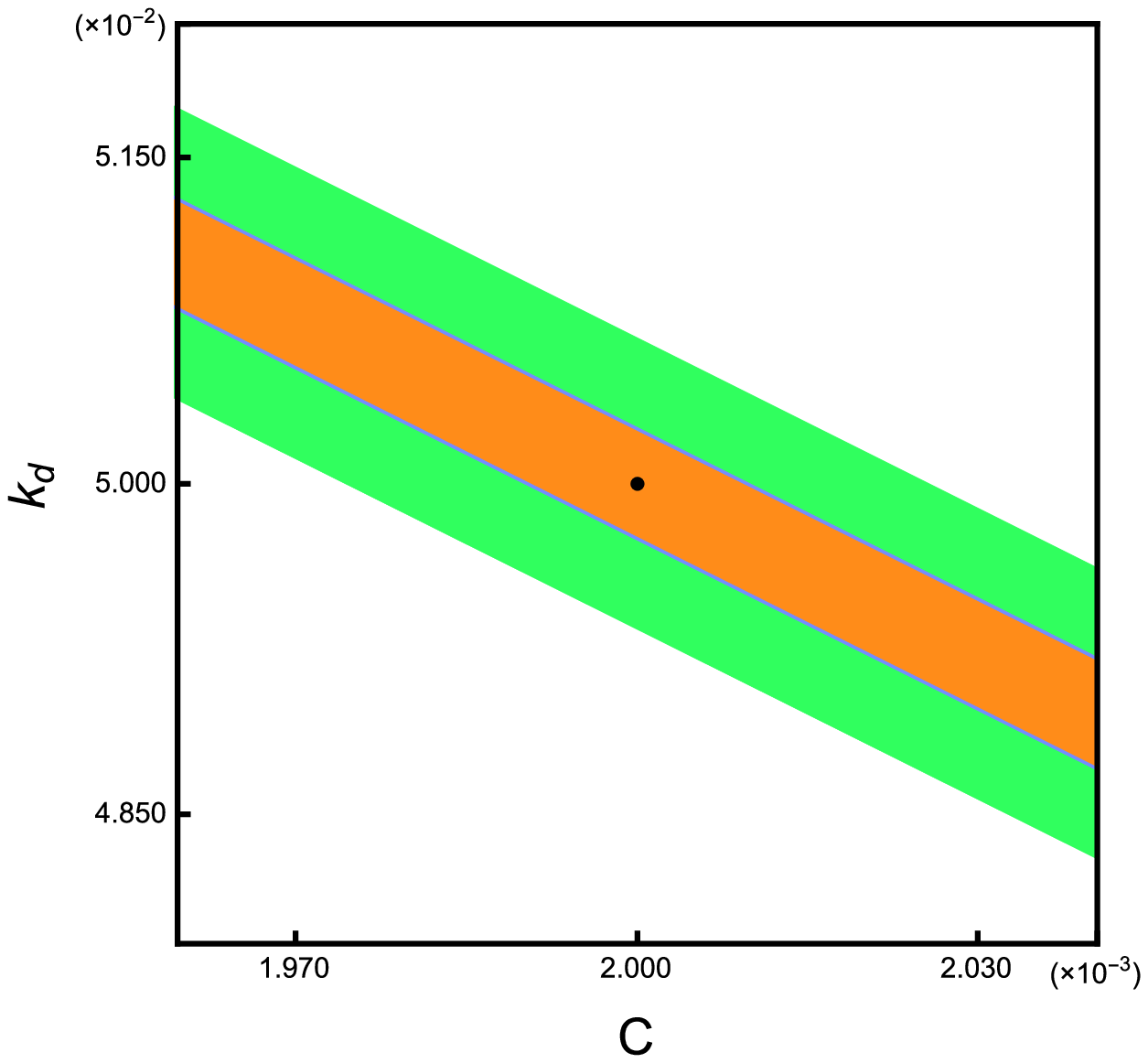,width=1.8in}
\epsfig{figure=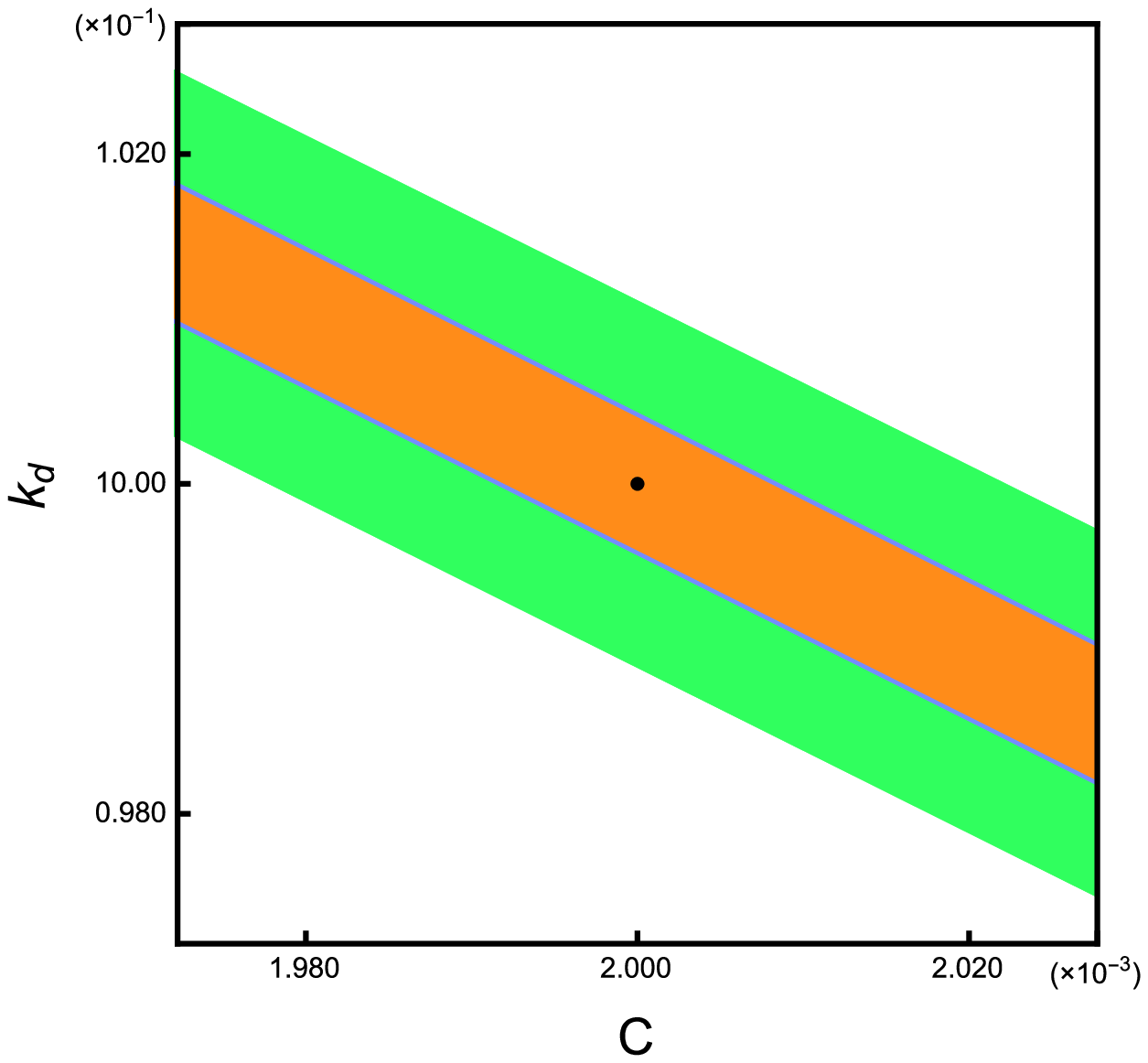,width=1.8in}
\epsfig{figure=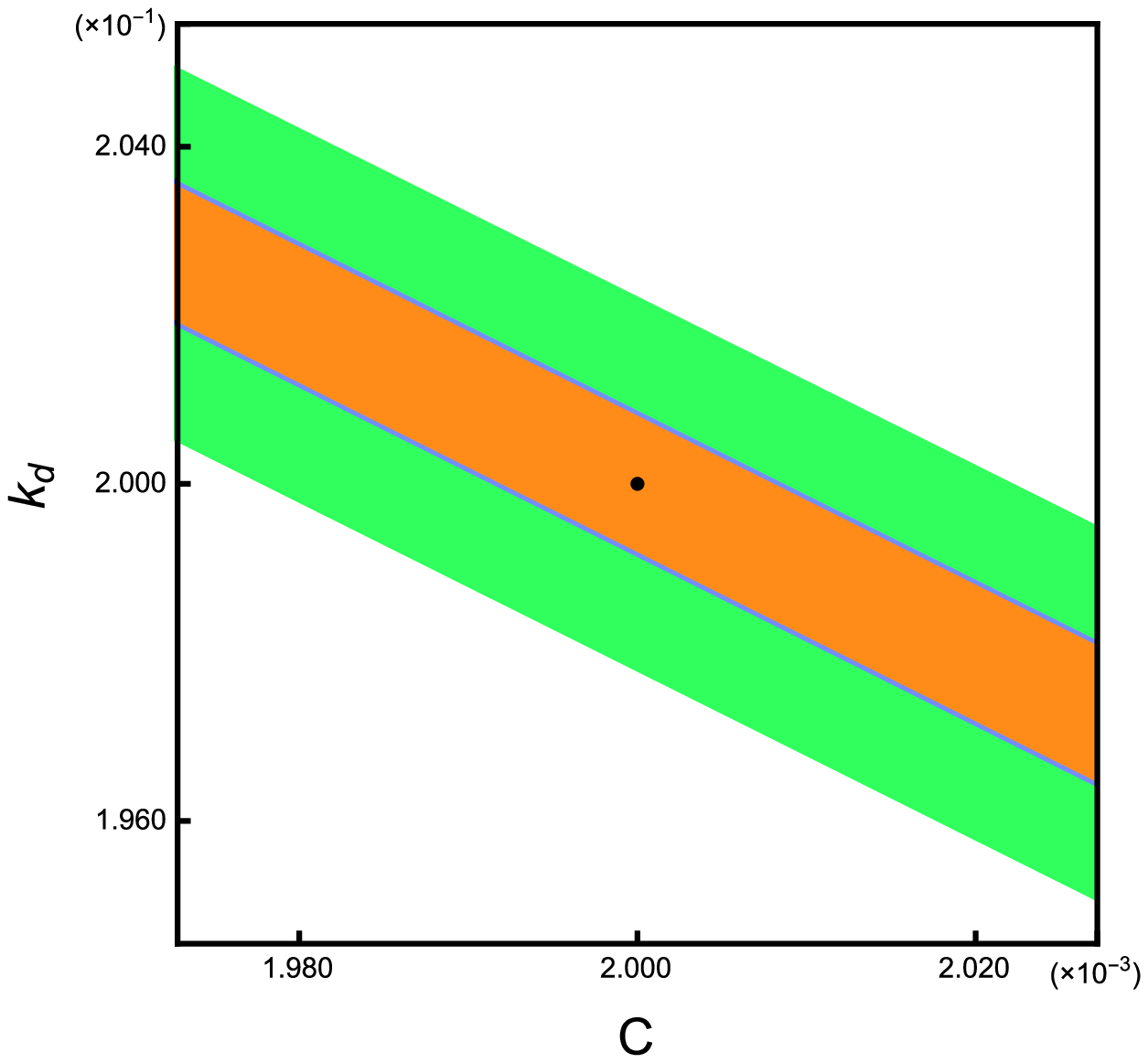,width=1.8in}
\caption{Here we show the sensitivity of the parameters entering in template III --Eq.~\eqref{bump}-- for the three observables. In green we plot the constraints from CMB+GC, in light blue from CMB+WL and orange is the sum of the three observables.}
\label{fig:model4}
\end{figure}
% 
%
%TAB Mod1 temp1
\begin{table}[H]
\begin{centering}\begin{tabular}{cccccc}
\toprule
& & & \multicolumn{3}{c}{\textbf{ Template 4} }  \tabularnewline
 \toprule
%\rowcolor{gray} \multicolumn{5}{c}{}  \tabularnewline
%\hline 
\multicolumn{2}{c}{}& $k_f^{\rm fid}$ & $\delta C$ & $\delta  k_{f}$ & $\delta \phi$   \tabularnewline
% \hline
% \multicolumn{5}{|c|}{ }  \tabularnewline
%\hline
 \rowcolor{gray}\multicolumn{1}{l}{\bf{CMB} + {\bf GC} }& &\multicolumn{4}{c}{ }  \tabularnewline
% \hline
  & & 0.004 & $4.2901 \cdot 10^{-4}$ & $2.4332 \cdot 10^{-6}$ & $4.5239 \cdot 10^{-2}$ \\ \hline
    & & 0.03 & $6.0884 \cdot 10^{-4}$ & $3.2235 \cdot 10^{-4}$ & $1.3316 \cdot 10^{-2}$ \\ \hline
      & &  0.1 & $1.6058 \cdot 10^{-3}$ & $1.9484 \cdot 10^{-3}$ & $6.0761 \cdot 10^{-2}$
    \tabularnewline
%\hline 
% \multicolumn{5}{c}{ }  \tabularnewline
%\hline
 \rowcolor{gray} \multicolumn{1}{l}{\bf{CMB} + {\bf WL} }& &\multicolumn{4}{c}{ }  \tabularnewline
%\hline 
& &0.004 & $3.5485 \cdot 10^{-3}$ & $8.0881 \cdot 10^{-6}$ & $1.0676 \cdot 10^{-1}$ \\\hline
& &0.03 & $9.01301 \cdot 10^{-4}$ & $5.0095 \cdot 10^{-4}$ & $2.4780 \cdot 10^{-2}$ \\\hline
& &0.1 & $9.5431 \cdot 10^{-4}$ & $1.8158 \cdot 10^{-3}$ & $2.3287 \cdot 10^{-2}$ \tabularnewline
%\hline 
%\rowcolor{gray} \multicolumn{5}{c}{ }  \tabularnewline
%\hline
 \rowcolor{gray} \multicolumn{1}{l}{ \bf{CMB} + \bf{GC}+\bf{WL} }& &\multicolumn{4}{c}{ }  \tabularnewline
 %\hline, 5.84194, 12.1132, 
  & &0.004 & $ 4.2677 \cdot 10^{-4}$ & $2.1838 \cdot 10^{-6}$ & $4.0459 \cdot 10^{-2}$ \\\hline
    & &0.03 & $ 4.1864 \cdot 10^{-4}$ & $2.7507 \cdot 10^{-4}$ & $1.0128 \cdot 10^{-2}$ \\\hline
      & &0.1 & $5.8419 \cdot 10^{-4}$ & $1.2113 \cdot 10^{-3}$ & $1.3087 \cdot 10^{-2}$
  \tabularnewline
\bottomrule	
\end{tabular}\par\end{centering}
\caption{Constraints on the parameters of the template of Eq.~\eqref{sharp} for all the observables.
\label{tab:model1-forecasts}}
\end{table}
%
%TAB Mod3 temp5
\begin{table}[h]
\begin{centering}\begin{tabular}{cccccc}
\toprule
& & & \multicolumn{3}{c}{\textbf{ Template 5} }  \tabularnewline
 \toprule
%\rowcolor{gray} \multicolumn{5}{c}{}  \tabularnewline
%\hline 
\multicolumn{2}{c}{}& $\Omega^{\rm fid}$ & $\delta C$ & $\delta  \Omega$ & $\delta \phi$   \tabularnewline
% \hline
% \multicolumn{5}{|c|}{ }  \tabularnewline
%\hline, 4.03403, 244.16, 142.034
 \rowcolor{gray}\multicolumn{1}{l}{\bf{CMB} + {\bf GC} }& &\multicolumn{4}{c}{ }  \tabularnewline
% \hline
  & & 5 & $4.1475 \cdot 10^{-4}$ & $3.3119 \cdot 10^{-3}$ & $1.5687 \cdot 10^{-2}$ \\ \hline
    & & 30 & $4.2242 \cdot 10^{-4}$ & $7.6280 \cdot 10^{-3}$ & $1.4770 \cdot 10^{-2}$ \\ \hline
      & & 100 & $4.0340 \cdot 10^{-4}$ & $2.4416 \cdot 10^{-2}$ & $1.4203 \cdot 10^{-2}$
    \tabularnewline
%\hline 
% \multicolumn{5}{c}{ }  \tabularnewline
%\hline
 \rowcolor{gray} \multicolumn{1}{l}{\bf{CMB} + {\bf WL} }& &\multicolumn{4}{c}{ }  \tabularnewline
%\hline 
& & 5 & $9.2838 \cdot 10^{-4}$ & $5.8846 \cdot 10^{-3}$ & $3.2115 \cdot 10^{-2}$ \\\hline
& & 30 & $5.7795 \cdot 10^{-3}$ & $3.1737 \cdot 10^{-1}$ & $1.8879 \cdot 10^{-1}$ \\\hline
& & 100 & $8.0314 \cdot 10^{-3}$ & $6.1294 \cdot 10^{-1}$ & $2.689 \cdot 10^{-1}$ \tabularnewline
%\hline 
%\rowcolor{gray} \multicolumn{5}{c}{ }  \tabularnewline
%\hline
 \rowcolor{gray} \multicolumn{1}{l}{ \bf{CMB} + \bf{GC}+\bf{WL} }& &\multicolumn{4}{c}{ }  \tabularnewline
 %\hline 9, , 137.206
  & & 5 & $3.8219 \cdot 10^{-4}$ & $2.6312 \cdot 10^{-3}$ & $1.3503 \cdot 10^{-2}$ \\\hline
    & & 30 & $4.2196 \cdot 10^{-4}$ & $7.5160 \cdot 10^{-3}$ & $1.4308 \cdot 10^{-2}$ \\\hline
      & & 100 & $4.0264 \cdot 10^{-3}$ & $2.3984 \cdot 10^{-2}$ & $1.3721 \cdot 10^{-2}$
  \tabularnewline
\bottomrule	
\end{tabular}\par\end{centering}
\caption{Constraints on the parameters of the template of Eq.~\eqref{res} for all the observables.
\label{tab:model3-forecasts}}
\end{table}
\begin{figure}[H]
\centering
\epsfig{figure=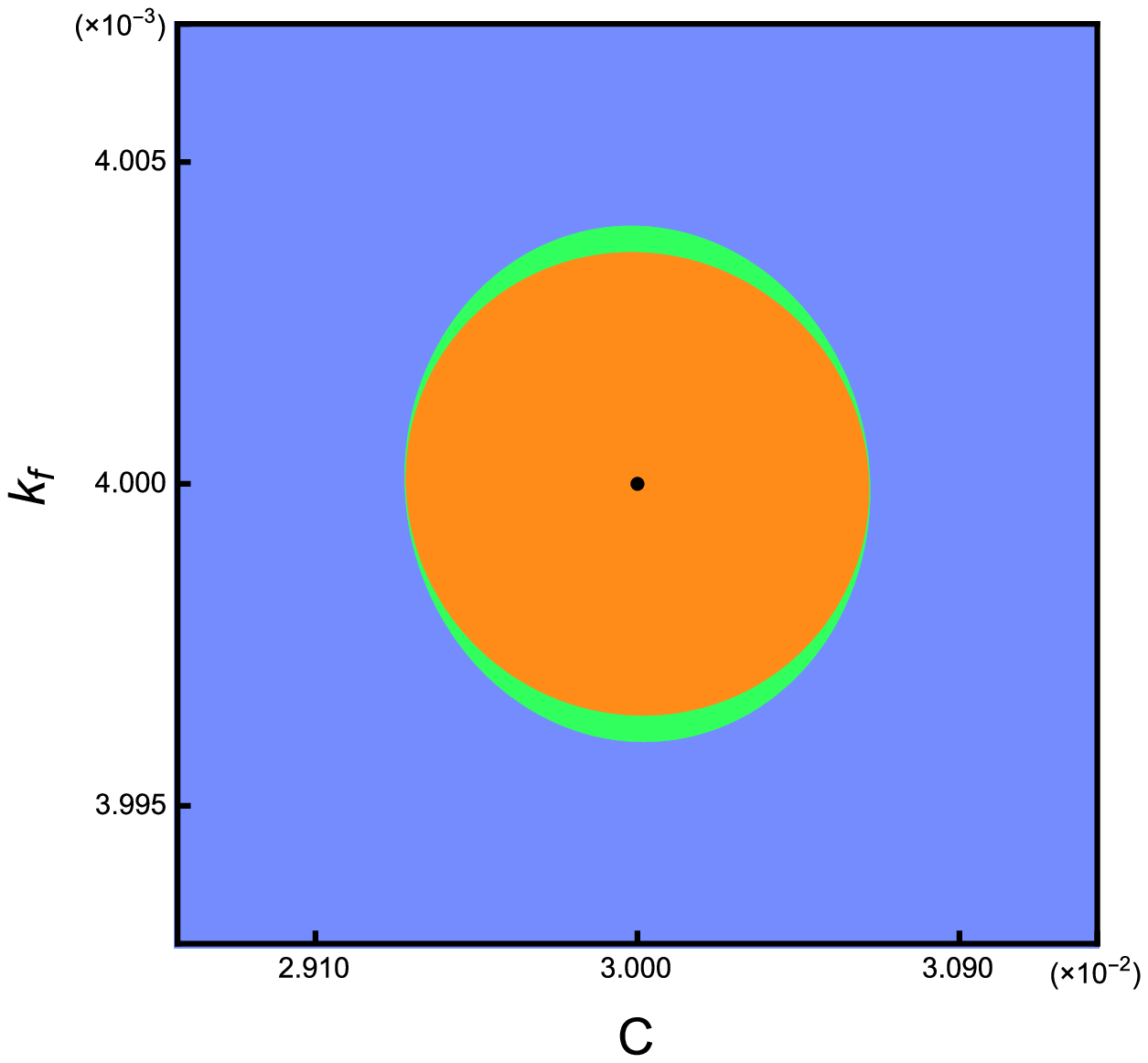,width=1.8in}
\epsfig{figure=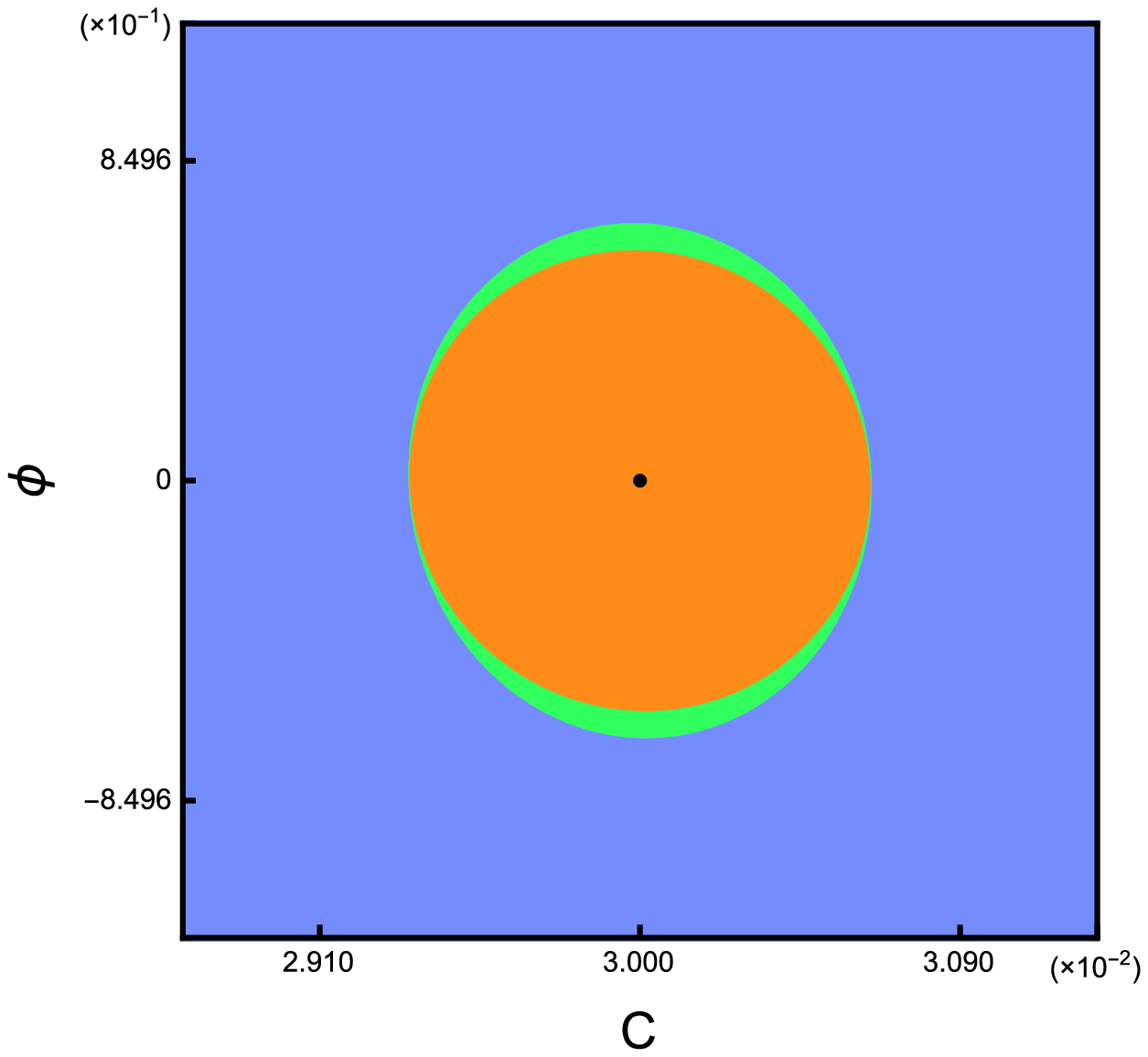,width=1.8in}
\epsfig{figure=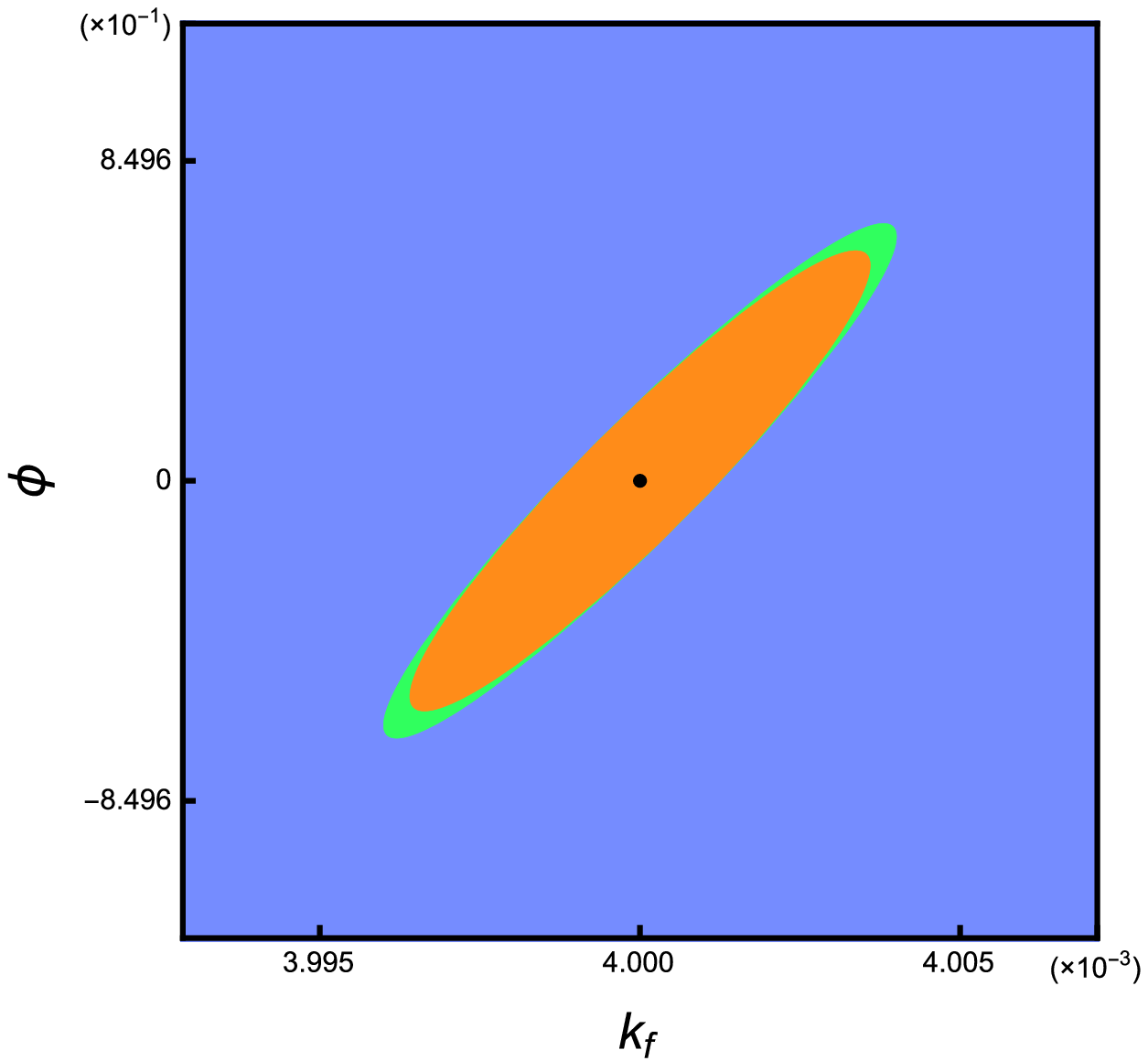,width=1.8in} \\
\epsfig{figure=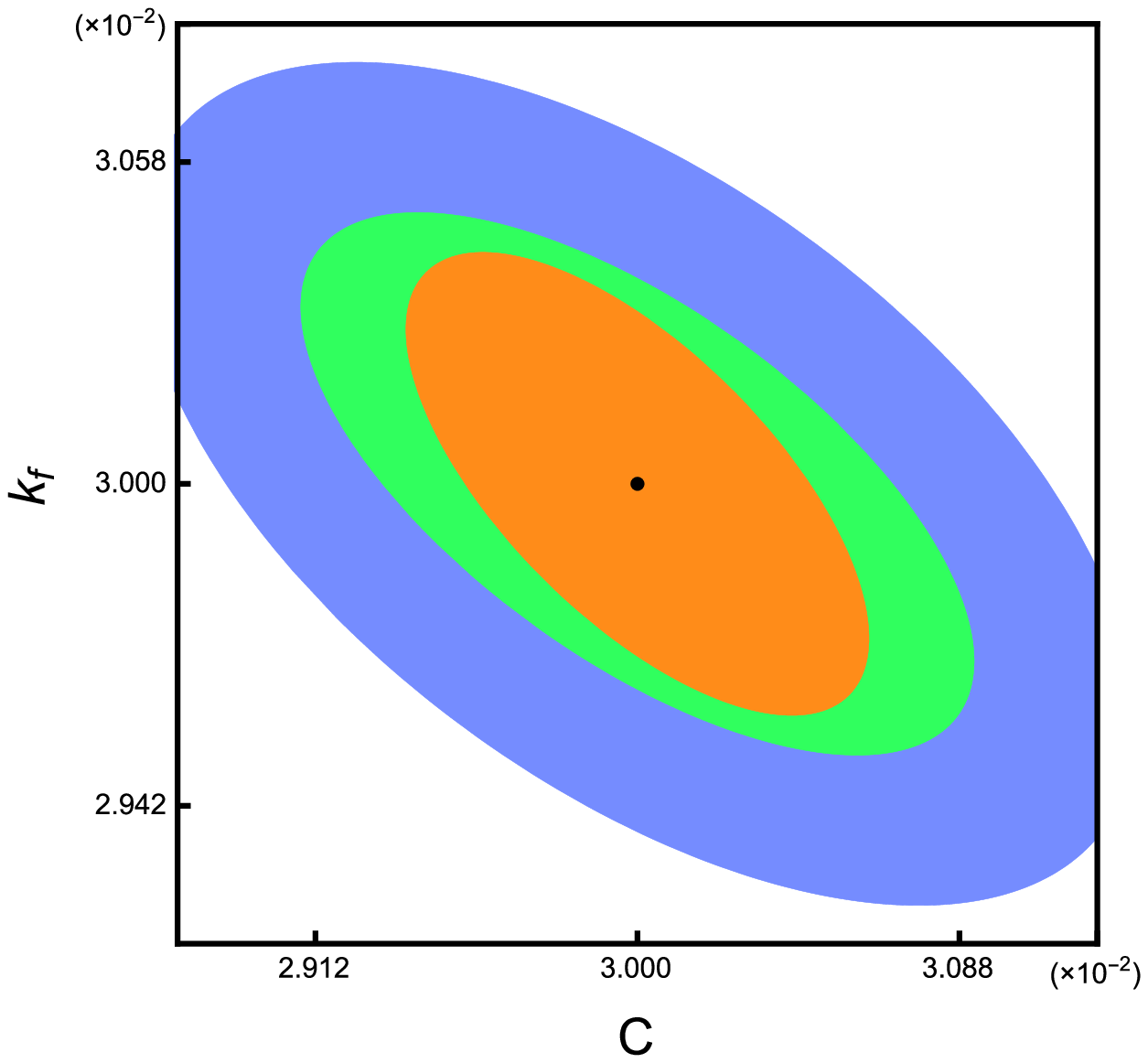,width=1.8in}
\epsfig{figure=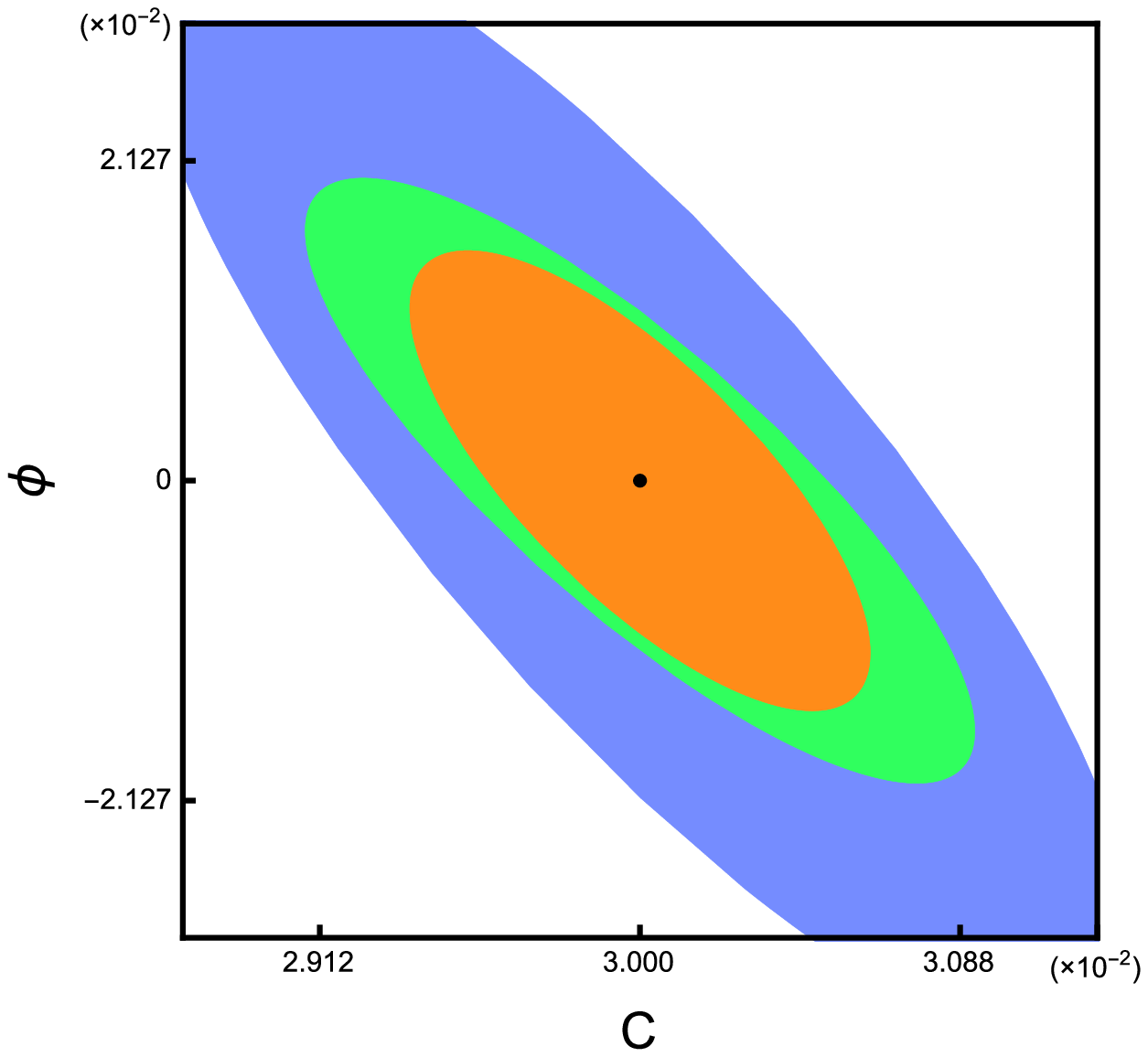,width=1.8in}
\epsfig{figure=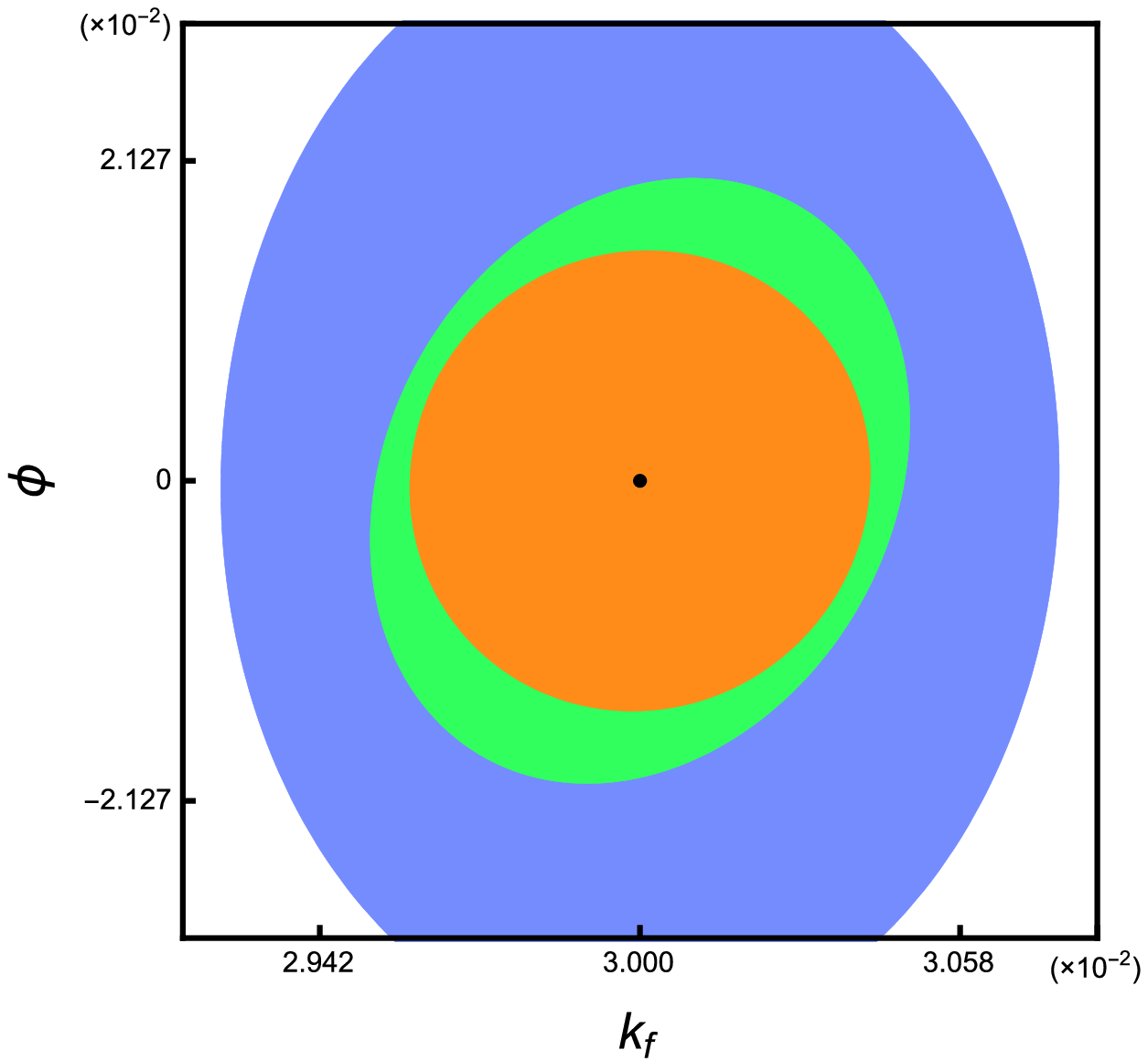,width=1.8in} \\
\epsfig{figure=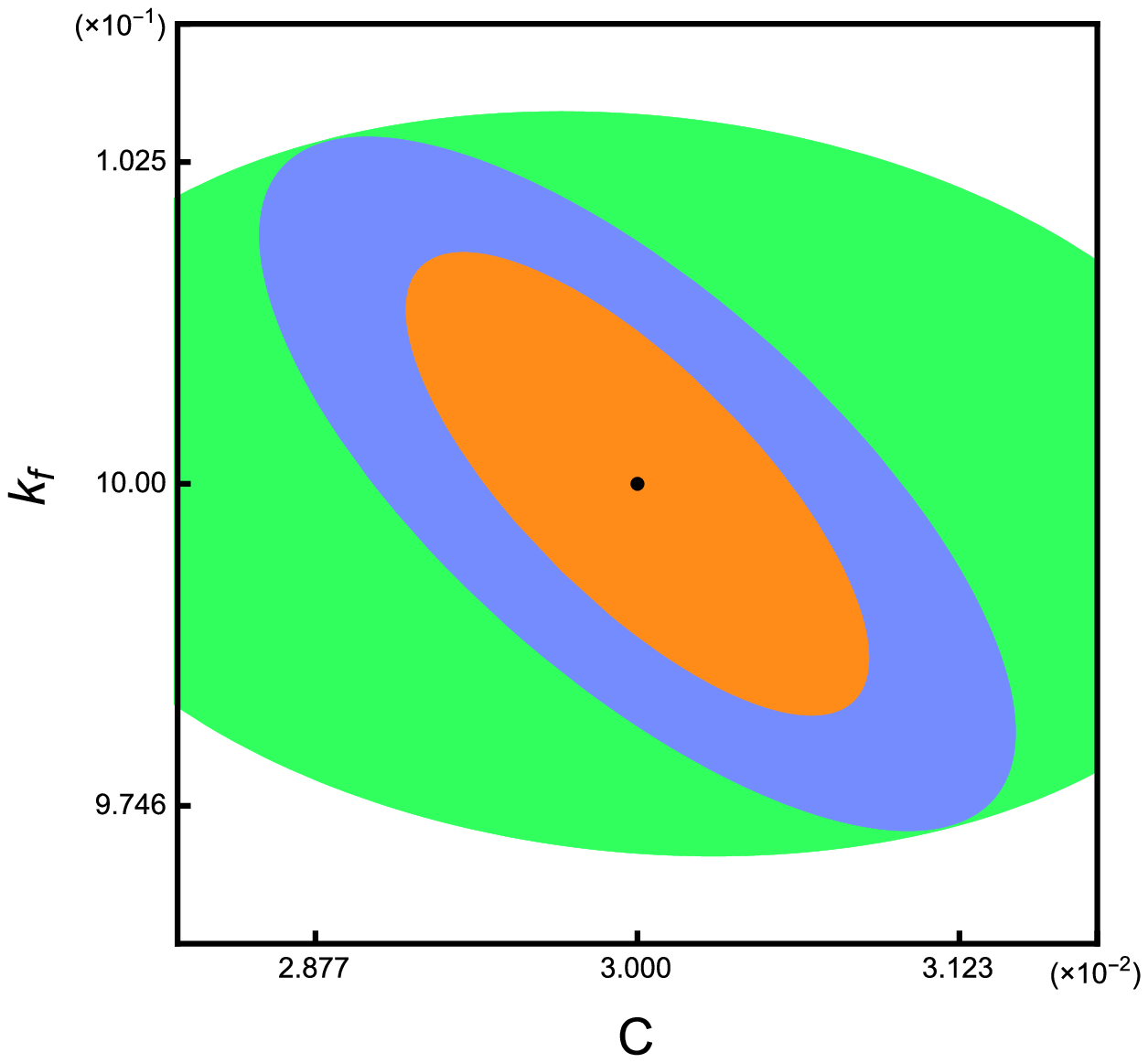,width=1.8in}
\epsfig{figure=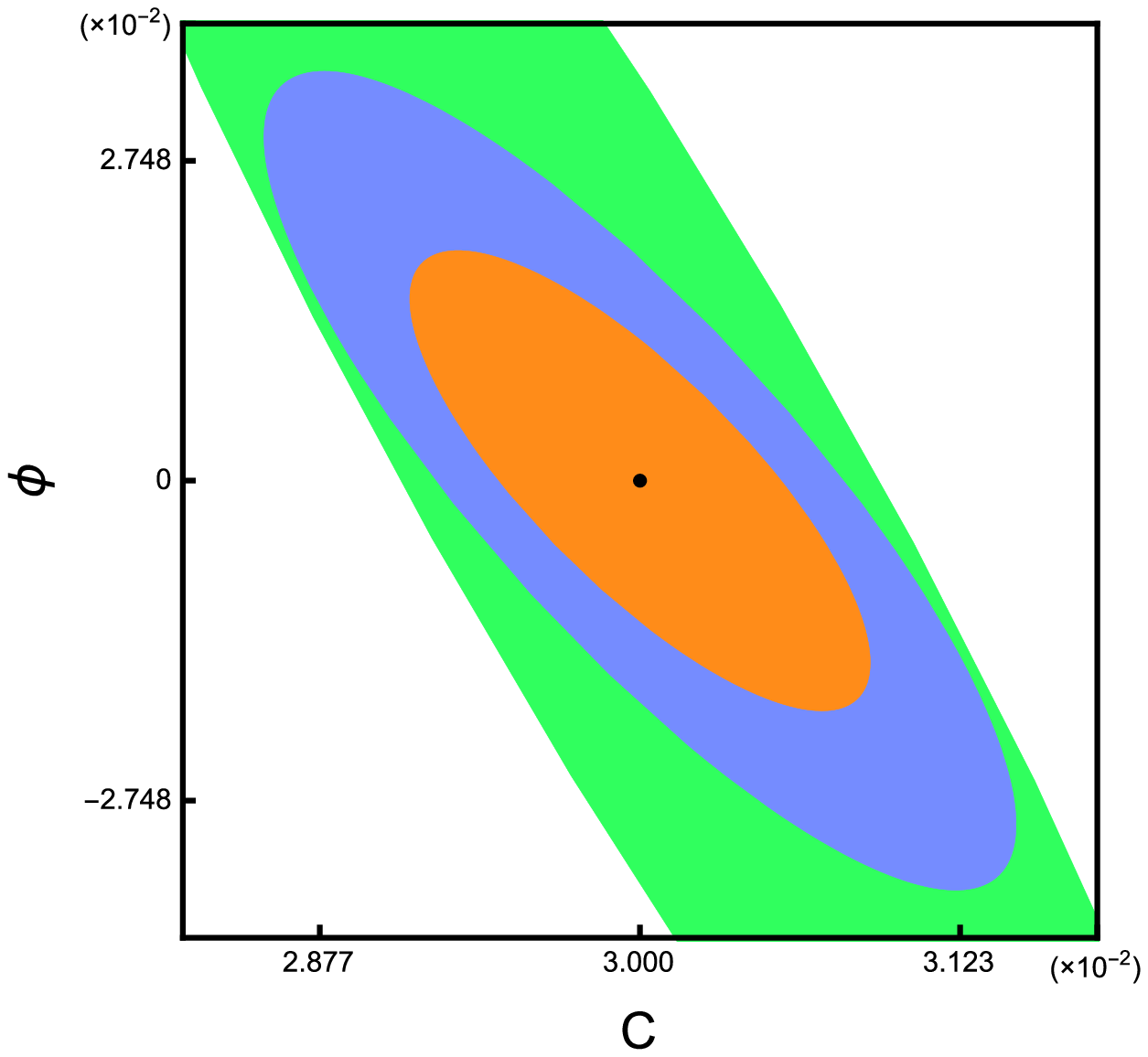,width=1.8in}
\epsfig{figure=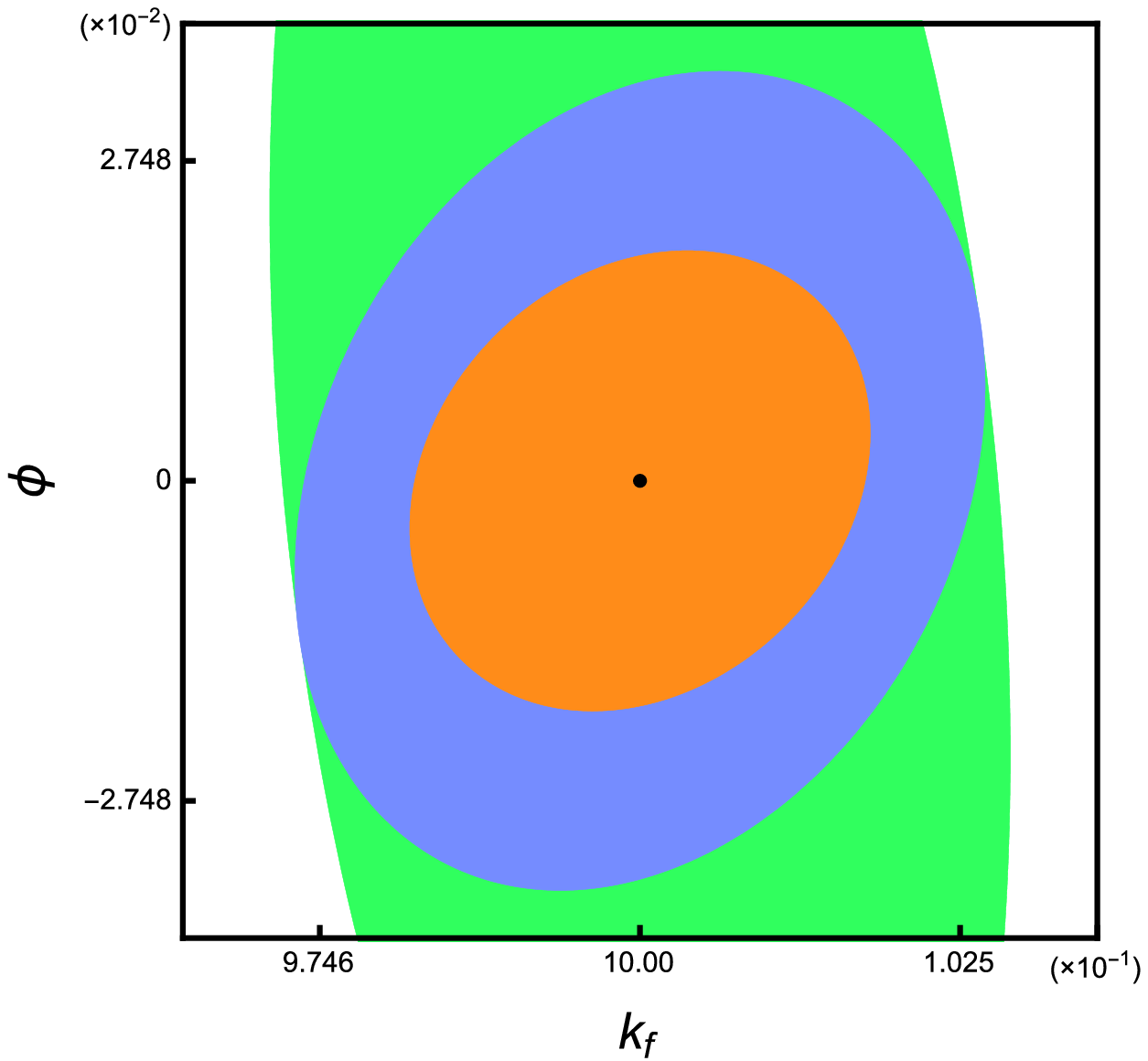,width=1.8in}
\caption{Here we show the sensitivity of the parameters entering in template IV --Eq.~\eqref{sharp}-- for the three observables. In green we plot the constraints from CMB+GC, in light blue from CMB+WL and orange is the sum of the three observables.}
\label{fig:model1}
\end{figure}
\begin{figure}[hp]
\centering
\epsfig{figure=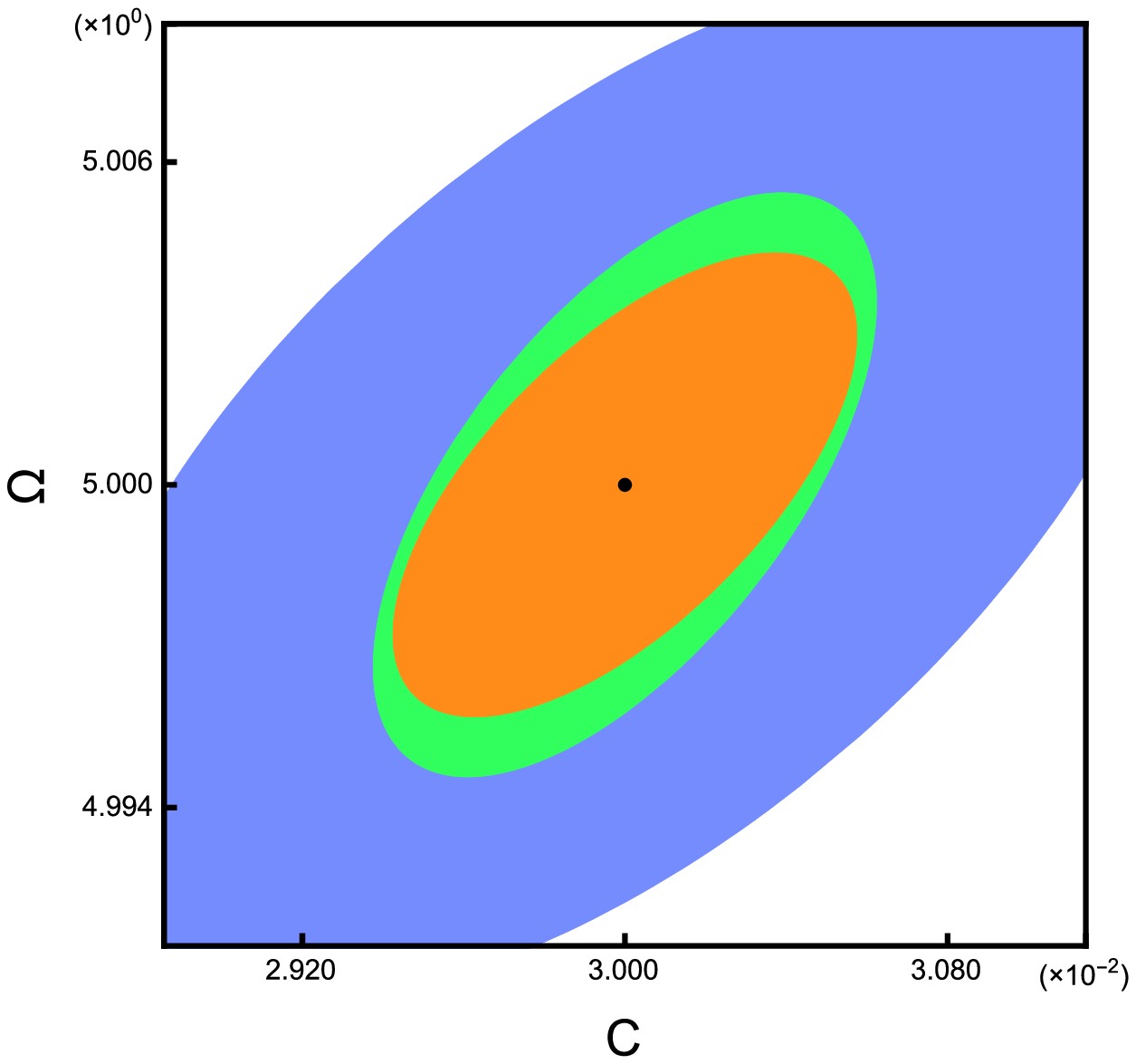,width=1.8in}
\epsfig{figure=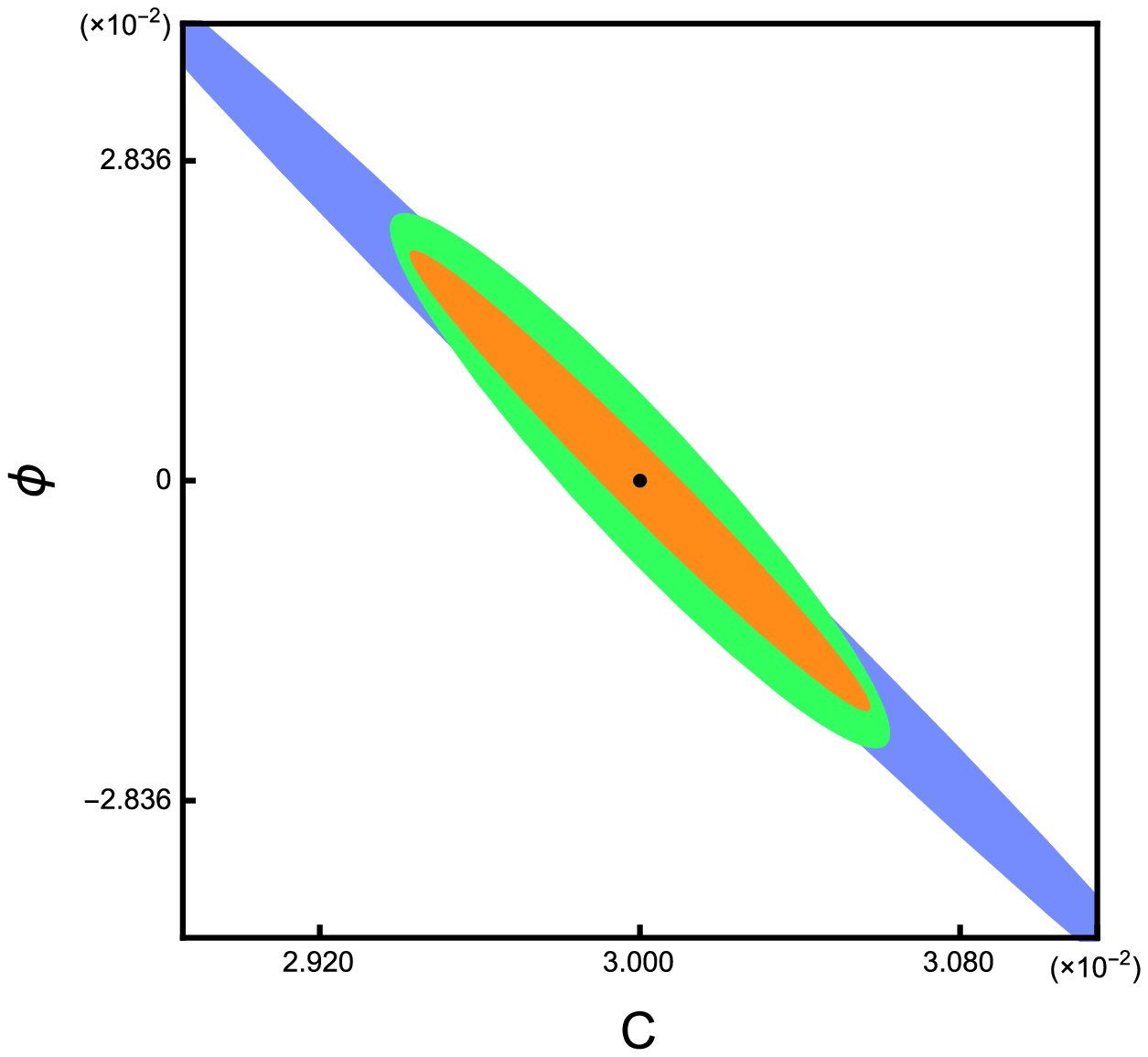,width=1.8in}
\epsfig{figure=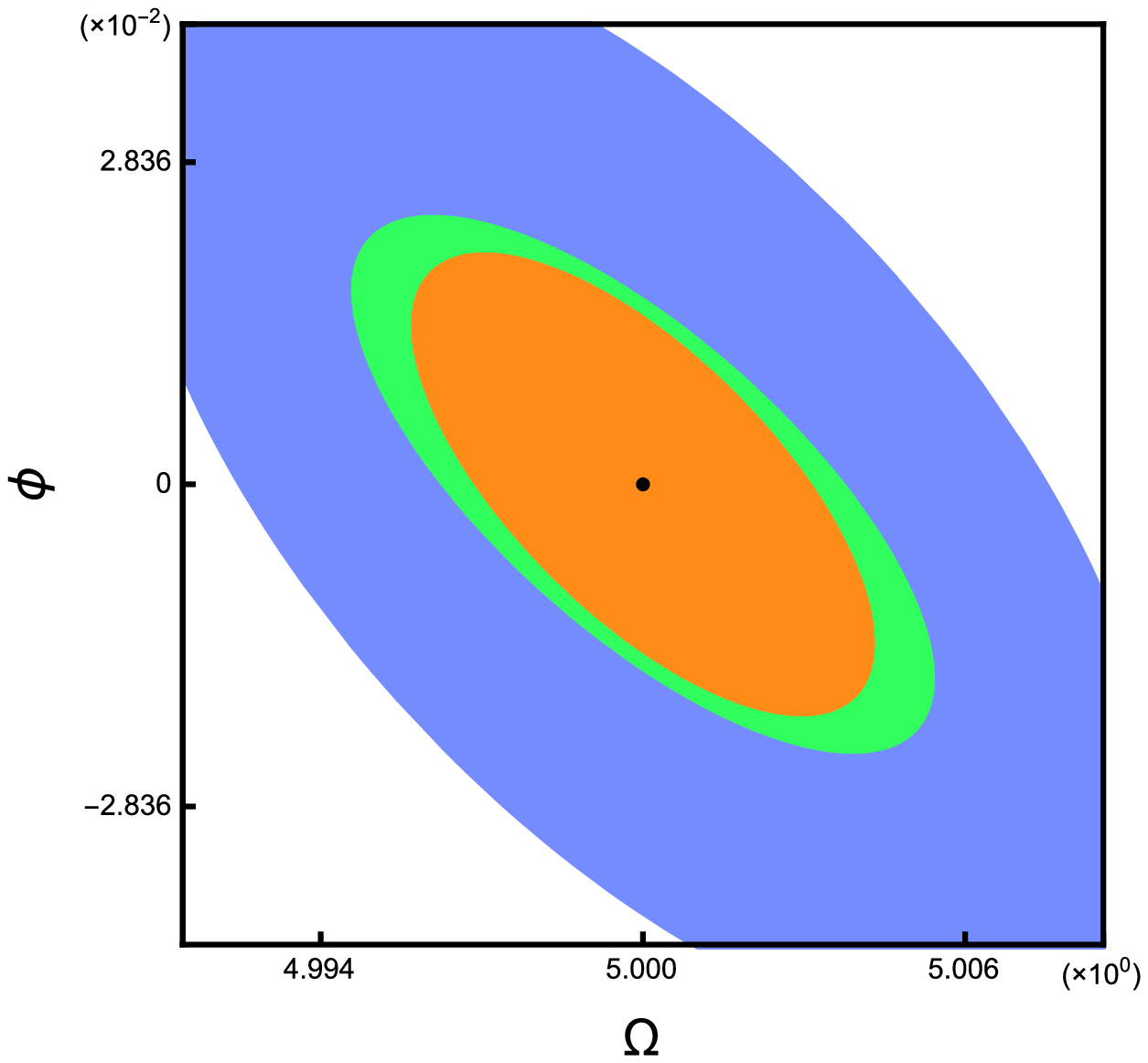,width=1.8in} \\
\epsfig{figure=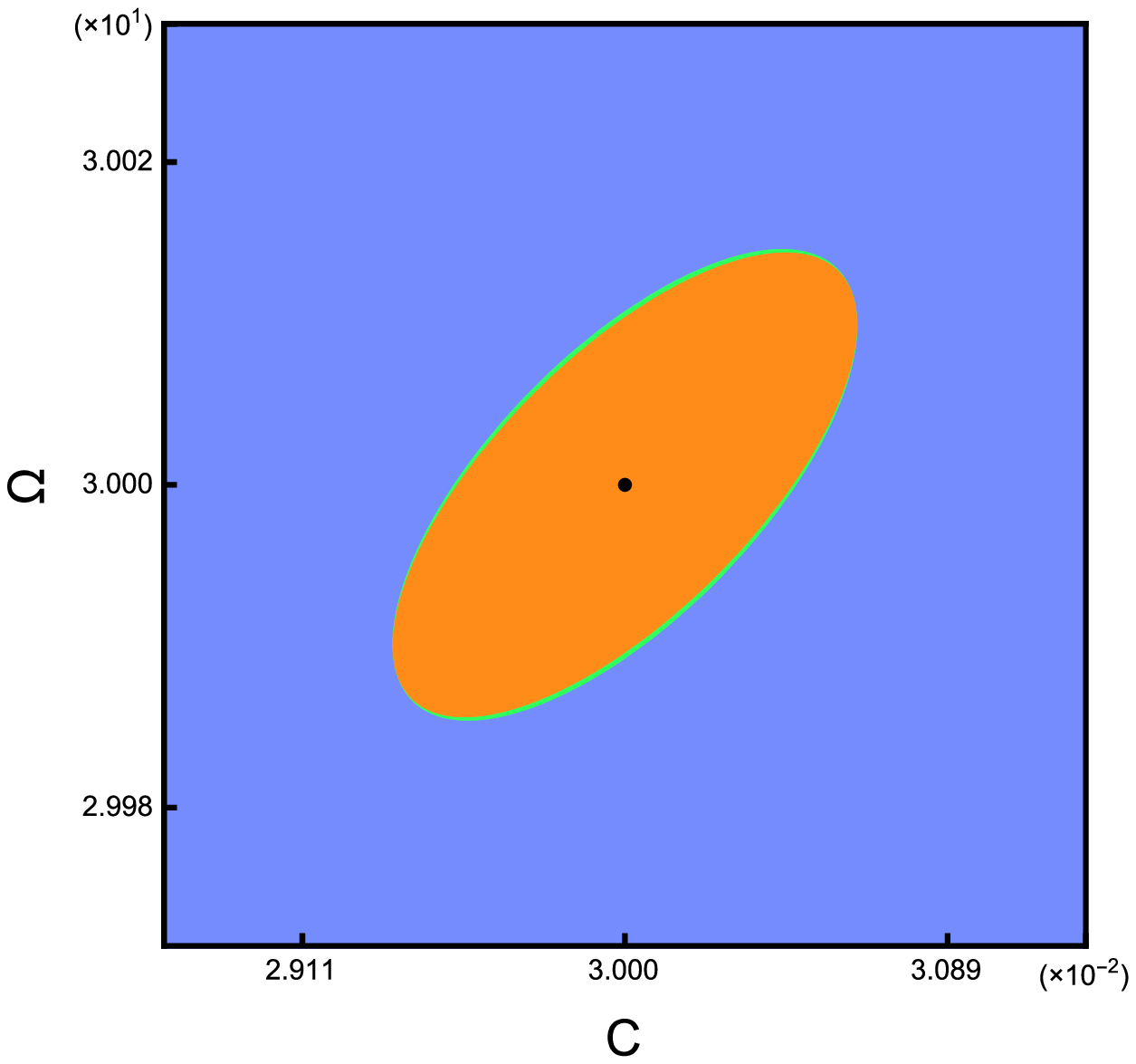,width=1.8in}
\epsfig{figure=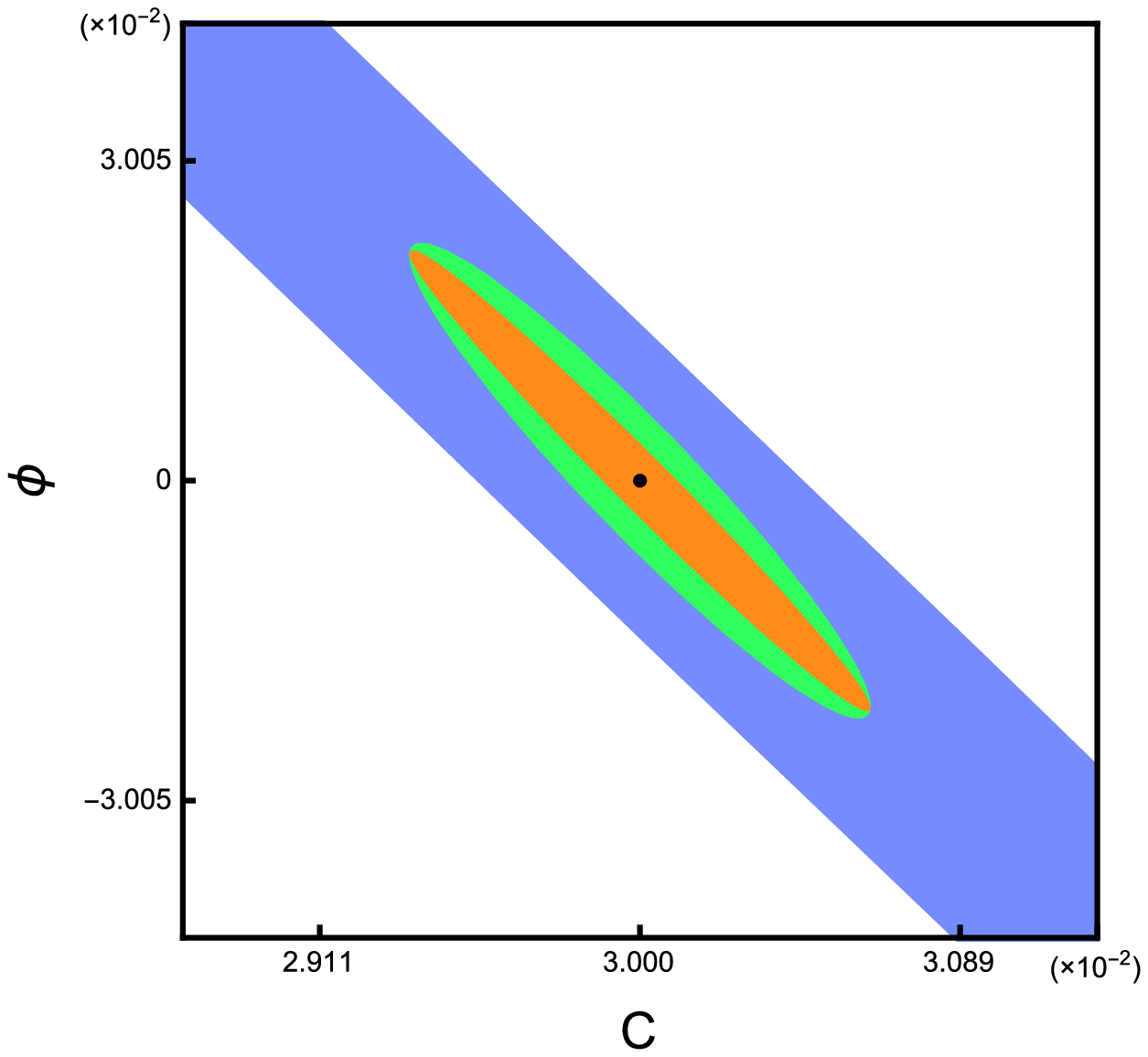,width=1.8in}
\epsfig{figure=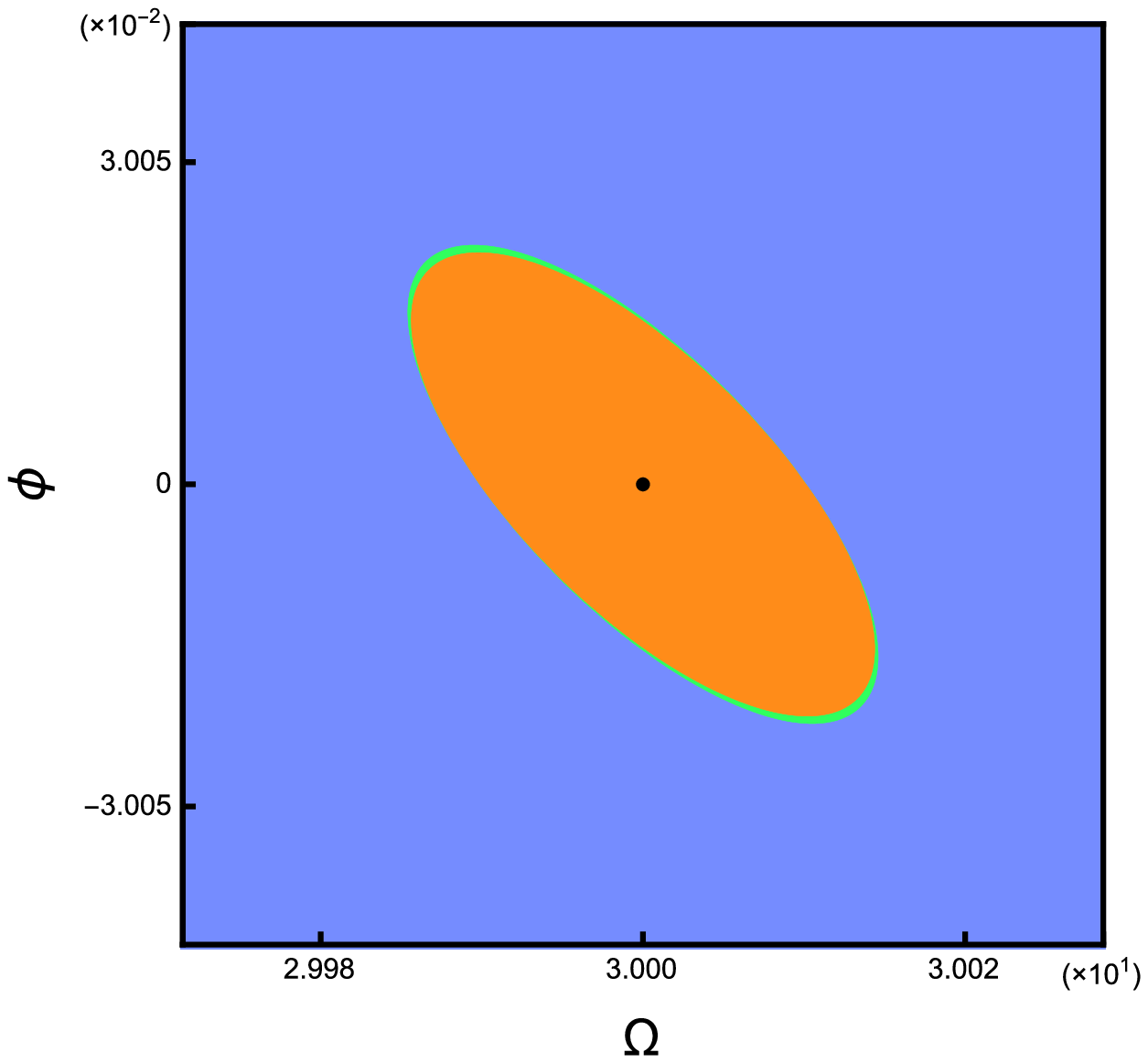,width=1.8in} \\
\epsfig{figure=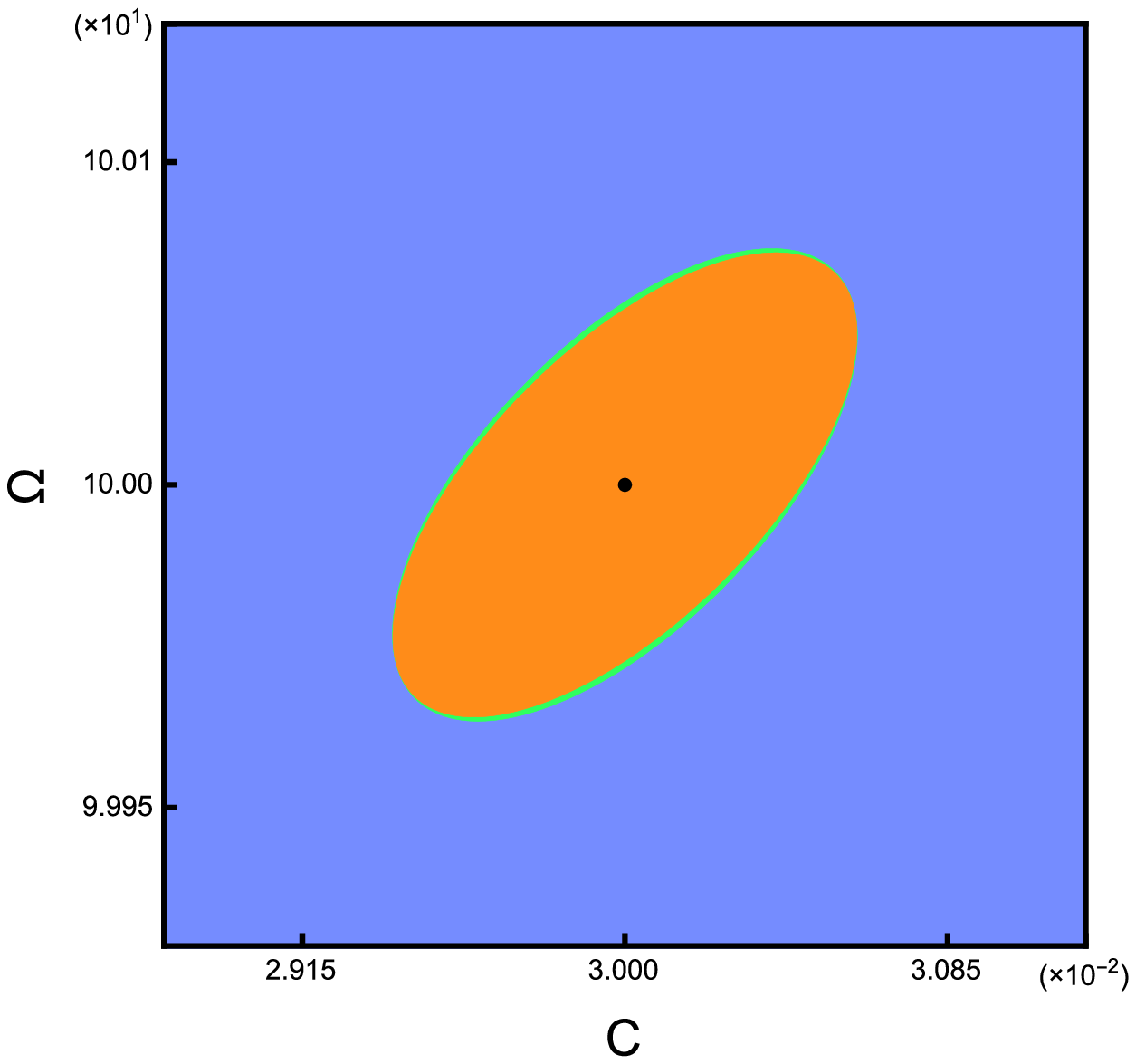,width=1.8in}
\epsfig{figure=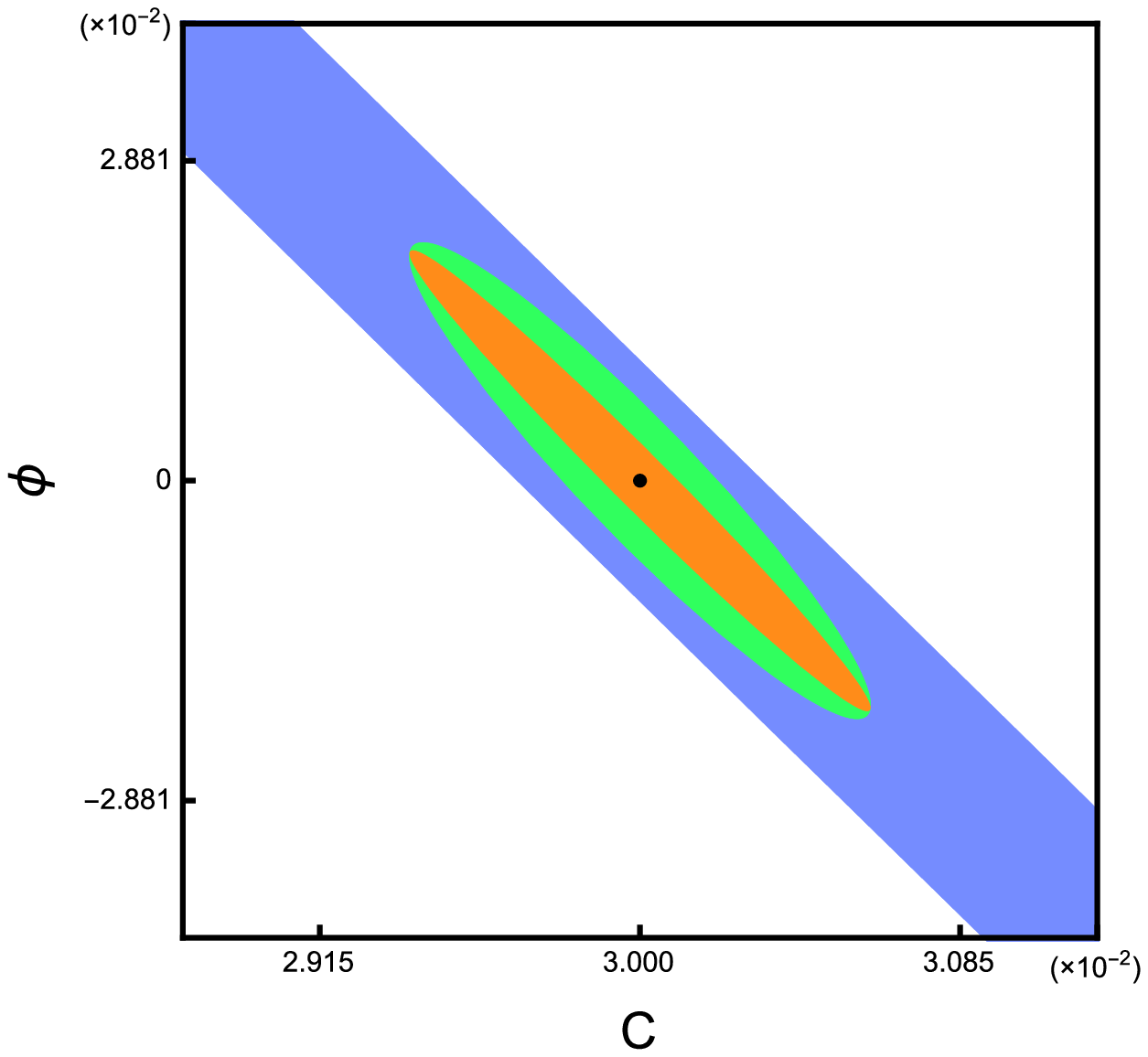,width=1.8in}
\epsfig{figure=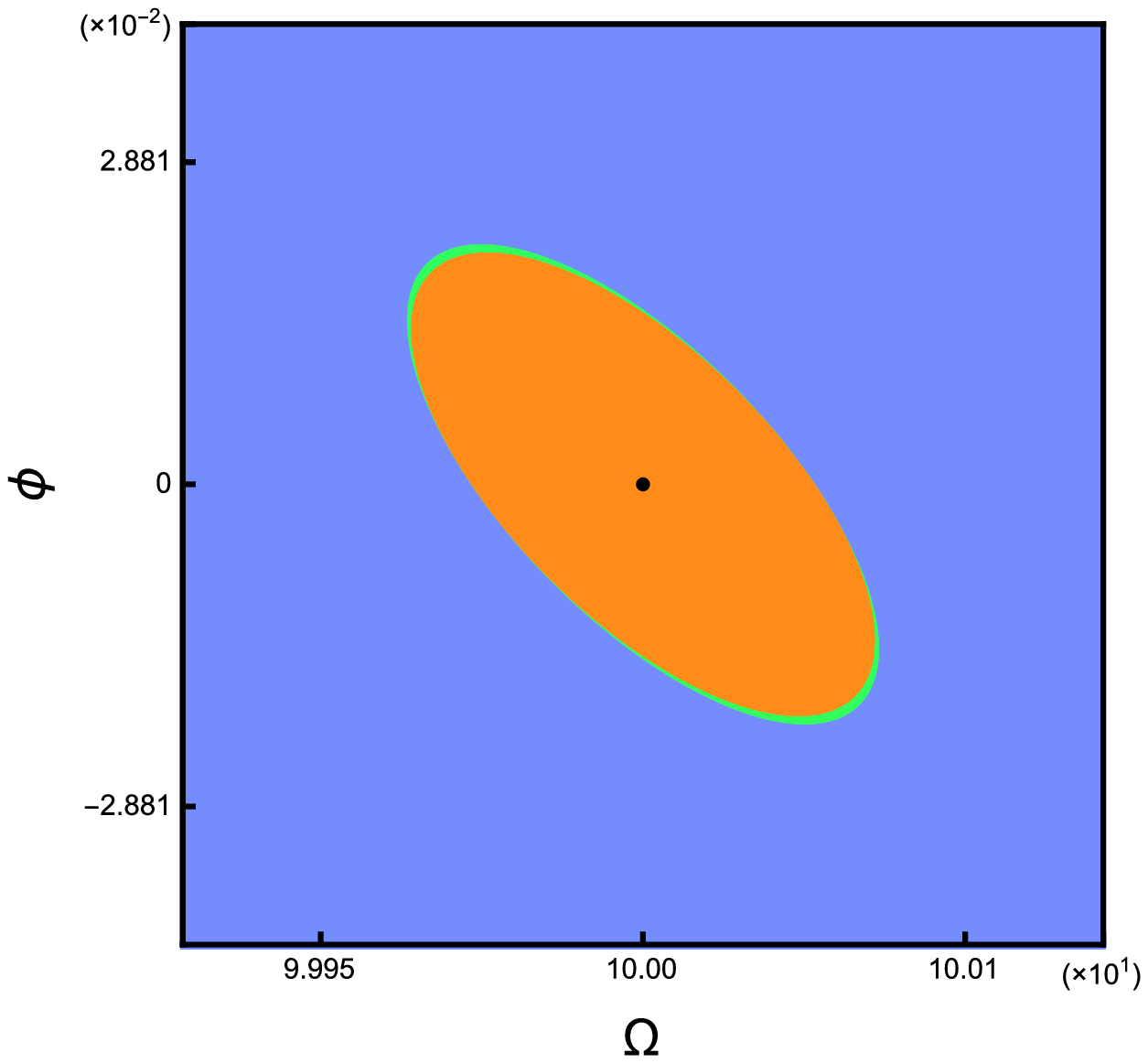,width=1.8in}
\caption{Here we show the sensitivity of the parameters entering in template V --Eq.~\eqref{res}-- for the three observables. In green we plot the constraints from CMB+GC, in light blue from CMB+WL and orange is the sum of the three observables.}
\label{fig:model3}
\end{figure}

In Fig.~\ref{fig:model5} we plot the marginalised confidence regions for Template I, given in Eq.~\eqref{temp1}, modeling a localised oscillatory feature, stemming from time dependence of the sound speed. As can be seen from the plots and Tab.~\ref{tab:model5-forecasts}, for this parametrisation of features, for low frequencies WL will be able to constrain better the parameters of the model. The gain is about $33\%$ with respect to the GC probe; however when we increase the frequency, GC probes better than WL the parameters of the model. 
The reason lies on the shape of the features and such a behaviour is seen in all of the templates: highly oscillatory profiles are better probed by GC. This is due to the integrals of the convergence power spectrum for WL, which are not able to capture fast oscillations in the matter power spectrum. 
However, combining all the three probes we are able to reach a sensitivity of about $5\%$ for the parameters of the model. 

In Fig.~\ref{fig:model2} we plot the marginalised confidence regions for Template II, shown in Eq.~\eqref{step}. For the fiducial parameters reflecting a sharp feature (Eq.~\eqref{step-fid-s}) this template is modeling a step in the inflationary potential. In this case, the high frequency oscillatory profile of the power spectrum makes WL a less efficient probe than GC, for the same reason explained above. The precision is set by CMB+GC to  $3\%$ for the amplitude, $2\%$ for $k_d$ and $0.03\%$ for the frequency.

In the regime of a mild feature\footnote{Note that, as explained in the text, in this case the template does not accurately describe a mild feature in the Hubble parameters.} corresponding to fiducial values written in Eq.~\eqref{step-fid-m}, as can be seen from the plots and Tab.~\ref{tab:model2-forecasts}, both GC and WL are comparable in size, i.e. they are both able to constrain the parameters with the same accuracy. In fact, adding WL the improvement is of about $20\%$ for the amplitude, $15\%$ for the position $k_d$, and $3\%$ for the frequency $k_f$. Therefore, combining all three observables the errors reduce to $15\%$ for the amplitude $C$, $0.2\%$ for the frequency $k_f$ and about $17\%$ for the position $k_d$, relative to their fiducial values. Here the situation is different, this class of models have a bump in the power spectrum of about $30\%$ of the total signal and they do not admit a highly oscillatory behaviour. Hence the LSS observables can equally constrain the parameters of the model.

In Fig.~\ref{fig:model4} we plot the full marginalised contours for Template III, given in Eq.~\eqref{bump}, modeling primordial features stemming from particle production during inflation. As can be seen in Tab.~\ref{tab:model4-forecasts}, for this particular case, WL considerably improves the forecasted errorbars especially for the high $k$ feature, for which we get a $40\%$ better error on both parameters compared to CMB+GC. This is easy to understand by just looking at the profile of the feature: for large values of $k_d$ the signal is peaked at scales way beyond the limit of GC observables; on the other hand, WL is also sensitive to small scale perturbations and hence it will be a better probe to test a feature with large $k_d$. For the middle and low $k_d$ fiducial values, the errors improve by $30\%$ and $20\%$, respectively. Combining all the three probes, the relative error for both parameters of the feature is of the order of $30\%$ for the low $k_d$, while it improves to $20\%$ for the higher multipoles. This relatively low precision is due to the fact that we are forced to have a very small amplitude in order to respect the Planck accuracy.

In Fig.~\ref{fig:model1}, we show the marginalised confidence regions for the template of Eq.~\eqref{sharp}, modeling a sharp feature.
As can be seen from the plots and Tab.~\ref{tab:model1-forecasts}, WL is able to better constrain the parameters only if the oscillatory behavior of feature is mild. 
This is especially the case for $k_f = 0.1$/Mpc,  where WL improves the errors of about $60\%$ for the amplitude, $40\%$ for the frequency and $80\%$ for the phase.  For the middle frequency the difference is only about $30\%$. On the other hand, for the highest frequency, $k_f = 0.004$/Mpc, CMB+GC is a better probe.

In Fig.~\ref{fig:model3}, we present the marginalised confidence regions corresponding to the template of Eq.~\eqref{res}, modeling a resonant feature in the inflationary potential. As can be seen from the plots and Tab.~\ref{tab:model3-forecasts}, WL and GC have the same sensitivity, only for the case with the lowest frequency $\Omega=5$, for which CMB+WL improves the errors on all parameters by $10\%$ compared to CMB+GC. For the other two cases, GC is a better probe and will capture the feature in the galaxy power spectrum  at least one order of magnitude better than WL alone. As previously argued, the convergence power spectrum from WL is less sensitive to highly oscillatory profiles.  

In summary, for oscillatory features with low frequencies, CMB+WL gives tighter constraints; the situation is inverted, i.e. CMB+GC is a better probe, when high values of the frequencies are chosen. For instance, this is the case for Template I, IV and V when we chose high values of the frequency. In Template II the parameters are chosen such that few oscillations appear on the power spectrum but with a large amplitude (up to $20\%$ of the total signal). These oscillations appear only at $k<0.01$ 1/Mpc which is well within the range of observation for both GC and WL and this is the reason why both GC ad WL probe equally the parameters of this model. Finally, Template III consists of a bump (without oscillations); in this case if the bump is centered at high $k$ then CMB+WL is able to give more stringent results with respect to CMB+GC because it has access to more modes.

\setcounter{equation}{0} 
\section{Concluding remarks} \label{s5:conclusions}

A possible hint of new high energy physics lies in the primordial CMB power spectrum. The Planck and WMAP data marginally support the presence of scale dependent features in the power spectrum which represent deviations from the $\Lambda$CDM predicted curve at several multipoles.  These features can be studied with templates that find motivation in models of inflation with different underlying UV mechanisms.

As usual in physics, when a signal appears in a data set, one needs to collect the maximum amount of data from several channels in order to increase the statistical significance of the claim. The observables dealt with in cosmology are early and late time spectra which include two- and three-point functions of temperature fluctuations at different redshifts, convoluted with several late time processes. These are the analogs of, say, scattering amplitudes of collisions in particle physics.
Moreover, the next generation of cosmological data is expected in the following few decades from LSS surveys like Euclid and LSST. 

Combined searches of primordial $n$-point functions, as well as primordial power spectrum/galaxy clustering power spectrum, have already been proposed in the literature to constrain the presence of inflationary features. In this work we have traced the primordial features in the late time observables of galaxy clustering and weak lensing power spectra. We examined the efficiency of an LSST-like survey to further constrain the parameters involved in the templates, providing a different channel where these features can be observed. 

We have focused on five templates motivated by well studied UV scenarios: variations of the sound speed of curvature perturbations (introduced, for instance, by the presence of heavy degrees of freedom coupled to inflation), features in the inflaton potential, features in the internal space of multifield models, particle production during inflation and axion inflation.  
We saw that LSS adds a large amount of information to cosmological observables since the corresponding surveys will be able to probe a much larger number of modes, at the cost, however, of increasing foreground uncertainties. We demonstrated that WL and GC are competitive probes and their addition to CMB significantly improves the forecasts. As cosmology enters the precision era, we should aim at the most stringent constraints that can be placed upon new physics and indeed, for some cases, WL data on
top of CMB+GC improve at most of about $80\%$. 

Specifically, we found that the forthcoming combined CMB+GC+WL data set will be able to increase the resolution on the amplitude, position, width, frequency and phase of the features in the primordial power spectrum. In the case of oscillatory profiles, the improvement is significant only for features with low frequency.

\subsection*{Acknowledgements}
We wish to thank Ana Ach\'ucarro, Xingang Chen, Eiichiro Komatsu and Martin Kunz for useful discussions and comments, and especially Vinicius Miranda, whose extensive and critical comments on the previous drafts of this work helped improve our paper. GAP acknowledges support from the Fondecyt Regular project number 1171811. DS acknowledges financial support from the Fondecyt project number 11140496. SS is supported by the Fondecyt project number 3160299.

\end{document}